\documentclass[12pt,a4paper]{article}
\usepackage{epsfig}
\newcommand{\be}{\begin{equation}}
\newcommand{\ee}{\end{equation}}
\newcommand{\bea}{\begin{eqnarray}}
\newcommand{\eea}{\end{eqnarray}}

\begin{document}
\begin{center}
{\hbox to\hsize{\hfill KEK-TH-960}}

\vspace{4\baselineskip}

\textbf{\Large SO(10) Group Theory for the Unified Model Building}\par
\bigskip
\vspace{2\baselineskip}
\textbf{\large Takeshi Fukuyama$^{\dagger}$
\footnote{E-Mail: fukuyama@se.ritsumei.ac.jp}},  
\textbf{\large Amon Ilakovac$^{\ddagger}$
\footnote{E-Mail: ailakov@rosalind.phy.hr}},   
\textbf{\large Tatsuru Kikuchi$^{\dagger}$
\footnote{E-Mail: rp009979@se.ritsumei.ac.jp}}, \\  
\textbf{\large Stjepan Meljanac$^{\star}$
\footnote{E-Mail: meljanac@irb.hr}}  
\textbf{\large and Nobuchika Okada$^{\diamond}$
\footnote{E-Mail: nobuchika.okada@kek.jp}} \\
\vspace{1\baselineskip}
\medskip{\it
$^\dagger$Department of Physics, Ritsumeikan University, \\
Kusatsu, Shiga, 525-8577 Japan\\
$^\ddagger$Department of Physics,
University of Zagreb,
P.O. Box 331,
Bijeni\v cka cesta 32,
HR-10002 Zagreb, Croatia\\
$^\star$Institut Rudjer Bo\v skovi\'c,
Bijeni\v cka cesta 54,
P.O. Box 180,
HR-10002 Zagreb, Croatia\\
$^\diamond$Theory Group, KEK, Oho 1-1, Tsukuba, Ibaraki 305-0801, Japan}\\
\vspace{2\baselineskip}
\textbf{\today}
\end{center}
\vspace{1\baselineskip}
\begin{abstract}

The complete tables of Clebsch-Gordan (CG) coefficients for a wide class
of $SO(10)$ SUSY grand unified theories (GUTs) are given. Explicit  
expressions of states of all corresponding multiplets  under standard model gauge    
group $G_{321} = SU(3)_{C} \times SU(2)_{L} \times U(1)_{Y}$, necessary for
evaluation of the CG coefficients are presented. The SUSY $SO(10)$ GUT model 
considered here includes most of the Higgs irreducible representations 
usually used in the literature:
${\bf 10}$, ${\bf 45}$, ${\bf 54}$, ${\bf 120}$, 
${\bf 126}$, ${\bf \overline{126}}$ and ${\bf 210}$.
Mass matrices of all $G_{321}$ multiplets are found for the most general superpotential.
These results are indispensable for the precision calculations of the gauge 
couplings unification and proton decay, etc.  
\end{abstract}
\vspace{\baselineskip}
%
%
\vspace{\baselineskip}
\newpage
\section{Introduction}
A particularly attractive idea for the physics beyond the standard model (SM) 
is the possible appearance of grand unified theories (GUTs) \cite{gut}.  
The idea of GUTs bears several profound features.
Perhaps the most obvious one is that GUTs have the potential to unify 
the diverse set of particle representations and parameters found in 
the SM into a single, comprehensive, and hopefully predictive framework.  
For example, through the GUT symmetry one might hope to explain 
the quantum numbers of the fermion spectrum, or even the origins of 
fermion mass.  Moreover, by unifying all $U(1)$ generators within a 
non-Abelian theory, GUTs would also provide an explanation for the 
quantization of electric charge.  
By combining GUTs with supersymmetry (SUSY), 
we hope to unify the attractive features of GUTs simultaneously 
with those of SUSY into a single theory, SUSY GUTs \cite{susygut}.  
The apparent gauge couplings unification of the 
minimal supersymmetric standard model (MSSM) is strong circumstantial 
evidence in favor of the emergence of a SUSY GUT near 
$M_{G} \simeq 2 \times 10^{16}$ GeV \cite{susygut2}.  

While there are {\it a priori} many choices for such possible groups, 
the list can be narrowed down by requiring groups of rank $\geq 4$ that have 
complex representations.  The smallest groups satisfying these requirements are
$SU(5)$, $SU(6)$, $SO(10)$, and $E_6$.  
Amongst these choices, $SO(10)$ is particularly attractive \cite{so10},  
because $SO(10)$ is the smallest simple Lie group 
for which a single anomaly-free irreducible representation (irrep) 
(namely the spinor ${\bf 16}$ representation) 
can accommodate the entire SM fermion content of each generation.  

Once we fix $SO(10)$ as the gauge group, we have also many choices 
of the Higgs fields though they are limited by the gauge symmetry.  
The Higgs fields play an essential role in the spontaneous symmetry breaking of the $SO(10)$ 
gauge group and as a source of the observed fermion masses.  
The $SO(10)$ gauge group must be broken  down to the standard model gauge group
$G_{321} = SU(3)_{C} \times SU(2)_{L} \times U(1)_{Y}$ gauge group, 
and each $SO(10)$ irrep has to be decomposed into $G_{321}$ multiplets.  
In this paper, we make an explicit construction of the  states of these $G_{321}$ multiplets.
Using these states, we calculate the CG coefficients appearing in 
the mass matrices for the states belonging to $G_{321}$ irreps and corresponding mass matrices
in a wide class of the $SO(10)$ models. 
The purpose of the present
paper is to give detailed structures of the $SO(10)$ GUTs based on a
general model as far as has been possible, and to serve a wide
range of unified model builders.

The paper is organized as follows.  
In Sec. \ref{class}, 
we give a class of SUSY $SO(10)$ GUTs and give an explicit form 
of the most general superpotential.  
In such a superpotential, we postulate a renormalizability 
in order to keep the predictability 
\cite{babu-mohapatra, f-o, goh-mohapatra-ng}.  
However, the result developed here is also applicable to 
the non-renormalizable models 
\cite{hall-rattazzi-sarid, anderson-raby-dimopoulos-hall-starkman, 
hall-raby, blazek-raby-tobe, albright-barr, shafi-tavartkilade, babu-pati-wilczek}.  
The symmetry breakings are considered in Sec. \ref{breakings}.  
Sec. \ref{states} is devoted to present explicit forms of the states 
in the $G_{321}$ multiplets for all $SO(10)$ irreps.  
This is the central part of the present paper.
Using these tables, we give in Sec. \ref{mass} the mass matrices with the CG coefficients for
a class of SUSY $SO(10)$ GUTs,
together with suitable tests and consistency checks for them.
In Sec. \ref{fermion}, we consider the quark and lepton mass matrices
in general SUSY SO(10) models.  Sec. \ref{conclusion} is devoted to
conclusion.  
We list the decompositions of each $SO(10)$ irreps 
under $G_{321}$ subgroup in Appendix \ref{decomp}.  
In Appendix \ref{CGC}, we present the complete list of the CG coefficients 
for the $G_{321}$ multiplets for all $SO(10)$ irreps.  
\section{A Class of SUSY SO(10) GUTs}\label{class}
In this section, we consider a class of renormalizable SUSY $SO(10)$ models.  
They include three families of matter fields $\Psi_{i}\, (i = 1,2,3)$ 
transforming as 16 dimensional fundamental spinor representation, ${\bf 16}$,
gauge fields contained in the adjoint representation, and set of $SO(10)$
multiplets of Higgs fields, enabling most general Yukawa couplings. 
The most general Yukawa couplings follow from decomposition of 
${\bf 16 \times 16 = 10 + 120 + 126}$, {\it i.e.} they include the Higgs fields in 
$H = {\bf 10}$, $D = {\bf 120}$, $\overline{\Delta} = {\bf \overline{126}}$ 
irreps, respectively. 
Furthermore, to consider as general case of the symmetry breaking 
of $SO(10)$ to the standard model 
gauge group $G_{321}$ as possible, we add several Higgs fields containing $G_{321}$ singlets.  
They are $A = {\bf 45}$, $\Delta = {\bf 126}$, $\Phi = {\bf 210}$ 
and $E = {\bf 54}$ irreps.  
Of course, that is not the most general case.
However, this set of Higgs fields is quite rich and gives rise to 
several realistic SUSY $SO(10)$ models.  
Our aim is to give a systematic method for treatment of models 
with complicated Higgs sectors.  
This is a generalization of the method proposed in Refs. \cite{he}, \cite{lee} and \cite{fuku}.  
We shall assume that the SUSY is preserved 
so that we consider the breaking of SUSY $SO(10)$ to the MSSM.  

Then the Yukawa couplings are 
\be
W_{Y} = Y_{10}^{ij} \, \Psi_{i} H \Psi_{j} + Y_{120}^{ij} \, \Psi_{i} D \Psi_{j} + 
Y_{126}^{ij} \,\Psi_{i} \overline{\Delta} \Psi_{j},
\label{Y}
\ee
where $i,j = 1,2,3$ denote the generation indices.  
Note that $H$ is a fundamental $SO(10)$ irrep, and $A$, $D$, $\Phi$ and 
$\Delta+ \overline{\Delta}$ are antisymmetric tensors of rank 2, 3, 4, and 5, 
respectively.  $E$ is a symmetric traceless tensor of rank 2.  

The most general Higgs superpotential is given by:
\bea
W &=& \frac{1}{2} m_{1} \Phi^2 + m_{2} \overline{\Delta} \Delta + \frac{1}{2} m_{3} H^2 
\nonumber\\
&+& \frac{1}{2} m_{4} A^2 + \frac{1}{2} m_{5} E^2 + \frac{1}{2} m_{6} D^2 
\nonumber\\
&+& \lambda_{1} \Phi^3 + \lambda_{2} \Phi \overline{\Delta} \Delta
+ \left(\lambda_3 \Delta + \lambda_4 \overline{\Delta} \right) H \Phi
\nonumber\\
&+&
\lambda_{5} A^2 \Phi -i \lambda_{6} A \overline{\Delta} \Delta
+ \frac{\lambda_7}{120} \varepsilon A \Phi^2
\nonumber\\
&+& E \left( \lambda_{8} E^2 + \lambda_{9} A^2 + \lambda_{10} \Phi^2 
+ \lambda_{11} \Delta^2 + \lambda_{12} \overline{\Delta}^2 + \lambda_{13} H^2 
\right)
\nonumber\\
&+& D^2 \left( \lambda_{14} E + \lambda_{15} \Phi \right)
\nonumber\\
&+& D \left\{ \lambda_{16} H A + \lambda_{17} H \Phi + \left( 
\lambda_{18} \Delta + \lambda_{19} \overline{\Delta} \right) A 
+ \left( \lambda_{20} \Delta + \lambda_{21} \overline{\Delta} \right) \Phi  
\right\},  
\label{potential}
\eea
where $SO(10)$ invariants are defined in the fundamental $SO(10)$ basis 
$1^{\prime},2^{\prime}, \cdots, 9^{\prime},0^{\prime}$ and in the $Y$ diagonal 
basis $1, 2, \cdots, 9, 0$ (which we will define in the next section) as follows:
\begin{eqnarray*}
\Phi^2
&\equiv& 
\Phi_{a'b'c'd'} \Phi_{a'b'c'd'}
\ =\ \Phi_{abcd}\Phi_{\overline{a}\overline{b}\overline{c}\overline{d}}
\nonumber\\
\overline{\Delta} \Delta 
&\equiv&
\overline{\Delta}_{a'b'c'd'e'} \Delta_{a'b'c'd'e'}
\ =\ \overline{\Delta}_{abcde} 
 \Delta_{\overline{a}\overline{b}\overline{c}\overline{d}\overline{e}}
\nonumber\\
H^2 
&\equiv&
H_{a'} H_{a'}
\ =\ H_{a} H_{\overline{a}},
\nonumber\\
A^2
&\equiv&
A_{a'b'} A_{a'b'}
\ =\ A_{ab} A_{\overline{a}\overline{b}},
\nonumber\\
E^2
&\equiv&
E_{a'b'} E_{a'b'}
\ =\ E_{ab} E_{\overline{a}\overline{b}},
\nonumber\\
D^2
&\equiv&
D_{a'b'c'} D_{a'b'c'}
\ =\ D_{abc} D_{\overline{a}\overline{b}\overline{c}},
\end{eqnarray*}
\bea 
\label{invariants}
\Phi^3
&\equiv& \Phi_{a'b'c'd'} \Phi_{a'b'e'f'} \Phi_{c'd'e'f'} 
\ =\ \Phi_{\overline{a}\overline{b}\overline{c}\overline{d}}
\Phi_{abef} \Phi_{cd\overline{e}\overline{f}},
\nonumber\\
\Phi \overline{\Delta} \Delta
&\equiv& \Phi_{a'b'c'd'} \overline{\Delta}_{a'b'e'f'g'} \Delta_{c'd'e'f'g'}
\ =\ \Phi_{\overline{a}\overline{b}\overline{c}\overline{d}} 
\overline{\Delta}_{abefg} \Delta_{cd\overline{e}\overline{f}\overline{g}},
\nonumber\\
\Delta H \Phi
&\equiv& \Delta_{a'b'c'd'e'} H_{a'}\Phi_{b'c'd'e'}
\ =\ \Delta_{\overline{a}\overline{b}\overline{c}\overline{d}\overline{e}} 
H_{a}\Phi_{bcde},
\nonumber\\
\overline{\Delta} H \Phi
&\equiv& \overline{\Delta}_{a'b'c'd'e'} H_{a'}\Phi_{b'c'd'e'}
\ =\ \overline{\Delta}_{\overline{a}\overline{b}\overline{c}\overline{d}\overline{e}} 
H_{a}\Phi_{bcde},
\nonumber\\
A^2 \Phi
&\equiv& A_{a'b'}A_{c'd'}\Phi_{a'b'c'd'}
\ =\ A_{\overline{a}\overline{b}}A_{cd}\Phi_{ab\overline{c}\overline{d}},
\nonumber\\
-iA \overline{\Delta} \Delta
&\equiv& -iA_{a'b'} \overline{\Delta}_{a'c'd'e'f'} \Delta_{b'c'd'e'f'}
\ =\ -iA_{\overline{a}\overline{b}} \overline{\Delta}_{acdef} 
\Delta_{b\overline{c}\overline{d}\overline{e}\overline{f}},
\nonumber\\
\frac{1}{120} \varepsilon A \Phi^2
&\equiv& \frac{1}{120}
\varepsilon_{a'_1 a'_2 a'_3 a'_4 a'_5 a'_6 a'_7 a'_8 a'_9 a'_0}
A_{a'_1 a'_2} \Phi_{a'_3 a'_4 a'_5 a'_6} \Phi_{a'_7 a'_8 a'_9 a'_0}
\nonumber\\
&=& \frac{1}{120}
\varepsilon_{\overline{a}_1 \overline{a}_2 \overline{a}_3 \overline{a}_4 \overline{a}_5 
\overline{a}_6 \overline{a}_7 \overline{a}_8 \overline{a}_9 \overline{a}_0}
A_{a_1 a_2} \Phi_{a_3 a_4 a_5 a_6} \Phi_{a_7 a_8 a_9 a_0},
\nonumber\\
E^3
&\equiv& E_{a'b'}E_{a'c'}E_{b'c'}
\ =\ E_{\overline{a}\overline{b}}E_{ac}E_{b\overline{c}},
\nonumber\\
EA^2
&\equiv& E_{a'b'}A_{a'c'}A_{b'c'}
\ =\ E_{\overline{a}\overline{b}}A_{ac}A_{b\overline{c}},
\nonumber\\
E\Phi^2
&\equiv& E_{a'b'}\Phi_{a'c'd'e'}\Phi_{b'c'd'e'}
\ =\ E_{\overline{a}\overline{b}}\Phi_{acde}\Phi_{b\overline{c}\overline{d}\overline{e}},
\nonumber\\
E\Delta^2
&\equiv& E_{a'b'}\Delta_{a'c'd'e'f'}\Delta_{b'c'd'e'f'}
\ =\ E_{\overline{a}\overline{b}} \Delta_{acdef} 
\Delta_{b\overline{c}\overline{d}\overline{e}\overline{f}},
\nonumber\\
E\overline{\Delta}^2
&\equiv& E_{a'b'}\overline{\Delta}_{a'c'd'e'f'}\overline{\Delta}_{b'c'd'e'f'}
\ =\ E_{\overline{a}\overline{b}} \overline{\Delta}_{acdef} 
\overline{\Delta}_{b\overline{c}\overline{d}\overline{e}\overline{f}},
\nonumber\\
EH^2
&\equiv& E_{a'b'}H_{a'}H_{b'}
\ =\ E_{\overline{a}\overline{b}}H_{a}H_{b},
\nonumber\\
ED^2
&\equiv& E_{a'b'}D_{a'c'd'}D_{b'c'd'}
\ =\ E_{\overline{a}\overline{b}}D_{acd}D_{b\overline{c}\overline{d}},
\nonumber\\
D^2\Phi
&\equiv& D_{a'b'c'}D_{a'd'e'}\Phi_{b'c'd'e'}
\ =\ D_{\overline{a}\overline{b}\overline{c}}D_{ade}\Phi_{bc\overline{d}\overline{e}},
\nonumber\\
DHA
&\equiv& D_{a'b'c'}H_{a'}A_{b'c'}
\ =\ D_{\overline{a}\overline{b}\overline{c}}H_{a}A_{bc},
\nonumber\\
DH\Phi
&\equiv& D_{a'b'c'}H_{d'}\Phi_{a'b'c'd'}
\ =\ D_{\overline{a}\overline{b}\overline{c}}H_{d}\Phi_{abc\overline{d}},
\nonumber\\
D\Delta A
&\equiv& D_{a'b'c'}\Delta_{a'b'c'd'e'} A_{d'e'}
\ =\ D_{\overline{a}\overline{b}\overline{c}}\Delta_{abcde} A_{\overline{d}\overline{e}},
\nonumber\\
D\overline{\Delta}A
&\equiv& D_{a'b'c'}\overline{\Delta}_{a'b'c'd'e'}A_{d'e'}
\ =\ D_{\overline{a}\overline{b}\overline{c}} \overline{\Delta}_{abcde} 
A_{\overline{d}\overline{e}},
\nonumber\\
D\Delta\Phi
&\equiv& D_{a'b'c'}\Delta_{a'b'd'e'f'}\Phi_{c'd'e'f'}
\ =\ D_{\overline{a}\overline{b}\overline{c}} \Delta_{abdef} 
\Phi_{c\overline{d}\overline{e}\overline{f}},
\nonumber\\
D\overline{\Delta}\Phi
&\equiv& D_{a'b'c'}\overline{\Delta}_{a'b'd'e'f'}\Phi_{c'd'e'f'}
\ =\ D_{\overline{a}\overline{b}\overline{c}} \overline{\Delta}_{abdef} 
\Phi_{c\overline{d}\overline{e}\overline{f}}.
\eea
Here $a^{\prime}, b^{\prime}, c^{\prime}, \cdots$ 
$\left(a,b,c, \cdots \right)$ run over all the $SO(10)$ vector 
($Y$ diagonal) indices 
and $\varepsilon$ is a totally antisymmetric $SO(10)$ invariant tensor with 
\be
\varepsilon_{1^{\prime} 2^{\prime} 3^{\prime} 4^{\prime} 5^{\prime}
6^{\prime} 7^{\prime} 8^{\prime} 9^{\prime} 0^{\prime}} =
i \varepsilon_{1234567890} = 1.
\ee
\section{Symmetry Breaking}\label{breakings}
Here we first introduce $Y$ diagonal basis (see also Ref. \cite{sato}): 
$1 = 1^{\prime} + 2^{\prime} i$, $2 = 1^{\prime} - 2^{\prime} i$, 
$3 = 3^{\prime} + 4^{\prime} i$, $4 = 3^{\prime} - 4^{\prime} i$, 
$5 = 5^{\prime} + 6^{\prime} i$, $6 = 5^{\prime} - 6^{\prime} i$, 
$7 = 7^{\prime} + 8^{\prime} i$, $8 = 7^{\prime} - 8^{\prime} i$, 
$9 = 9^{\prime} + 0^{\prime} i$, $0 = 9^{\prime} - 0^{\prime} i$, 
up to the normalization factor 
$\frac{1}{\sqrt{2}}$.  It is more convenient since 
$\left(1, 3, 5, 7, 9 \right)$ 
transforms as ${\bf 5}$-plet and 
$\left(2, 4, 6, 8, 0 \right)$ 
transforms as ${\bf \overline{5}}$-plet under $SU(5) \times U(1)_{X}$
(for that reason $Y$ diagonal basis could also be called $SU(5)$ basis).  
Consequently, 
$\left(1, 3 \right)$ and 
$\left(2, 4 \right)$ are $SU(2)_{L}$ doublets with definite 
hypercharges $Y=\frac{1}{2}$ and $Y=-\frac{1}{2}$, respectively.  
Similarly, 
$\left(5, 7, 9 \right)$ and 
$\left(6, 8, 0 \right)$ transform under $SU(3)_{C}$ as 
${\bf 3}$ and ${\bf \overline{3}}$ with definite hypercharges 
$Y=-\frac{1}{3}$ and $Y=\frac{1}{3}$, respectively.  
Note that under the complex conjugation ($c.c.$), 
$\overline{1} = 2$, 
$\overline{3} = 4$, 
$\overline{5} = 6$, 
$\overline{7} = 8$, 
$\overline{9} = 0$, and vice versa.  
The $SO(10)$ invariants are build in such a way that an index $a$ 
is contracted (summed) with the corresponding $c.c.$ index $\overline{a}$, 
for example, $T_{\cdots a \cdots} T_{\cdots \overline{a} \cdots}$. 
 
The basis in $A = {\bf 45}$, $D = {\bf 120}$, $\Phi = {\bf 210}$ and 
$\Delta + \overline{\Delta} = {\bf 126 + \overline{126}}$ dimensional spaces 
are defined by totally antisymmetric (unit) tensors 
$(a^{\prime} b^{\prime})$, 
$(a^{\prime} b^{\prime} c^{\prime})$, 
$(a^{\prime} b^{\prime} c^{\prime} d^{\prime})$ and 
$(a^{\prime} b^{\prime} c^{\prime} d^{\prime} e^{\prime})$, 
respectively,  
and similarly in $a$, $b$, $c$, $d$, $e$ indices in $Y$ diagonal basis. 
The states of the $\Delta$ and $\overline{\Delta}$
have additional properties,
\bea
i \varepsilon_{\bar{a}_1\bar{a}_2\bar{a}_3\bar{a}_4\bar{a}_5
 \bar{a}_6\bar{a}_7\bar{a}_8\bar{a}_9\bar{a}_{10}}
 \overline{\Delta}_{a_6a_7a_8a_9a_{10}} &=&
 \overline{\Delta}_{\bar{a}_1\bar{a}_2\bar{a}_3\bar{a}_4\bar{a}_5},
\nonumber \\
i \varepsilon_{\bar{a}_1\bar{a}_2\bar{a}_3\bar{a}_4\bar{a}_5
 \bar{a}_6\bar{a}_7\bar{a}_8\bar{a}_9\bar{a}_{10}}
 \Delta_{a_6a_7a_8a_9a_{10}} &=&
 -\Delta_{\bar{a}_1\bar{a}_2\bar{a}_3\bar{a}_4\bar{a}_5},
\eea
that allow one to project out the $\Delta$ and $\overline{\Delta}$ states, respectively,
from the $256$ antisymmetric states $(abcde)$.
The explicit expressions for antisymmetric tensors are, for example, 
\bea
(ab) &=& ab - ba,
\nonumber\\
(abc) &=& abc +cab + bca - bac - acb - cba
\eea
{\it etc}.  
Important relations are 
\bea
(12) &=& -i (1^{\prime} 2^{\prime}),
\nonumber\\
(34) &=& -i (3^{\prime} 4^{\prime}),
\nonumber\\
(56) &=& -i (5^{\prime} 6^{\prime}),
\nonumber\\
(78) &=& -i (7^{\prime} 8^{\prime}),
\nonumber\\
(90) &=& -i (9^{\prime} 0^{\prime})
\eea
Symmetric $E = {\bf 54}$ dimensional space is spanned by traceless symmetric states
$\left\{a'b' \right\} \equiv a'b' + b'a' \,\, 
\left(a^{\prime}, b^{\prime} = 
1^{\prime}, 2^{\prime}, \cdots, 9^{\prime}, 0^{\prime} \right)$ 
and $\sum_{a^{\prime}} c_{a^{\prime}}\, \{a^{\prime} a^{\prime}\}$ 
with $\sum_{a^{\prime}} c_{a^{\prime}} \equiv 0$.  
Also, important relations are 
\bea
\left\{12 \right\} = 
   1^{\prime} 1^{\prime} 
 + 2^{\prime} 2^{\prime},
\nonumber\\
\left\{34 \right\} = 
   3^{\prime} 3^{\prime} 
 + 4^{\prime} 4^{\prime},
\nonumber\\
\left\{56 \right\} =
   5^{\prime} 5^{\prime} 
 + 6^{\prime} 6^{\prime},
\nonumber\\
\left\{78 \right\} =
   7^{\prime} 7^{\prime} 
 + 8^{\prime} 8^{\prime},
\nonumber\\
\left\{90 \right\} =
   9^{\prime} 9^{\prime} 
 + 0^{\prime} 0^{\prime}.
\eea
Now, the Higgs fields $A$, $E$, $\Delta$, $\overline{\Delta}$, and $\Phi$ 
contain 8 directions of singlets under the $G_{321}$ subgroup 
(see Appendix \ref{decomp}). The 
corresponding vacuum expectation values (VEVs) are defined by: 
\bea
\langle A \rangle &=& \sum_{i=1}^2 A_i \, \widehat{A}_i,
\\
\langle E \rangle &=& E \, \widehat{E}, 
\\
\langle \Delta \rangle &=& v_R \, \widehat{v_R}, 
\\
\langle \, \overline{\Delta} \, \rangle &=& 
\overline{v_R} \, \widehat{\overline{v_R}}, 
\\
\langle \Phi \rangle &=& \sum_{i=1}^3 \Phi_i \, \widehat{\Phi}_i,  
\eea
where unit directions $\widehat{A}_i \,\, (i=1,2)$, $\widehat{E}$, 
$\widehat{v_R}$, $\widehat{{\overline{v_R}}}$ and 
$\widehat{\Phi}_i \,\, (i=1,2,3)$ in the $Y$ diagonal basis are: 
\bea
\label{ud1}
\widehat{A}_1 
&=& \widehat{A}^{(1,1,0)}_{(1,1,3)}\ =\ \frac{i}{2}(12+34), 
\label{singlet1}
\\
\label{ud2}
\widehat{A}_2 
&=& \widehat{A}^{(1,1,0)}_{(15,1,1)}\ =\ \frac{i}{\sqrt{6}}(56+78+90), 
\\
\label{ud3}
\widehat{E} 
&=& \widehat{E}^{(1,1,0)}_{(1,1,1)}\ =\ 
\frac{1}{\sqrt{60}}\{{\it 3\times}[12+34]-{\it 2\times}[56+78+90]\}, 
\\
\label{ud4}
\widehat{v_R} 
&=& \widehat{\Delta}^{(1,1,0)}_{(\overline{10},1,3)}\ =\ 
\frac{1}{\sqrt{120}}(24680), 
\\
\label{ud5}
\widehat{\overline{v_R}} 
&=& \widehat{\overline{\Delta}}^{(1,1,0)}_{(10,1,3)}\ =\ 
\frac{1}{\sqrt{120}}(13579), 
\\
\label{ud6}
\widehat{\Phi}_1 &=& \widehat{\Phi}^{(1,1,0)}_{(1,1,1)}\ =\ 
-\frac{1}{\sqrt{24}}(1234), 
\\
\label{ud7}
\widehat{\Phi}_2 &=& \widehat{\Phi}^{(1,1,0)}_{(15,1,1)}\ =\ 
-\frac{1}{\sqrt{72}}(5678+5690+7890), 
\\
\label{ud8}
\widehat{\Phi}_3 &=& \widehat{\Phi}^{(1,1,0)}_{(15,1,3)}\ =\ 
-\frac{1}{12}([12+34][56+78+90]).  
\label{singlet8}
\eea
Here and hereafter, the upper and the lower indices indicate the 
$SU(3)_C \times SU(2)_L \times U(1)_Y$, 
$SU(4) \times SU(2)_L \times SU(2)_R$ quantum numbers, respectively 
in the case of double indices. A word about notation: the square brackets
are used for grouping of indices. This grouping of indices 
is used to emphasize the $SU(2)_L$ and $SU(3)_C$ structures within the state vectors.
The square brackets satisfy usual distributive law with respect
to summation of indices and tensor product of indices, e.g.
\bea
&&([12+34][56+78+90])
\nonumber\\ 
&&=\ (1256+1278+1290+3456+3478+3490)
\nonumber\\ 
&&=\ (1256)+(1278)+(1290)+(3456)+(3478)+(3490),
\nonumber\\
&&([1,3][5[78+90],7[56+90],9[56+78]])
\nonumber\\
&&=\ (1578+1590,1756+1790,1956+1978,
\nonumber\\
&&\qquad  3578+3590,3756+3790,3956+3978)
\nonumber\\
&&=\ (1578,1756,1956,3578,3756,3956)
\nonumber\\
&&+\ \ (1590,1790,1978,3590,3790,3978).
\eea 
Further, the numerical factors which could be misinterpreted as additional $SO(10)$ 
indices are written in italics.

The unit directions appearing in VEVs satisfy the following orthonormality relations
\bea
\widehat{A_i} \cdot \widehat{A_j} &=& 
\delta_{ij} \,\, \left(i,j = 1,2 \right),  
\nonumber\\
\widehat{E}^2 &=& 1, 
\nonumber\\
\widehat{v_R} \cdot \widehat{v_R} &=& 
\widehat{\overline{v_R}} \cdot \widehat{\overline{v_R}} = 0, 
\nonumber\\
\widehat{v_R} \cdot \widehat{\overline{v_R}} &=& 1,
\nonumber\\
\widehat{\Phi_i} \cdot \widehat{\Phi_j} &=& 
\delta_{ij} \,\, \left(i,j = 1,2,3 \right).  
\eea
Due to the D-flatness condition the absolute values of the VEVs, 
$v_R$ and ${\overline{v_R}}$ are equal, 
\be
|v_R| = |\overline{v_R}|.
\ee

The superpotential of Eq. (\ref{potential}) calculated at the VEVs in 
Eqs. (\ref{singlet1})--(\ref{singlet8}) is: 
\bea
\left< W \right> &=& 
\frac{1}{2} m_{1} \left<\Phi \right>^2 
+ m_{2} \left<\overline{\Delta} \right> \left<\Delta \right>
\nonumber\\
&+& \frac{1}{2} m_{4} \left<A \right>^2 
+ \frac{1}{2} m_{5} \left<E \right>^2 
\nonumber\\
&+& \lambda_{1} \left<\Phi \right>^3 
+ \lambda_{2} \left<\Phi \right>\left<\overline{\Delta} \right>\left<\Delta \right> 
\nonumber\\
&+&
\lambda_{5} \left<A \right>^2 \left<\Phi \right> 
-i \lambda_{6} \left<A \right> \left<\overline{\Delta} \right> \left<\Delta \right> 
+ \frac{\lambda_7}{120} \varepsilon \left<A \right>\left<\Phi \right>^2
\nonumber\\
&+& \left<E \right> \left[ \lambda_{8} \left<E \right>^2 
+ \lambda_{9} \left<A \right>^2 + \lambda_{10} \left<\Phi \right>^2 
\right]. 
\eea
Inserting the VEVs from Eqs. (\ref{singlet1})--(\ref{singlet8}), one obtains 
\bea
\left< W \right> &=& 
\frac{1}{2} m_{1} \left[\Phi_1^2 + \Phi_2^2 +\Phi_3^2 \right] 
+ m_{2} v_R \overline{v_R}
\nonumber\\
&+& \frac{1}{2} m_{4} \left(A_1^2 + A_2^2 \right) 
+ \frac{1}{2} m_{5} E^2 
\nonumber\\
&+& \lambda_{1} \left[ \Phi_2^3 \, \frac{1}{9 \sqrt{2}}
+ 3 \,\Phi_1 \Phi_3^2 \, \frac{1}{6 \sqrt{6}}
+ 3 \,\Phi_2 \Phi_3^2 \, \frac{1}{9 \sqrt{2}} \right]
\nonumber\\
&+& \lambda_{2} \left[\Phi_1 \, \frac{1}{10 \sqrt{6}}
+ \Phi_2 \, \frac{1}{10 \sqrt{2}}
+ \Phi_3 \, \frac{1}{10} \right]
v_R \overline{v_R}
\nonumber\\
&+&
\lambda_5 \left[A_1^2 \Phi_1 \, \frac{1}{\sqrt{6}}
+ A_2^2 \Phi_2 \, \frac{\sqrt{2}}{3} 
+ A_1 A_2 \Phi_3 \, \frac{2}{\sqrt{6}} \right]
\nonumber\\
&+& \lambda_6 \left[A_1 \, \left( -\frac{1}{5} \right) 
+ A_2 \, \left( -\frac{3}{5\sqrt{6}} \right) \right] 
v_R \overline{v_R} 
\nonumber\\ 
&+& \lambda_7 
\left[2\, A_2 \Phi_1 \Phi_2 \, \frac{\sqrt{2}}{5} 
+ A_2 \Phi_3^2 \, \frac{2\sqrt{2}}{5 \sqrt{3}} 
+ 2 \, A_1 \Phi_2 \Phi_3 \, \frac{\sqrt{2}}{5} \right]
\nonumber\\
&+& \lambda_{8} E^3 \, \frac{1}{2\sqrt{15}} 
+ \lambda_{9} E \left[A_1^2 \, \frac{\sqrt{3}}{2 \sqrt{5}} 
+ A_2^2 \, \left(-\frac{1}{\sqrt{15}} \right) \right] 
\nonumber\\
&+& \lambda_{10} E \left[ \Phi_1^2 \, \frac{\sqrt{3}}{2\sqrt{5}} 
+ \Phi_2^2 \, \left(-\frac{1}{\sqrt{15}} \right) 
+ \Phi_3^2 \, \frac{1}{4 \sqrt{15}} \right]. 
\eea
The VEVs are determined by the following equation: 
\be
\left\{
\frac{\partial}{\partial{\Phi_1}}, \, 
\frac{\partial}{\partial{\Phi_2}}, \, 
\frac{\partial}{\partial{\Phi_3}}, \, 
\frac{\partial}{\partial{v_R}}, \, 
\frac{\partial}{\partial{\overline{v_R}}}, \, 
\frac{\partial}{\partial{A_1}}, \, 
\frac{\partial}{\partial{A_2}}, \, 
\frac{\partial}{\partial{E}} 
\right\} \left< W \right> =0. 
\label{VEVs}
\ee
From Eq. (\ref{VEVs}), we obtain seven equations for
$\Phi_1$, $\Phi_2$, $\Phi_3$, $A_1$, $A_1$, $E$ and $v_R\overline{v_R}$.
They are:
\bea
0 &=&
m_1\Phi_1
+\frac{\lambda_1 \Phi_3^2}{2\sqrt{6}}
+\frac{\lambda_2 v_R \overline{v_R}}{10\sqrt{6}}
+\frac{\lambda_5 A_1^2}{\sqrt{6}}
+\frac{2\sqrt{2} \lambda_7 A_2 \Phi_2}{5}
+\frac{\sqrt{3} \lambda_{10} \Phi_1 E}{\sqrt{5}},
\nonumber \\
0 &=&
m_1 \Phi_2 
+\frac{\lambda_1\Phi_2^2}{3\sqrt{2}}
+\frac{\lambda_1\Phi_3^2}{3\sqrt{2}}
+\frac{\lambda_2 v_R\overline{v_R}}{10\sqrt{2}}
+\frac{\sqrt{2}}{3} \lambda_5 A_2^2
+\frac{2\sqrt{2}\lambda_7 \Phi_1 A_2}{5}
\nonumber \\
&&
+\frac{2\sqrt{2} \lambda_7 A_1 \Phi_3}{5} 
-\frac{2 \lambda_{10} \Phi_2 E}{\sqrt{15}},
\nonumber \\
0 &=& 
m_1 \Phi_3
+\frac{\lambda_1\Phi_1 \Phi_3}{\sqrt{6}}
+\frac{\sqrt{2}\lambda_1 \Phi_2 \Phi_3}{3} 
+\frac{\lambda_2 v_R\overline{v_R}}{10}
+\frac{\sqrt{2}\lambda_5 A_1 A_2}{\sqrt{3}} 
+\frac{2\sqrt{2}\lambda_7 A_1 \Phi_2}{5} 
\nonumber \\
&&
+\frac{4\sqrt{2}\lambda_7 A_2 \Phi_3}{5\sqrt{3}}
+\frac{\lambda_{10} \Phi_3 E}{2\sqrt{15}},
\nonumber \\
0 &=&
v_R\overline{v_R} 
\left[
m_2
+\frac{\lambda_2\Phi_1}{10\sqrt{6}}
+\frac{\lambda_2\Phi_2}{10\sqrt{2}}
+\frac{\lambda_2\Phi_3}{10}
-\frac{\lambda_6 A_1}{5}
-\frac{\sqrt{3}\lambda_6 A_2}{5\sqrt{2}}
\right],
\nonumber \\
0 &=&
m_4 A_1
+\frac{\sqrt{2}\lambda_5 A_1\Phi_1}{\sqrt{3}}
+\frac{\sqrt{2}\lambda_5 A_2\Phi_3}{\sqrt{3}}
-\frac{\lambda_6 v_R\overline{v_R}}{5}
+\frac{2\sqrt{2}\lambda_7\Phi_2\Phi_3}{5}
+\frac{\sqrt{3}\lambda_9 A_1 E}{\sqrt{5}},
\nonumber \\
0 &=&
m_4 A_2
+\frac{\sqrt{2}\lambda_5 A_1\Phi_3}{\sqrt{3}}
+\frac{2\sqrt{2}\lambda_5 A_2\Phi_2}{3}
-\frac{\sqrt{3}\lambda_6 v_R\overline{v_R}}{5\sqrt{2}}
+\frac{2\sqrt{2}\lambda_7\Phi_3^2}{5\sqrt{3}}
\nonumber \\
&&
+\frac{2\sqrt{2}\lambda_7 \Phi_1\Phi_2}{5}
-\frac{2\lambda_9 A_2 E}{\sqrt{15}},
\nonumber \\
0 &=&
m_5 E
+\frac{\sqrt{3}\lambda_8 E^2}{2\sqrt{5}}
+\frac{\sqrt{3}\lambda_9 A_1^2}{2\sqrt{5}}
-\frac{\lambda_9 A_2^2}{\sqrt{15}}
+\frac{\sqrt{3}\lambda_{10} \Phi_1^2}{2\sqrt{5}}
-\frac{\lambda_{10}\Phi_2^2}{\sqrt{15}}
+\frac{\lambda_{10}\Phi_3^2}{4\sqrt{15}}
\label{VEVeqs}.
\eea
If we assume $v_R\overline{v_R}\neq 0$, we obtain five quadratic equations and 
one linear equation for $\Phi_i$, $A_i$ and $E$. 
For that case there are $32$ solutions. Two of them 
correspond to $SU(5)$ symmetry and remaining $30$ solutions to $G_{321}$ standard
gauge group symmetry solutions. If we set $v_R=0$, we find six quadratic 
equations for $\Phi_1$, $\Phi_2$, $\Phi_3$, $A_1$, $A_2$ and $E$ with $64$ 
solutions with symmetry groups having rank $5$. They are isomorphic to 
$G_{3211}\equiv SU(3)_C\times SU(2)_L\times U(1)_R\times U(1)_{B-L}$.
However, there are solutions with higher symmetries. They are 
$G_{3221}$, $G_{421}$, $G_{422}$ and $G_{51}$.
\footnote{
For reader's convenience, we list the decompositions
of each representation in Appendix \ref{decomp}.  }
For general coupling 
constants $\lambda_1,\cdots,\lambda_{21}$, $m_1,\cdots,m_8$, the solutions 
with higher symmetries are specified by following relations.
Solutions with higher symmetries are characterized by:
\begin{enumerate}
\item
$SU(5) \times U(1)_{X}$ and $(SU(5) \times U(1))^{\mathrm{flipped}}$ symmetry solutions
\bea
\label{G51vac}
\left\{
\begin{array}{lll}
E  &=& v_{R} \ =\ 0,
\\
\Phi_1 & = & \frac{\varepsilon}{\sqrt{6}} \,\Phi_3, \quad
\Phi_2 \ =\ \frac{\varepsilon}{\sqrt{2}} \, \Phi_3,
A_{1} \ = \ \frac{2\varepsilon}{\sqrt{6}} A_{2},  
\end{array}
\right. 
\eea
where $\varepsilon = 1$ and $\varepsilon = -1$ correspond to the 
$SU(5)\times U(1)_{X}$ symmetric vacua and 
$(SU(5) \times U(1))^{\mathrm{flipped}}$ symmetric vacua, respectively.
\item
$SU(5)$ symmetry solutions
\bea
\label{G5vac}
\left\{
\begin{array}{lll}
E  &=& 0,
\\
\Phi_1 & = & \frac{1}{\sqrt{6}} \,\Phi_3, \quad
\Phi_2  \ =\ \frac{1}{\sqrt{2}}\, \Phi_3, \quad
A_{1}\ = \ \frac{2}{\sqrt{6}}\, A_{2}, \quad
v_R \ \neq\ 0.
\end{array}
\right.
\eea
\item
$G_{422} \equiv SU(4) \times SU(2)_{L} \times SU(2)_{R}$ symmetry solutions
\bea
\label{G422vac}
\left\{
\begin{array}{lll}
\Phi_2 & =& \Phi_3 \ =\ A_{1}\ =\ A_{2} \ =\ v_{R} \ =\ 0, 
\\
\Phi_1 & \neq& 0, \quad E \ \neq\ 0. 
\end{array}
\right.
\eea
\item
$G_{3221} \equiv SU(3)_{C} \times SU(2)_{L} \times SU(2)_{R} \times U(1)_{B-L}$ symmetry solutions
\bea
\label{G3221vac}
\left\{
\begin{array}{lll}
\Phi_3 &=& A_1 \ =\ v_R \ =\ 0,
\\
\Phi_1  &\neq& 0,  \quad \Phi_2 \ \neq\ 0,  \quad A_2 \ \neq\ 0,  \quad E \ \neq\ 0.
\end{array}
\right. 
\eea
\item
$G_{421} \equiv SU(4) \times SU(2)_{L} \times U(1)$ symmetry solutions
\bea
\label{G421vac}
\left\{
\begin{array}{lll}
\Phi_2 &=& \Phi_3\ =\ A_2\ =\ v_R\ =\ 0 
\\
\Phi_1  &\neq& 0,  \quad A_1 \ \neq\ 0, \quad  E \ \neq\ 0.
\end{array}
\right.
\eea
\item
$G_{3211} \equiv SU(3)_{C} \times SU(2)_{L} \times U(1)_{R} \times U(1)_{B-L}$ 
symmetry solutions
\bea
\label{G3211vac}
\left\{
\begin{array}{lll}
v_{R} &=&0, 
\\
\Phi_i &\neq& 0\ (i=1,2,3), \quad A_i\ \neq\ (i=1,2), \quad E\ \neq\ 0. 
\end{array}
\right. 
\eea
\end{enumerate}
The higher symmetry solutions given in Eqs. (\ref{G51vac})-(\ref{G3211vac}) lead
to the crucial consistency checks for all results in this paper.
\section{The states and Clebsch-Gordan coefficients}\label{states}
\subsection{The states in the $G_{321}$ multiplets}
In order to obtain and study the mass matrices, it is convenient to decompose 
the Higgs representations under the $G_{321}$ gauge group.  
The explicit decompositions of ${\bf 10}$, ${\bf 45}$, ${\bf 54}$, 
${\bf 120}$, ${\bf 126}$, ${\bf \overline{126}}$, and ${\bf 210}$ representations 
in $Y$ diagonal basis are presented according to the $G_{321}$ multiplets with the same 
quantum numbers which generally mix among themselves.  
The 8 singlets ${\bf(1,1},0)$ are already given in Eqs. (\ref{singlet1})--(\ref{singlet8}).  
There are $45 - 12 = 33$ would-be NG modes.  They are in the following multiplets: 
${\bf(1,1},0)$, [${\bf(3,2},-\frac{5}{6}) +c.c.$], [${\bf(1,1},1)+c.c.$], 
[${\bf(3,1},{\frac{2}{3}}) +c.c.$], and [${\bf(3,2},{\frac{1}{6}}) +c.c.$].  
The corresponding orthonormal states are listed in Table \ref{tabw} in $Y$-diagonal basis.  
The physically important modes, so called, Higgs doublets 
[${\bf(1,2},{\frac{1}{2}}) +c.c.$] and color triplets 
[${\bf(3,1},{-\frac{1}{3}}) +c.c.$] states are listed 
in Tables \ref{tabd} and \ref{tabt}, respectively.  
The remaining $G_{321}$ multiplets are listed in Tables \ref{tabo1} and \ref{tabo2}. 
There are altogether 691 states accommodated in $G_{321}$ multiplets 
(see Table \ref{tabstates}).  
There are 26 mass matrices, 5 containing NG modes, 1 containing doublets, 
1 containing color triplets, 19 containing the other modes.  
There are $5$ multiplets (33 states) with zero mass and 69 $G_{321}$ multiplets with masses 
different from zero (containing $691-33=658$ states).  Hence the mass spectrum 
contains 70 different mass eigenvalues.  
For the $SU(5)$ solutions, there are only 21 different masses 
and $G_{321}$ multiplets are grouped into multiplets transforming under the $SU(5)$ group. 
For the $G_{422}$ solutions, there are $27$ different 
mass eigenvalues 
and $G_{321}$ multiplets are grouped into multiplets transforming under the $G_{422}$ group
(see Appendix \ref{decomp}). 
The higher symmetries serve as a strong consistency check of our mass matrices and 
CG coefficients.

We point out that the main basic blocks in all $961$ states are $SU(2)_L$ irreps 
${\bf 1}$, ${\bf 2}$ and ${\bf 3}$, and $SU(3)$ irreps ${\bf 1}$, ${\bf 3}$, 
${\bf \bar{3}}$, ${\bf 6}$, ${\bf \bar{6}}$ and ${\bf 8}$:
\bea
&&
\label{su2}
\begin{tabular}{cl}
$SU(2)_L$ & \\
${\bf 1}$ & $[12+34]$, $(13)$, $(24)$, $(1234)$,\\
${\bf 2}$ & $[1,3]$, $[-3,1]$, $[2,4]$, $[-4,2]$,\\
${\bf 3}$ & $\left[14,32,\frac{12-34}{\sqrt{2}}\right]$,
      $\left\{\frac{11}{2},-\frac{33}{2},-\frac{13}{\sqrt{2}}\right\}$,
      $\left\{\frac{22}{2},-\frac{44}{2},-\frac{24}{\sqrt{2}}\right\}$
\end{tabular}
\\
&&
\label{su3}
\begin{tabular}{cl}
$SU(3)_C$ & \\
${\bf 1}$ & $[56+78+90]$, $(579)$, $(680)$, $(5678+5690+7890)$\\
${\bf 3}$ & $[5,7,9]$, $(80,06,68)$, $(5680,7806,9068)$\\
${\bf \bar{3}}$ & $[6,8,0]$, $(79,95,57)$, $(5697,7859,9075)$\\
${\bf 6}$ & $\left(580,670,689,\frac{6[90-78]}{\sqrt{2}},\frac{8[56-90]}{\sqrt{2}},
       \frac{0[78-56]}{\sqrt{2}}\right)$,\\
    & $\left\{\frac{55}{2},\frac{77}{2},\frac{99}{2},
       \frac{79}{\sqrt{2}},\frac{95}{\sqrt{2}},
       \frac{57}{\sqrt{2}}\right\}$,\\
${\bf \bar{6}}$ & $\left(679,589,570,\frac{5[09-78]}{\sqrt{2}},\frac{7[65-09]}{\sqrt{2}},
       \frac{9[87-65]}{\sqrt{2}}\right)$,\\
    & $\left\{\frac{66}{2},\frac{88}{2},\frac{00}{2},
       \frac{\{80\}}{\sqrt{2}},\frac{\{06\}}{\sqrt{2}},
       \frac{\{68\}}{\sqrt{2}}\right\}$,\\
${\bf 8}$ & $\left[58,50,70,76,96,98,\frac{56-78}{\sqrt{2}},
       \frac{56+78-{\it 2}\times 90}{\sqrt{6}}\right]$,\\
    & $\left(5890,5078,7056,7690,9678,9856,\frac{5690-7890}{\sqrt{2}},
       \frac{{\it 2}\times 5678-5690-7890}{\sqrt{6}}\right)$.
\end{tabular}
\eea 
All states can be constructed combining and antisymmetrizing or symmetrizing
the basic blocks.
The basic blocks (\ref{su2}) and (\ref{su3}) which appear only in antisymmetric tensors
are embraced by parenthesis, the  basic blocks which appear only in the 
symmetric tensors are embraced by  
curly brackets ($\{aa\}/2\ =\ aa$),  
while the basic blocks that appear both in symmetric and antisymmetric tensors 
are embraced by square brackets.

\begin{table}[p]
\caption{States in the would-be NG modes}
\label{tabw}
\begin{center}
\begin{tabular}{|c|c|c|}
\hline \hline
${\bf(1,1},{1}) +c.c.$
& $\widehat{A}^{(1,1,1)}_{(1,1,3)}$ 
& $\frac{i}{\sqrt{2}} \left(13 \right) +c.c. $
\\
& $\widehat{D}^{(1,1,1)}_{(\overline{10},1,1)}$ 
& $\frac{1}{\sqrt{6}} \left(680 \right) +c.c. $
\\
& $\widehat{\Delta}^{(1,1,1)}_{(\overline{10},1,3)}$ 
& $\frac{1}{\sqrt{240}} \left( \left[12+34 \right]  680 \right) +c.c. $
\\
& $\widehat{\Phi}^{(1,1,1)}_{(15,1,3)}$ 
& $\frac{1}{\sqrt{72}} (13  \left[56+78+90 \right]) +c.c. $
\\
\hline
${\bf(3,1},{\frac{2}{3}}) +c.c.$
& $\widehat{A}^{(3,1,\frac{2}{3})}_{(15,1,1)}$  
& $\frac{i}{\sqrt{2}} \left(80,06,68 \right) +c.c.$
\\
& $\widehat{D}^{(3,1,\frac{2}{3})}_{(6,1,3)}$ 
& $\frac{1}{\sqrt{6}} (13 \left[5,7,9 \right])+c.c.$
\\
& $\widehat{\overline{\Delta}}^{(3,1,\frac{2}{3})}_{(10,1,3)}$ 
& $\frac{1}{\sqrt{240}} \left( 13 \left[5 \left[78+90 \right], 
7 \left[56+90 \right], 9 \left[56+78 \right] \right]\right)+c.c.$
\\
& $\widehat{\Phi}^{(3,1,\frac{2}{3})}_{(15,1,1)}$ 
& $\frac{1}{\sqrt{24}} \left(5680,7806,9068 \right)+c.c. $
\\
& $\widehat{\Phi}^{(3,1,\frac{2}{3})}_{(15,1,3)}$ 
& $\frac{1}{\sqrt{48}} (\left[12+34 \right] \left[80,06,68 \right])+c.c.$
\\
\hline
${\bf(3,2},{-\frac{5}{6}}) +c.c.$  
& $\widehat{A}^{(3,2,-\frac{5}{6})}_{(6,2,2)}$  
& $\frac{i}{\sqrt{2}} \left(\left[2,4 \right] \left[5,7,9 \right] \right)+c.c.$
\\
& $\widehat{E}^{(3,2,-\frac{5}{6})}_{(6,2,2)}$ 
& $\frac{1}{\sqrt{2}} \left\{\left[2,4 \right] \left[5,7,9\right] \right\}+c.c.$
\\
& $\widehat{\Phi}^{(3,2,-\frac{5}{6})}_{(6,2,2)}$
& $\frac{1}{\sqrt{24}} \left( \left[234,124 \right] \left[5,7,9 \right] \right)+c.c. $
\\
& $\widehat{\Phi}^{(3,2,-\frac{5}{6})}_{(10,2,2)}$ 
& $\frac{1}{\sqrt{48}} \left( \left[2,4 \right] 
\left[5 \left[78+90 \right], 
7 \left[56+90 \right],9 \left[56+78 \right] \right] \right)+c.c.$
\\
\hline
${\bf(3,2},{\frac{1}{6}}) +c.c.$ 
& $\widehat{A}^{(3,2,\frac{1}{6})}_{(6,2,2)}$ 
& $\frac{i}{\sqrt{2}} (\left[1,3 \right] \left[5,7,9 \right])+c.c.$
\\
& $\widehat{E}^{(3,2,\frac{1}{6})}_{(6,2,2)}$ 
& $\frac{1}{\sqrt{2}} \left\{\left[1,3 \right] \left[5,7,9 \right] \right\}+c.c.$
\\
& $\widehat{D}^{(3,2,\frac{1}{6})}_{(15,2,2)}$ 
& $\frac{1}{\sqrt{6}} \left( \left[-4,2 \right] \left[80,06,68 \right] \right)+c.c.$
\\
& $\widehat{\Delta}^{(3,2,\frac{1}{6})}_{(15,2,2)}$
& $\frac{1}{\sqrt{240}} \Big[
\left( \left[-124,234 \right]
\left[80,06,68 \right] \right) 
+  \left(\left[-4,2 \right]
\left[5680,7806,9068 \right] \right)
\Big] +c.c. $
\\
& $\widehat{\overline{\Delta}}^{(3,2,\frac{1}{6})}_{(15,2,2)}$
& $\frac{1}{\sqrt{240}} \Big[ 
\left( \left[-124,234 \right]
\left[80,06,68 \right] \right)
-  \left(\left[-4,2 \right]
\left[5680,7806,9068 \right] \right)
\Big] +c.c. $
\\
& $\widehat{\Phi}^{(3,2,\frac{1}{6})}_{(6,2,2)}$
& $\frac{1}{\sqrt{24}} \left( \left[134,123 \right] \left[5,7,9 \right] \right)+c.c. $
\\
& $\widehat{\Phi}^{(3,2,\frac{1}{6})}_{(10,2,2)}$
& $\frac{1}{\sqrt{48}} \left( \left[1,3 \right]
\left[5 \left[78+90 \right], 
7 \left[56+90 \right], 9 \left[56+78 \right] \right] \right)+c.c.$
\\
\hline \hline
\end{tabular}
\end{center}
\end{table}
\begin{table}[p]
\caption{States in the doublets [${\bf(1,2},{\frac{1}{2}}) +c.c.$]}
\label{tabd}
\begin{center}
\begin{tabular}{|c|c|c|}
\hline \hline
${\bf(1,2},{\frac{1}{2}}) +c.c.$
& $\widehat{H}^{(1,2,\frac{1}{2})}_{(1,2,2)}$
& $\left[1,3 \right] + c.c. $
\\
& $\widehat{D}^{(1,2,\frac{1}{2})}_{(1,2,2)}$
& $\frac{1}{\sqrt{6}} \left(134,123 \right)+c.c.$
\\
& $\widehat{D}^{(1,2,\frac{1}{2})}_{(15,2,2)}$
& $\frac{1}{\sqrt{18}} \left( \left[1,3 \right]\left[56+78+90 \right] \right)+c.c.$
\\
& $\widehat{\Delta}^{(1,2,\frac{1}{2})}_{(15,2,2)}$
& $\frac{1}{\sqrt{720}} \Big[ 
   \left( \left[134,123 \right] \left[56+78+90 \right] \right)
-  \left( \left[1,3 \right] \left[5678+5690+7890 \right] \right) \Big] +c.c. $
\\
& $\widehat{\overline{\Delta}}^{(1,2,\frac{1}{2})}_{(15,2,2)}$
& $\frac{1}{\sqrt{720}} \Big[ 
   \left( \left[134,123 \right] \left[56+78+90 \right] \right)
+  \left( \left[1,3 \right] \left[5678+5690+7890 \right] \right) \Big] +c.c. $
\\
& $\widehat{\Phi}^{(1,2,\frac{1}{2})}_{(\overline{10},2,2)}$
& $\frac{1}{\sqrt{24}} \left( \left[-4,2 \right] 680 \right)+c.c. $
\\
\hline \hline
\end{tabular}
\end{center}
\end{table}
\begin{table}[p]
\caption{States in the color triplets
[${\bf(3,1},{-\frac{1}{3}}) +c.c.$]}
\label{tabt}
\begin{center}
\begin{tabular}{|c|c|c|}
\hline \hline
${\bf(3,1},{-\frac{1}{3}}) +c.c.$
& $\widehat{H}^{(3,1,-\frac{1}{3})}_{(6,1,1)}$
& $\left[5,7,9 \right]+c.c.$ 
\\
& $\widehat{D}^{(3,1,-\frac{1}{3})}_{(6,1,3)}$
& $\frac{1}{\sqrt{12}}\left( \left[12+34 \right] \left[5,7,9 \right] \right)+c.c.$ 
\\
& $\widehat{D}^{(3,1,-\frac{1}{3})}_{(10,1,1)}$
& $\frac{1}{\sqrt{12}}
\left(5\left[78+90 \right], 7\left[56+90 \right], 9\left[56+78 \right] \right)+c.c.$
\\
& $\widehat{\Delta}^{(3,1,-\frac{1}{3})}_{(6,1,1)}$
& $\frac{1}{\sqrt{240}} \Big[ \left(1234 \left[5,7,9 \right] \right)
-  \left(57890,56790,56789 \right) \Big] +c.c. $
\\
& $\widehat{\overline{\Delta}}^{(3,1,-\frac{1}{3})}_{(6,1,1)}$
& $\frac{1}{\sqrt{240}} \Big[ \left(1234 \left[5,7,9 \right] \right)
+  \left(57890,56790,56789 \right) \Big] +c.c. $
\\
& $\widehat{\overline{\Delta}}^{(3,1,-\frac{1}{3})}_{(10,1,3)}$
&$\frac{1}{2 \sqrt{120}} \left( \left[12+34 \right]
\left[5\left[78+90 \right], 7\left[56+90 \right], 9\left[56+78 \right] \right]
\right) +c.c. $
\\
& $\widehat{\Phi}^{(3,1,-\frac{1}{3})}_{(15,1,3)}$
&$\frac{1}{\sqrt{24}} \left(24 \left[80,06,68 \right] \right)+c.c. $
\\
\hline \hline
\end{tabular}
\end{center}
\end{table}
\begin{table}[p]
\caption{States in the other $G_{321}$ multiplets including $\Delta$ 
or $\overline{\Delta}$}
\label{tabo1}
\begin{center}
\begin{tabular}{|c|c|c|}
\hline \hline
${\bf(1,1},{2}) +c.c.$
& $\widehat{\Delta}^{(1,1,2)}_{(\overline{10},1,3)}$
& $\frac{1}{\sqrt{120}} \left(13680 \right)+c.c. $
\\ \hline
${\bf(1,3},{1}) +c.c.$
& $\widehat{E}^{(1,3,1)}_{(1,3,3)}$
& $\left[11,-33,-\frac{\left\{13 \right\}}{\sqrt{2}} \right]+c.c$
\\
& $\widehat{\overline{\Delta}}^{(1,3,1)}_{(\overline{10},3,1)}$
& $\frac{1}{\sqrt{120}} \left(\left[14,32, \frac{12-34}{\sqrt{2}} \right]
 680 \right) +c.c.$
\\ \hline
${\bf(3,1},{-\frac{4}{3}}) +c.c.$ 
& $\widehat{D}^{(3,1,-\frac{4}{3})}_{(6,1,3)}$
& $\frac{1}{\sqrt{6}} \left( 24 \left[5,7,9 \right] \right)+c.c.$
\\
& $\widehat{\overline{\Delta}}^{(3,1,-\frac{4}{3})}_{(10,1,3)}$
& $\frac{1}{\sqrt{240}} \left( 24  
\left[5 \left[78+90 \right],7 \left[56+90 \right],9 \left[56+78 \right] \right]
\right)
+c.c.$
\\ \hline
${\bf(3,2},{\frac{7}{6}}) +c.c.$
& $\widehat{D}^{(3,2,\frac{7}{6})}_{(15,2,2)}$
& $\frac{1}{\sqrt{6}} \left( \left[1,3 \right] 
 \left[80, 06, 68 \right] \right) +c.c.$
\\
& $\widehat{\Delta}^{(3,2,\frac{7}{6})}_{(15,2,2)}$
& $\frac{1}{\sqrt{240}} \left( \left[134,123 \right] \left[80,06,68 \right]
- \left[1,3 \right] \left[5680, 7806, 9068 \right] \right)$ +c.c
\\
& $\widehat{\overline{\Delta}}^{(3,2,\frac{7}{6})}_{(15,2,2)}$
& $\frac{1}{\sqrt{240}} \left( \left[134,123 \right] \left[80,06,68 \right]
+ \left[1,3 \right]\left[5680, 7806, 9068 \right] \right)$ +c.c
\\ \hline
${\bf(3,3},{-\frac{1}{3}}) +c.c.$
& $\widehat{D}^{(3,3,-\frac{1}{3})}_{(6,3,1)}$
& $\frac{1}{\sqrt{6}} \left( \left[14,32,\frac{12-34}{\sqrt{2}} \right]
\left[5,7,9 \right] \right) +c.c.$
\\
& $\widehat{\Delta}^{(3,3,-\frac{1}{3})}_{(10,3,1)}$
& $\frac{1}{\sqrt{240}} \left( \left[14,32, \frac{12-34}{\sqrt{2}} \right] \right.$
\\
&& $\left. \left[5 \left[78+90 \right],7 \left[56+90 \right],9 \left[56+78 \right] 
 \right] \right)
+c.c.$
\\ \hline
${\bf(6,1},{-\frac{2}{3}}) +c.c.$
& $\widehat{E}^{(6,1,-\frac{2}{3})}_{(20',1,1)}$
& $\left[55,77,99, 
      \frac{\{79\}}{\sqrt{2}},\frac{\{95\}}{\sqrt{2}},\frac{\{57\}}{\sqrt{2}} \right] +c.c.$
\\
& $\widehat{\overline{\Delta}}^{(6,1,-\frac{2}{3})}_{(10,1,3)}$
& $\frac{1}{\sqrt{120}} \left( 24 
\left[580, 670,689,
\frac{6 \left[90-78 \right]}{\sqrt{2}},
\frac{8 \left[56-90 \right]}{\sqrt{2}},
\frac{0 \left[78-56 \right]}{\sqrt{2}} \right] \right) +c.c.$
\\ \hline
${\bf(6,1},{\frac{1}{3}}) +c.c.$ 
& $\widehat{D}^{(6,1,\frac{1}{3})}_{(10,1,1)}$
& $\frac{1}{\sqrt{6}} \left(580, 670, 689, 
\frac{6 \left[90-78 \right]}{\sqrt{2}},
\frac{8 \left[56-90 \right]}{\sqrt{2}},
\frac{0 \left[78-56 \right]}{\sqrt{2}} \right)+c.c.$
\\
& $\widehat{\overline{\Delta}}^{(6,1,\frac{1}{3})}_{(10,1,3)}$
& $\frac{1}{\sqrt{240}} \left( \left[12+34 \right]
\left[580, 670, 689,
\frac{6 \left[90-78 \right]}{\sqrt{2}},
\frac{8 \left[56-90 \right]}{\sqrt{2}},
\frac{0 \left[78-56 \right]}{\sqrt{2}} \right] \right)+c.c.$
\\ \hline
${\bf(6,1},{\frac{4}{3}}) +c.c.$ 
& $\widehat{\overline{\Delta}}^{(6,1,\frac{4}{3})}_{(10,1,3)}$
& $\frac{1}{\sqrt{120}} \left(13 
\left[580, 670, 689, 
\frac{6 \left[90-78 \right]}{\sqrt{2}},
\frac{8 \left[56-90 \right]}{\sqrt{2}},
\frac{0 \left[78-56 \right]}{\sqrt{2}} \right] \right)+c.c.$
\\ \hline
${\bf(6,3},{\frac{1}{3}}) +c.c.$
& $\widehat{\Delta}^{(6,3,\frac{1}{3})}_{(10,3,1)}$
& $\frac{1}{\sqrt{120}} \left( \left[14,32, \frac{12-34}{\sqrt{2}} \right] \right.$
\\
&& $\left. \left[580, 670, 689, 
\frac{6 \left[90-78 \right]}{\sqrt{2}},
\frac{8 \left[56-90 \right]}{\sqrt{2}},
\frac{0 \left[78-56 \right]}{\sqrt{2}} \right] \right)+c.c.$
\\ \hline
${\bf(8,2},{\frac{1}{2}}) +c.c.$
& $\widehat{D}^{(8,2,\frac{1}{2})}_{(15,2,2)}$
& $\frac{1}{\sqrt{6}} \left( \left[1,3 \right] \left[58, 50, 70, 76, 96, 98,
\frac{56-78}{\sqrt{2}}, \frac{56+78-{\it 2}[90]}{\sqrt{6}} \right] \right)+c.c.$
\\
& $\widehat{\Delta}^{(8,2,\frac{1}{2})}_{(15,2,2)}$
& $\frac{1}{\sqrt{240}} \left( 
\left[134,123 \right] \left[58, 50, 70, 76, 96, 98,
\frac{56-78}{\sqrt{2}}, \frac{56+78-{\it 2}[90]}{\sqrt{6}} \right] \right.$
\\
&& $ +    \left[1,3 \right]
\left[5890, 5078, 7056, 7690, 9678, 9856, \right. $
\\
&& $\left.\left. \frac{5690-7890}{\sqrt{2}}, 
\frac{{\it 2}[5678]-5690-7890}{\sqrt{6}} \right] \right) + c.c.$
\\
& $\widehat{\overline{\Delta}}^{(8,2,\frac{1}{2})}_{(15,2,2)}$
& $\frac{1}{\sqrt{240}} \left(
\left[134,123 \right] \left[58, 50, 70, 76, 96, 98,
\frac{56-78}{\sqrt{2}}, \frac{56+78-{\it 2}[90]}{\sqrt{6}} \right] \right.$
\\ 
&& $ -    \left[1,3 \right]
\left[5890, 5078, 7056, 7690, 9678, 9856, \right. $
\\
&& $\left.\left. \frac{5690-7890}{\sqrt{2}}, 
\frac{{\it 2}[5678]-5690-7890}{\sqrt{6}} \right] \right) + c.c.$
\\
\hline \hline
\end{tabular}
\end{center}
\end{table}
\begin{table}[p]
\caption{States in the other $G_{321}$ multiplets including $\Phi$}
\label{tabo2}
\begin{center}
\begin{tabular}{|c|c|c|}
\hline \hline
${\bf(1,2},{\frac{3}{2}}) +c.c.$ 
& $\widehat{\Phi}^{(1,2,\frac{3}{2})}_{(\overline{10},2,2)}$
& $\frac{1}{\sqrt{24}} \left( \left[1,3 \right]  680 \right)+c.c.$
\\
\hline
${\bf(1,3},{0})$ 
& $\widehat{A}^{(1,3,0)}_{(1,3,1)}$ 
& $\frac{i}{\sqrt{2}} \left(14,32, \frac{12-34}{\sqrt{2}} \right)$
\\
& $\widehat{E}^{(1,3,0)}_{(1,3,3)}$ 
& $\frac{1}{\sqrt{2}} \left\{14,32, \frac{12-34}{\sqrt{2}} \right\}$
\\
&$\widehat{\Phi}^{(1,3,0)}_{(15,3,1)}$
&$\frac{1}{\sqrt{72}} \left( \left[14,32, \frac{12-34}{\sqrt{2}} \right]
\left[56+78+90 \right] \right)$
\\ \hline
${\bf(3,1},{\frac{5}{3}}) +c.c.$
& $\widehat{\Phi}^{(3,1,\frac{5}{3})}_{(15,1,3)}$
& $\frac{1}{\sqrt{24}} \left(13  \left[80,06,68 \right] \right)+c.c. $
\\
\hline
${\bf(3,3},{\frac{2}{3}}) +c.c.$ 
& $\widehat{\Phi}^{(3,3,\frac{2}{3})}_{(15,3,1)}$ 
& $\frac{1}{\sqrt{24}} \left( \left[14,32, \frac{12-34}{\sqrt{2}} \right] \right)
\left[80,06,68 \right] +c.c.$
\\
\hline
${\bf(6,2},{-\frac{1}{6}}) +c.c.$ 
& $\widehat{\Phi}^{(6,2,-\frac{1}{6})}_{(10,2,2)}$
& $\frac{1}{\sqrt{24}} \left(  \left[2,4 \right]
\left[580, 670, 689, 
\frac{6 \left(90-78 \right)}{\sqrt{2}},
\frac{8 \left(56-90 \right)}{\sqrt{2}},
\frac{0 \left(78-56 \right)}{\sqrt{2}} \right] \right) +c.c.$
\\
\hline
${\bf(6,2},{\frac{5}{6}}) +c.c.$ 
& $\widehat{\Phi}^{(6,2,\frac{5}{6})}_{(10,2,2)}$
& $\frac{1}{\sqrt{24}} \left(  \left[1,3 \right]
\left[580, 670, 689, 
\frac{6 \left(90-78 \right)}{\sqrt{2}},
\frac{8 \left(56-90 \right)}{\sqrt{2}},
\frac{0 \left(78-56 \right)}{\sqrt{2}} \right] \right) +c.c.$
\\ \hline
${\bf(8,1},{0})$ 
& $\widehat{A}^{(8,1,0)}_{(15,1,1)}$
& $\frac{i}{\sqrt{2}}
\left(58,50,70,76,96,98,\frac{1}{\sqrt{2}}\left[56-78 \right],
\frac{1}{\sqrt{6}}\left[56+78-{\it 2}[90] \right] \right) $
\\
& $\widehat{E}^{(8,1,0)}_{(20',1,1)}$
& $\frac{1}{\sqrt{2}}
\left\{58,50,70,76,96,98,\frac{1}{\sqrt{2}}\left[56-78 \right],
\frac{1}{\sqrt{6}}\left[56+78-{\it 2}[90] \right] \right\} $
\\
& $\widehat{\Phi}^{(8,1,0)}_{(15,1,1)}$
& $\frac{1}{\sqrt{24}}
\left(5890,5078,7056,7690,9678,9856, \frac{1}{\sqrt{2}}\left[5690-7890 \right],
\right. $\\ 
& &$ \left.
\frac{1}{\sqrt{6}}\left[{\it 2}[5678]-5690-7890] \right] \right) $
\\
& $\widehat{\Phi}^{(8,1,0)}_{(15,1,3)}$
& $\frac{1}{\sqrt{48}} \left( \left[12+34 \right]
\left[58,50,70,76,96,98, \frac{1}{\sqrt{2}}\left[56-78 \right], \right.\right. $
\\
&&$ \left.\left. 
\frac{1}{\sqrt{6}}\left[56+78-{\it 2}[90] \right] \right] \right) $
\\
\hline
${\bf(8,1},{1}) +c.c.$ 
& $\widehat{\Phi}^{(8,1,1)}_{(15,1,3)}$
& $\frac{1}{\sqrt{24}} \left(13 
\left[58,50,70,76,96,98, \frac{1}{\sqrt{2}}\left[56-78 \right], \right.\right. $
\\
&&$ \left.\left. 
\frac{1}{\sqrt{6}}\left[56+78-{\it 2}[90] \right] \right] \right) +c.c. $
\\
\hline
${\bf(8,3},{0})$ 
& $\widehat{\Phi}^{(8,3,0)}_{(15,3,1)}$
& $\frac{1}{\sqrt{24}} \left( \left[14,32, \frac{12-34}{\sqrt{2}} \right]
\left[58,50,70,76,96,98, \frac{1}{\sqrt{2}}\left[56-78 \right], \right.\right. $
\\
&&$ \left.\left. 
\frac{1}{\sqrt{6}}\left[56+78 - {\it 2} [90] \right] \right] \right)$
\\
\hline \hline
\end{tabular}
\end{center}
\end{table}
\begin{table}[p]
\caption{Number of $G_{321}$ multiplets}
\label{tabstates}
\begin{center}
\begin{tabular}{|cc|c|}
\hline \hline
&8 & ${\bf(1,1},{0})$ \\
&4 & [${\bf(1,1},{1}) +c.c.$] \\ 
&1 & [${\bf(1,1},{2}) +c.c.$] \\
&6 & [${\bf(1,2},{\frac{1}{2}}) +c.c.$] \\
&1 & [${\bf(1,2},{\frac{3}{2}}) +c.c.$] \\
&3 & ${\bf(1,3},{0})$ \\
&2 & [${\bf(1,3},{1}) +c.c.$] \\
&2 & [${\bf(3,1},{-\frac{4}{3}}) +c.c.$] \\
&7 & [${\bf(3,1},{-\frac{1}{3}}) +c.c.$] \\
&5 & [${\bf(3,1},{\frac{2}{3}}) +c.c.$] \\
&1 & [${\bf(3,1},{\frac{5}{3}}) +c.c.$] \\
&4 & [${\bf(3,2},{-\frac{5}{6}}) +c.c.$] \\
&7 & [${\bf(3,2},{\frac{1}{6}}) +c.c.$] \\
&3 & [${\bf(3,2},{\frac{7}{6}}) +c.c.$] \\
&2 & [${\bf(3,3},{-\frac{1}{3}}) +c.c.$] \\
&1 & [${\bf(3,3},{\frac{2}{3}}) +c.c.$] \\
&2 & [${\bf(6,1},{-\frac{2}{3}}) +c.c.$] \\
&2 & [${\bf(6,1},{\frac{1}{3}}) +c.c.$] \\
&1 & [${\bf(6,1},{\frac{4}{3}}) +c.c.$] \\
&1 & [${\bf(6,2},{-\frac{1}{6}}) +c.c.$] \\
&1 & [${\bf(6,2},{\frac{5}{6}}) +c.c.$] \\
&1 & [${\bf(6,3},{\frac{1}{3}}) +c.c.$] \\
&4 & ${\bf(8,1},{0})$ \\
&1 & [${\bf(8,1},{1}) +c.c.$] \\
&3 & [${\bf(8,2},-{\frac{1}{2}}) +c.c.$] \\
&1 & ${\bf(8,3},{0})$ \\
\hline
sum & 74 & multiplets\\
$-$ & 5 & NG multiplets \\
\hline 
& 69 & massive multiplets \\
\hline \hline
\end{tabular}
\end{center}
\end{table}
\clearpage
\subsection{${\cal H}$ operators and Clebsch-Gordan coefficients}
Let us denote by $R$ the sum of all representations
${\bf 10}$, ${\bf 45}$, ${\bf 54}$,
${\bf 120}$, ${\bf 126}$, ${\bf \overline{126}}$, and ${\bf 210}$,
\be
\label{RRI}
R = \sum_{I} R^{I}, \quad dim\, R\ =\ 961.
\ee
There are $21$ cubic invariants (see Eq. (\ref{invariants})),
\be
{\cal I}(R^I,R^J,R^{\overline{K}})\ \equiv\ {\cal I}^{IJ\overline{K}},
\qquad
\overline{R^K}\ \equiv\ R^{\overline{K}}
\ee
where $R^I \times R^J = \sum_{K} R^K$.
 
Let us denote ${\cal H}$-operators (see Refs. \cite{he} and \cite{lee})
\be
{\cal H}_K(R^I,R^J)\ =\ {\cal H}_K(R^J,R^I)\ \sim\ R^K
\ee
transforming as $R^K$ and 
\be
{\cal H}_K(R^I,R^J)\ =\ 
  \frac{1}{N}\frac{\partial {\cal I}^{IJ\overline{K}}}{\partial R^{\overline{K}}}.
\ee
The normalization factor $N$ is chosen so that 
\be
{\cal H}_K(R^I,R^J)\ R^{\overline K}\ =\ {\cal I}^{IJ\overline{K}}.
\ee
For example, in the $Y$-diagonal basis
\bea
\left[{\cal H}_\Phi(\Phi_1,\Phi_2) \right]_{abcd}
&=& \frac{1}{6}
\left[
(\Phi_1)_{abef}(\Phi_2)_{cd\bar{e}\bar{f}}  
- (\Phi_1)_{acef}(\Phi_2)_{bd\bar{e}\bar{f}} \right.
\nonumber \\
&&+ (\Phi_1)_{adef}(\Phi_2)_{bc\bar{e}\bar{f}} 
+ (\Phi_1)_{cdef}(\Phi_2)_{ab\bar{e}\bar{f}} 
\nonumber \\
&&
\left. 
- (\Phi_1)_{bdef}(\Phi_2)_{ac\bar{e}\bar{f}} 
+ (\Phi_1)_{bcef}(\Phi_2)_{ad\bar{e}\bar{f}}
\right]
\eea
For invariants of the type ${\cal I}^{III}$, there is only one ${\cal H}$ operator, ${\cal H}_I$, 
for invariants of the type ${\cal I}^{II\overline{K}}$ 
there are two ${\cal H}$ operators, ${\cal H}_{\overline{I}}$ 
and ${\cal H}_K$ and for invariants of the type ${\cal I}^{IJ\overline{K}}$ 
there are three ${\cal H}$ operators,
${\cal H}_{\overline{I}}$, ${\cal H}_{\overline{J}}$ and 
${\cal H}_K$. The ${\cal H}$ operators are symmetric 
in $R^I$ and $R^J$, ${\cal H}(R^I,R^J)={\cal H}(R^J,R^I)$. More generally,
\be
{\cal I}^{IJ\overline{K}}\ =\ {\cal I}^{JI\overline{K}}\ =\ {\cal I}^{I\overline{K}J}\ =\ 
{\cal I}^{J\overline{K}I}\ =\ {\cal I}^{\overline{K}IJ}\ =\ {\cal I}^{\overline{K}JI}.
\ee

Specially, we are interested in CG coefficients when at least one of the states transforms 
as a singlet under the $G_{321}$ group. That requires the decomposition of each 
$SO(10)$ irrep $R^I$ into $G_{321}$ irreps,
\be
R^I\ =\ \sum_i R^I_i,
\ee
where indices "$i$" just enumerate the $G_{321}$ irreps with fixed $Y$ contained in a specific
$R^I$ irrep.
There are eight singlets, denoted here shortly by $S^I_i$, 
defined in Eqs. (\ref{ud1})-(\ref{ud8}). 
For this choice of $G_{321}$-singlet states, 
the action of 
${\cal H}$ operators 
are reduced to
the invariant subspaces of states with fixed $G_{321}$ 
quantum numbers, which are listed in Tables 
\ref{tabw}--\ref{tabo2},
\be
{\cal H}_K(\hat{S}^I_i,R^J_j)\ =\ \sum_k C^{IJK}_{ijk} R^K_k,
\ee
where $R^J_j$ and $R^K_k$ transform as identical $G_{321}$ irreps (have the same $G_{321}$
quantum numbers). For example, for the evaluation of the first 
column in the $({\bf 1,3,}1)$
mass
matrix
(see Eq. (\ref{M131})), 
the following ${\cal H}$ operators have to be evaluated,
\bea
{\cal H}_E(\widehat{E},\widehat{E}^{(1,3,1)}_{(1,3,3)}) &=&
 \frac{\sqrt{3}}{2\sqrt{5}}\ \widehat{E}^{(1,3,1)}_{(1,3,3)},
\nonumber\\
{\cal H}_{\overline{\Delta}}(\widehat{v_R},\widehat{E}^{(1,3,1)}_{(1,3,3)}) &=& 
 \frac{1}{5}\ \widehat{\overline{\Delta}}^{(1,3,1)}_{(\overline{10},3,1)}.
\eea
 Note that
\be
\label{CIJKijk}
C^{IJK}_{ijk}\ =\ {\cal I}(\hat{S}^I_i,\hat{R}^J_j,\hat{R}^{\overline K}_{\overline k})
\ \equiv\  \hat{S}^I_i \hat{R}^J_j \hat{R}^{\overline K}_{\overline k},
\ee
(the second part of Eq. (\ref{CIJKijk}) defines shortand notation) where 
\be
\hat{R}^{\overline I}_{\overline{i}}\ \equiv\ \overline{\hat{R}^I_i}
\ee
is the complex conjugated irrep of $\hat{R}^I_i$.
The CG coefficients are listed in $25$ tables in Appendix \ref{CGC}.
Note that the CG coefficients depend only on the indices $i,j,\cdots$, 
that represent $G_{321}$ multiplets and not on the specific states within
the $G_{321}$ multiplets. That can be used as a consistency check of the states belonging to 
the specific $G_{321}$ multiplets.

We choose all $SO(10)$ invariants in Eq. (\ref{invariants})
to be real, except for invariants 
containing
the $\Delta$ or the $\overline{\Delta}$
field separately (for example, $\Delta H\Phi$ or $E\overline{\Delta}^2$). 
In this last case, the sum of the two invariants, one containing
$\Delta$ field and the other having $\overline{\Delta}$ field in place of $\Delta$, are real.
Generally, CG coefficients are complex (in our case they are either real or pure imaginary).
Starting with cubic invariants in Eq. (\ref{invariants}), our choice of phases of states in 
Tables \ref{tabw} -- \ref{tabo2} is such that it leads to the minimal number of 
imaginary terms in the mass matrices.

Symmetry relations imply
\be
C^{IJK}_{ijk}\ =\ C^{JIK}_{jik}.
\ee

CG coefficients also satisfy hermiticity relations,
\be
\label{herm}
\overline{C^{IJK}_{ijk}}\ =\ C^{\overline{I}KJ}_{\overline{i}kj}.
\ee
Here the $\overline{i}$ represents the label assigned to the irrep complex conjugated
to the irrep designated by $i$.

Furthermore, following relations are valid,
\bea
\label{sumrule1}
&&\sum_j C^{IJ\overline{J}}_{ij\overline{j}} dim\, R^J_j\ =\ 0, \qquad \sum_j dim\, R^J_j\ =\ dim\, R^J
\\
\label{sumrule2}
&&\sum_i \sum_j C^{IJK}_{ijk} 
  C^{\overline{I}\overline{J}\overline{K}}_{\overline{i}\overline{j}\overline{l}} 
  dim\, R^I_i dim\, R^J_j\ =\ C(I,J,K)\delta_{k\overline{l}}, 
\nonumber \\
&&\sum_i dim\, R^I_i\ =\ dim\, R^I, \qquad \sum_j dim\, R^J_j\ =\ dim\, R^J,
\eea
where $C(I,J,K)$ are constants depending on irreps $R^I$, $R^J$ and $R^K$.
\newpage
\section{Mass Matrices}\label{mass}
\subsection{Mass Matrices}
For the sum of all representations $R$ (see Eq. (\ref{RRI})),
we define fluctuations of the Higgs field around the VEVs as 
\be
R = \left<R \right> + \sum_{I,i} r^{I}_{i}.  
\ee
Then the matrix element of the mass matrix corresponding to 
$G_{321}$ multiplets $i,j$ 
($R^I_i$ and $R^J_j$ transform identically under $G_{321}$) is: 
\bea
{\mathcal M}_{ij}^{IJ} &=& \left(\frac{\partial^{2} W}
{\partial r^{I}_{i} \partial \overline{r^{J}_{j}}} 
\right)_{R = \left<R \right>},
\\
{\mathcal M}_{ij}^{IJ} &=& m_{I} \delta_{IJ} \delta_{ij}
+ \sum_{K,k} \lambda^{IJK} S^{I}_{i} C^{IJK}_{ijk},
\eea
where $\lambda^{III} = 6 \lambda_{p}$, $\lambda^{IIK} = 2 \lambda_{p}$ and 
$\lambda^{IJK} = \lambda_{p}$ for invariants containing three identical representations,
two identical representations (different from the third one), and three different
representations, respectively. Specifically, $\{\lambda^{IJK}\}\ =\ $ 
$\{6 \lambda_{1}$, $6 \lambda_{8}$, $2 \lambda_{5}$, $2 \lambda_{7}$, $2 \lambda_{9}$,  
$2 \lambda_{10}$,  $2 \lambda_{11}$,  $2 \lambda_{12}$,  $2 \lambda_{13}$, 
$2 \lambda_{14}$ , $2 \lambda_{15}$,  
$\lambda_{2}$, $\lambda_{3}$, $\lambda_{4}$,
$\lambda_{6}$, $\lambda_{16}$, $\lambda_{17}$, $\lambda_{18}$,
$\lambda_{19}$, $\lambda_{20}$, $\lambda_{21}\}$.

According to Eq. (\ref{sumrule1})  the trace of the total mass matrix over all $dim\, R$ states in $R$ is 
\be
\label{TrMtot}
{\mathrm{Tr}}\,{\mathcal{M}} = \sum_{I} m_I\, {dim\, R^{I}}.  
\ee

The mass matrices are generally non-hermitian.  So the squares of the physical masses 
are equal to the absolute values of eigenvalues of matrices ${\mathcal{M}}^\dagger {\mathcal{M}}$
and ${\mathcal{M}} {\mathcal{M}}^\dagger$. 
(One can obtain hermitian matrices 
${\mathcal{M}}^{\dagger}{\mathcal M}$, {\it i.e.} 
${\mathcal M}{\mathcal{M}}^{\dagger}$ with the same spectra.) For a real superpotential,
that is for $\lambda_3\ =\ \lambda_4$,  
$\lambda_{11}\ =\ \lambda_{12}$,
$\lambda_{18}\ =\ \lambda_{19}$ and
$\lambda_{20}\ =\ \lambda_{21}$ and all coupling constants and VEVs real, the matrices
are hermitian due to the hermiticity relation (\ref{herm}).

Now we are ready to present the explicit forms of the mass matrices calculated 
from the superpotential of Eq. (\ref{potential}). Every matrix is
designated with the corresponding $G_{321}$ multiplet and appears $dim\, R^{I}_{i}$ 
times in the total mass matrix ${\cal M}$. A mass matrix  associated with 
a $G_{321}$ multiplet and the mass matrix associated with the 
corresponding complex conjugated $G_{321}$ multiplet are 
equal up to transposition, and therefore for multiplets with $Y\ne 0$ we 
list only one of the two mass matrices. Of course, when enumerating the total 
degrees of freedom, one has to include all mass eigenvalues ($961$ in total).
The basis designating the  columns ({\bf c:}) of the mass matrices listed below 
is given in the same way as shown in 
Tables \ref{tabw}, \ref{tabd}, \ref{tabt}, \ref{tabo1} and \ref{tabo2}, while the
rows are designated by the corresponding complex conjugated $G_{321}$ multiplets 
({\bf r:}) :

\noindent
\begin{minipage}{16cm}
$({\bf 1,1,}0)$\\[.2cm]
{\bf c:}
$\widehat{A}^{(1,1,0)}_{(1,1,3)}$,
$\widehat{A}^{(1,1,0)}_{(15,1,1)}$,
$\widehat{E}^{(1,1,0)}_{(1,1,1)}$,
$\widehat{\Delta}^{(1,1,0)}_{(\overline{10},1,3)}$,
$\widehat{\overline{\Delta}}^{(1,1,0)}_{(10,1,3)}$,
$\widehat{\Phi}^{(1,1,0)}_{(1,1,1)}$,
$\widehat{\Phi}^{(1,1,0)}_{(15,1,1)}$,
$\widehat{\Phi}^{(1,1,0)}_{(15,1,3)}$\\[.15cm]
{\bf r:}
$\widehat{A}^{(1,1,0)}_{(1,1,3)}$,
$\widehat{A}^{(1,1,0)}_{(15,1,1)}$,
$\widehat{E}^{(1,1,0)}_{(1,1,1)}$,
$\widehat{\overline{\Delta}}^{(1,1,0)}_{(10,1,3)}$,
$\widehat{\Delta}^{(1,1,0)}_{(\overline{10},1,3)}$,
$\widehat{\Phi}^{(1,1,0)}_{(1,1,1)}$,
$\widehat{\Phi}^{(1,1,0)}_{(15,1,1)}$,
$\widehat{\Phi}^{(1,1,0)}_{(15,1,3)}$\\[.2cm]
\be
\left(
\begin{array}{ccccccccc}
\begin{array}{c}
m_{11}^{(1,1,0)} \\
\sqrt{\frac{2}{3}} \lambda_5 \Phi_3 \\
\sqrt{\frac{3}{5}} \lambda_9 A_1 \\
- \frac{\lambda_6 v_R}{5} \\
- \frac{\lambda_6 \overline{v_R}}{5} \\
\sqrt{\frac{2}{3}} \lambda_5 A_1 \\
\frac{2 \sqrt{2} \lambda_7 \Phi_3}{5} \\
m_{81}^{(1,1,0)}
\end{array}
\begin{array}{c}
\sqrt{\frac{2}{3}} \lambda_5 \Phi_3  \\
m_{22}^{(1,1,0)} \\
- \frac{2 \lambda_9 A_2}{\sqrt{15}} \\
- \frac{1}{5} \sqrt{\frac{3}{2}} \lambda_6 v_R \\
- \frac{1}{5} \sqrt{\frac{3}{2}} \lambda_6 \overline{v_R} \\
\frac{2 \sqrt{2} \lambda_7 \Phi_2}{5} \\
m_{72}^{(1,1,0)} \\
m_{82}^{(1,1,0)} 
\end{array}
\begin{array}{c}
\sqrt{\frac{3}{5}} \lambda_9 A_1 \\
- \frac{2 \lambda_9 A_2}{\sqrt{15}} \\
m_{33}^{(1,1,0)} \\
0 \\
0 \\
\sqrt{\frac{3}{5}} \lambda_{10} \Phi_1 \\ 
- \frac{2 \lambda_{10} \Phi_2}{\sqrt{15}} \\
\frac{\lambda_{10} \Phi_3}{2 \sqrt{15}}
\end{array}
\begin{array}{c}
- \frac{\lambda_6 \overline{v_R}}{5} \\
- \frac{1}{5} \sqrt{\frac{3}{2}} \lambda_6 \overline{v_R} \\
0 \\
m_{44}^{(1,1,0)} \\
0 \\
\frac{\lambda_2 \overline{v_R}}{10 \sqrt{6}} \\
\frac{\lambda_2 \overline{v_R}}{10 \sqrt{2}} \\
\frac{\lambda_2 \overline{v_R}}{10}
\end{array}
\begin{array}{c}
- \frac{\lambda_6 v_R}{5} \\
- \frac{1}{5} \sqrt{\frac{3}{2}} \lambda_6 v_R \\
0 \\
0 \\
m_{44}^{(1,1,0)} \\
\frac{\lambda_2 v_R}{10 \sqrt{6}} \\
\frac{\lambda_2 v_R}{10 \sqrt{2}} \\
\frac{\lambda_2 v_R}{10}
\end{array}
\begin{array}{c}
\sqrt{\frac{2}{3}} \lambda_5 A_1 \\
\frac{2 \sqrt{2} \lambda_7 \Phi_2}{5} \\
\sqrt{\frac{3}{5}} \lambda_{10} \Phi_1 \\
\frac{\lambda_2 v_R}{10 \sqrt{6}} \\
\frac{\lambda_2 \overline{v_R}}{10 \sqrt{6}} \\
m_{66}^{(1,1,0)} \\
\frac{2 \sqrt{2} \lambda_7 A_2}{5} \\
\frac{\lambda_1 \Phi_3}{\sqrt{6}}
\end{array}
\begin{array}{c}
\frac{2 \sqrt{2} \lambda_7 \Phi_3}{5} \\
m_{72}^{(1,1,0)} \\
- \frac{2 \lambda_{10} \Phi_2}{\sqrt{15}} \\
\frac{\lambda_2 v_R}{10 \sqrt{2}} \\
\frac{\lambda_2 \overline{v_R}}{10 \sqrt{2}} \\
\frac{2 \sqrt{2} \lambda_7 A_2}{5} \\
m_{77}^{(1,1,0)} \\
m_{87}^{(1,1,0)} 
\end{array}
\begin{array}{c}
m_{81}^{(1,1,0)} \\
m_{82}^{(1,1,0)} \\
\frac{\lambda_{10} \Phi_3}{2 \sqrt{15}} \\
\frac{\lambda_2 v_R}{10} \\
\frac{\lambda_2 \overline{v_R}}{10} \\
\frac{\lambda_1 \Phi_3}{\sqrt{6}} \\
m_{87}^{(1,1,0)} \\
m_{88}^{(1,1,0)}
\end{array}
\end{array}
\right),
\ee
\end{minipage}
where 
\bea
m_{11}^{(1,1,0)} &\equiv& 
m_4 + \sqrt{\frac{2}{3}} \lambda_5 \Phi_1 
+ \sqrt{\frac{3}{5}} \lambda_9 E, \nonumber\\ 
m_{22}^{(1,1,0)} &\equiv& 
m_4 + \frac{2 \sqrt{2} \lambda_5 \Phi_2}{3} 
- \frac{2 \lambda_9 E}{\sqrt{15}}, \nonumber\\
m_{33}^{(1,1,0)} &\equiv& 
m_5 + \sqrt{\frac{3}{5}} \lambda_8 E, 
\nonumber\\
m_{44}^{(1,1,0)} &\equiv&  
m_2 
+ \frac{\lambda_2\Phi_1}{10\sqrt{6}}
+ \frac{\lambda_2\Phi_2}{10\sqrt{2}} 
+ \frac{\lambda_2\Phi_3}{10}
- \frac{\lambda_6 A_1}{5}
- \frac{\sqrt{3} \lambda_6 A_2}{5\sqrt{2}} 
\nonumber \\
m_{66}^{(1,1,0)} &\equiv& 
m_{1} + \sqrt{\frac{3}{5}} \lambda_{10} E, \nonumber\\
m_{72}^{(1,1,0)} &\equiv& 
\frac{2 \sqrt{2} \lambda_5 A_2}{3} 
+ \frac{2 \sqrt{2} \lambda_7 \Phi_1}{5}, \nonumber\\
m_{77}^{(1,1,0)} &\equiv& 
m_1 + \frac{\sqrt{2} \lambda_1 \Phi_2}{3} 
- \frac{2 \lambda_{10} E}{\sqrt{15}}, \nonumber\\
m_{81}^{(1,1,0)} &\equiv& 
\sqrt{\frac{2}{3}} \lambda_5 A_2 
+ \frac{2 \sqrt{2} \lambda_7 \Phi_2}{5}, \nonumber\\
m_{82}^{(1,1,0)} &\equiv& 
\sqrt{\frac{2}{3}} \lambda_5 A_1 
+ \frac{4}{5}\sqrt{\frac{2}{3}} \lambda_7 \Phi_3, \nonumber\\
m_{87}^{(1,1,0)} &\equiv& 
\frac{\sqrt{2} \lambda_1 \Phi_3}{3} 
+ \frac{2 \sqrt{2} \lambda_7 A_1}{5}, \nonumber\\
m_{88}^{(1,1,0)} &\equiv& 
m_1 + \frac{\lambda_1 \Phi_1}{\sqrt{6}} 
+ \frac{\sqrt{2} \lambda_1 \Phi_2}{3} 
+ \frac{4}{5} \sqrt{\frac{2}{3}} \lambda_7 A_2 
+ \frac{\lambda_{10} E}{2 \sqrt{15}}.  
\eea\\[.2cm]
\begin{minipage}{16cm}
$\left[{\bf{(1,1}},1) +c.c. \right]$\\[.2cm]
{\bf c:}
$\widehat{A}^{(1,1,1)}_{(1,1,3)}$,
$\widehat{D}^{(1,1,1)}_{(\overline{10},1,1)}$,
$\widehat{\Delta}^{(1,1,1)}_{(\overline{10},1,3)}$\\[.15cm]
{\bf r:}
$\widehat{A}^{(1,1,-1)}_{(1,1,3)}$,
$\widehat{D}^{(1,1,-1)}_{(10,1,1)}$,
$\widehat{\Delta}^{(1,1,-1)}_{(10,1,3)}$\\[.2cm]
\be
\left(
\begin{array}{cccc}
\begin{array}{c}
m_4 + \frac{\sqrt{2} \lambda_5 \Phi_1}{\sqrt{3}} + \frac{\sqrt{3} \lambda_9 E}{\sqrt{5}} \\
\frac{i \lambda_{18} v_R}{\sqrt{10}} \\
- \frac{\lambda_6 v_R}{5} \\ 
-\sqrt{\frac{2}{3}}\lambda_5 A_2 - \frac{2 \sqrt{2} \lambda_7 \Phi_2}{5} 
\end{array}
\begin{array}{c}
-\frac{i \lambda_{19} \overline{v_R}}{\sqrt{10}} \\
m_6 - \frac{2 \lambda_{14} E}{\sqrt{15}} + \frac{\sqrt{2} \lambda_{15} \Phi_2}{3}  \\
\frac{i \lambda_{19} A_1}{\sqrt{10}} - \frac{\lambda_{21} \Phi_3}{2 \sqrt{10}} \\ 
-\frac{\lambda_{21} \overline{v_R}}{\sqrt{10}}
\end{array}
\begin{array}{c}
- \frac{\lambda_6 \overline{v_R}}{5} \\
-\frac{i \lambda_{18} A_1}{\sqrt{10}} - \frac{\lambda_{20} \Phi_3}{2 \sqrt{10}} \\
m_{33}^{(1,1,1)} \\ 
- \frac{\lambda_2 \overline{v_R}}{10}
\end{array}
\begin{array}{c}
-\sqrt{\frac{2}{3}}\lambda_5 A_2 - \frac{2 \sqrt{2} \lambda_7 \Phi_2}{5}  \\
-\frac{\lambda_{20} v_R}{\sqrt{10}} \\
- \frac{\lambda_2 v_R}{10} \\ 
m_{44}^{(1,1,1)}
\end{array}
\end{array}
\right),
\ee
\end{minipage}
where 
\bea
m_{33}^{(1,1,1)} &\equiv& 
m_2 + \frac{\lambda_2 \Phi_1}{10 \sqrt{6}} + \frac{\lambda_2 \Phi_2}{10 \sqrt{2}} 
- \frac{1}{5}\sqrt{\frac{3}{2}} \lambda_6 A_2, \nonumber\\
m_{44}^{(1,1,1)} &\equiv& 
m_1 + \frac{\lambda_1 \Phi_1}{\sqrt{6}} + \frac{\sqrt{2} \lambda_1 \Phi_2}{3} 
+ \frac{4}{5}\sqrt{\frac{2}{3}} \lambda_7 A_2
+ \frac{\lambda_{10} E}{2 \sqrt{15}}.  
\eea\\[.2cm]
\begin{minipage}{16cm}
$\left[{\bf{(3,1}}, \frac{2}{3}) +c.c. \right]$\\[.2cm]
{\bf c:}
$\widehat{A}^{(3,1,\frac{2}{3})}_{(15,1,1)}$,
$\widehat{D}^{(3,1,\frac{2}{3})}_{(6,1,3)}$,
$\widehat{\overline{\Delta}}^{(3,1,\frac{2}{3})}_{(10,1,3)}$,
$\widehat{\Phi}^{(3,1,\frac{2}{3})}_{(15,1,1)}$,
$\widehat{\Phi}^{(3,1,\frac{2}{3})}_{(15,1,3)}$\\[.15cm]
{\bf r:}
$\widehat{A}^{(\overline{3},1,-\frac{2}{3})}_{(15,1,1)}$,
$\widehat{D}^{(\overline{3},1,-\frac{2}{3})}_{(6,1,3)}$,
$\widehat{\Delta}^{(\overline{3},1,-\frac{2}{3})}_{(\overline{10},1,3)}$,
$\widehat{\Phi}^{(\overline{3},1,-\frac{2}{3})}_{(15,1,1)}$,
$\widehat{\Phi}^{(\overline{3},1,-\frac{2}{3})}_{(15,1,3)}$\\
\be
\left(
\begin{array}{ccccc}
\begin{array}{c}
m_{11}^{(3,1,\frac{2}{3})} \\
\frac{i \lambda_{19} \overline{v_R}}{\sqrt{10}} \\
- \frac{\lambda_6 \overline{v_R}}{5} \\
- \frac{\sqrt{2}}{3} \lambda_5 A_2 - \frac{2 \sqrt{2} \lambda_7 \Phi_1}{5} \\
- \sqrt{\frac{2}{3}} \lambda_5 A_1 - \frac{2}{5} \sqrt{\frac{2}{3}} \lambda_7 \Phi_3
\end{array}
\begin{array}{c}
-\frac{i \lambda_{18} v_R}{\sqrt{10}} \\
m_{22}^{(3,1,\frac{2}{3})} \\
m_{32}^{(3,1,\frac{2}{3})} \\
\frac{\lambda_{20} v_R}{2 \sqrt{30}} \\
\frac{\lambda_{20} v_R}{2 \sqrt{15}}
\end{array}
\begin{array}{c}
- \frac{\lambda_6 v_R}{5} \\
m_{23}^{(3,1,\frac{2}{3})} \\
m_{33}^{(3,1,\frac{2}{3})} \\
- \frac{\lambda_{2} v_R}{10 \sqrt{3}} \\
- \frac{\lambda_{2} v_R}{5 \sqrt{6}}
\end{array}
\begin{array}{c}
- \frac{\sqrt{2}}{3} \lambda_5 A_2 - \frac{2 \sqrt{2} \lambda_7 \Phi_1}{5}  \\
\frac{\lambda_{21} \overline{v_R}}{2 \sqrt{30}} \\
- \frac{\lambda_{2} \overline{v_R}}{10 \sqrt{3}} \\
m_{44}^{(3,1,\frac{2}{3})} \\
m_{45}^{(3,1,\frac{2}{3})}
\end{array}
\begin{array}{c}
- \sqrt{\frac{2}{3}} \lambda_5 A_1 - \frac{2}{5} \sqrt{\frac{2}{3}} \lambda_7 \Phi_3 \\
\frac{\lambda_{21} \overline{v_R}}{2 \sqrt{15}} \\
- \frac{\lambda_{2} \overline{v_R}}{5 \sqrt{6}} \\
m_{45}^{(3,1,\frac{2}{3})} \\
m_{55}^{(3,1,\frac{2}{3})}
\end{array}
\end{array}
\right),
\ee
\end{minipage}
where 
\bea
m_{11}^{(3,1,\frac{2}{3})} &\equiv& 
m_4 + \frac{\sqrt{2} \lambda_5 \Phi_2}{3} - \frac{2 \lambda_9 E}{\sqrt{15}}, 
\nonumber\\
m_{22}^{(3,1,\frac{2}{3})} &\equiv& 
m_6 + \frac{4 \lambda_{14} E}{3 \sqrt{15}} 
+ \frac{1}{3} \sqrt{\frac{2}{3}} \lambda_{15} \Phi_1 
+ \frac{2 \lambda_{15} \Phi_3}{9}, \nonumber\\
m_{23}^{(3,1,\frac{2}{3})} &\equiv& 
- \frac{i \lambda_{19} A_2}{\sqrt{15}} 
+ \frac{\lambda_{21} \Phi_2}{6 \sqrt{5}} 
+ \frac{\lambda_{21} \Phi_3}{3 \sqrt{10}}, \nonumber\\
m_{32}^{(3,1,\frac{2}{3})} &\equiv& 
\frac{i \lambda_{18} A_2}{\sqrt{15}} 
+ \frac{\lambda_{20} \Phi_2}{6 \sqrt{5}} 
+ \frac{\lambda_{20} \Phi_3}{3 \sqrt{10}}, \nonumber\\
m_{33}^{(3,1,\frac{2}{3})} &\equiv& 
m_2 + \frac{\lambda_2 \Phi_1}{10 \sqrt{6}} + \frac{\lambda_2 \Phi_2}{30 \sqrt{2}} 
+ \frac{\lambda_2 \Phi_3}{30} 
- \frac{\lambda_6 A_1}{5} - \frac{\lambda_6 A_2}{5 \sqrt{6}}, \nonumber\\
m_{44}^{(3,1,\frac{2}{3})} &\equiv& 
m_1 + \frac{\lambda_1 \Phi_2}{3 \sqrt{2}} 
- \frac{2 \lambda_{10} E}{\sqrt{15}}, \nonumber\\
m_{45}^{(3,1,\frac{2}{3})} &\equiv& 
\frac{\lambda_1 \Phi_3}{3 \sqrt{2}} 
+ \frac{2 \sqrt{2} \lambda_7 A_1}{5}, \nonumber\\
m_{55}^{(3,1,\frac{2}{3})} &\equiv& 
m_1 + \frac{\lambda_1 \Phi_1}{\sqrt{6}} 
+ \frac{\lambda_1 \Phi_2}{3 \sqrt{2}} 
+ \frac{2}{5} \sqrt{\frac{2}{3}} \lambda_7 A_2 
+ \frac{\lambda_{10} E}{2 \sqrt{15}}.  
\eea\\[.2cm]
\begin{minipage}{16cm}
$\left[{\bf{(3,2}}, -\frac{5}{6}) +c.c. \right]$\\[.2cm]
{\bf c:}
$\widehat{A}^{(3,2,-\frac{5}{6})}_{(6,2,2)}$,
$\widehat{E}^{(3,2,-\frac{5}{6})}_{(6,2,2)}$,
$\widehat{\Phi}^{(3,2,-\frac{5}{6})}_{(6,2,2)}$
$\widehat{\Phi}^{(3,2,-\frac{5}{6})}_{(10,2,2)}$\\[.15cm]
{\bf r:}
$\widehat{A}^{(\overline{3},2,\frac{5}{6})}_{(6,2,2)}$,
$\widehat{E}^{(\overline{3},2,\frac{5}{6})}_{(6,2,2)}$,
$\widehat{\Phi}^{(\overline{3},2,\frac{5}{6})}_{(6,2,2)}$
$\widehat{\Phi}^{(\overline{3},2,\frac{5}{6})}_{(\overline{10},2,2)}$\\[.2cm]
\be
\left(
\begin{array}{cccc}
\begin{array}{c}
m_4 -\frac{\lambda_5 \Phi_3}{3} + \frac{\lambda_9 E}{2 \sqrt{15}} \\
- \frac{\lambda_9 A_1}{2} - \frac{\lambda_9 A_2}{\sqrt{6}} \\
- \frac{\lambda_5 A_1}{\sqrt{3}} + \frac{2}{5} \sqrt{\frac{2}{3}} \lambda_7 \Phi_2 \\
- \frac{2 \lambda_5 A_2}{3} + \frac{2}{5} \sqrt{\frac{2}{3}} \lambda_7 \Phi_3 \\
\end{array}
\begin{array}{c}
-\frac{\lambda_9 A_1}{2} - \frac{\lambda_9 A_2}{\sqrt{6}} \\
m_5 + \frac{1}{2} \sqrt{\frac{3}{5}} \lambda_8 E \\
\frac{\lambda_{10} \Phi_1}{2 \sqrt{2}} + \frac{\lambda_{10} \Phi_3}{4 \sqrt{3}} \\
\frac{\lambda_{10} \Phi_2}{2 \sqrt{3}} + \frac{\lambda_{10} \Phi_3}{2 \sqrt{6}} \\
\end{array}
\begin{array}{c}
- \frac{\lambda_5 A_1}{\sqrt{3}} + \frac{2}{5} \sqrt{\frac{2}{3}} \lambda_7 \Phi_2 \\
\frac{\lambda_{10} \Phi_1}{2 \sqrt{2}} + \frac{\lambda_{10}  \Phi_3 }{4 \sqrt{3}} \\
m_1 - \frac{\lambda_1 \Phi_3}{6} + \frac{7 \lambda_{10} E}{4 \sqrt{15}} \\
\frac{\lambda_1 \Phi_3}{3 \sqrt{2}} - \frac{4 \lambda_7 A_2 }{5 \sqrt{3}} \\
\end{array}
\begin{array}{c}
- \frac{2 \lambda_5 A_2}{3} + \frac{2}{5} \sqrt{\frac{2}{3}} \lambda_7 \Phi_3 \\
\frac{\lambda_{10} \Phi_2}{2 \sqrt{3}} + \frac{\lambda_{10} \Phi_3}{2 \sqrt{6}} \\
\frac{\lambda_1 \Phi_3}{3 \sqrt{2}} - \frac{4 \lambda_7 A_2}{5 \sqrt{3}} \\
m_{44}^{(3,2,-\frac{5}{6})}  \\
\end{array}
\end{array}
\right),
\ee
\end{minipage}\\[.2cm]
where 
\bea
m_{44}^{(3,2,-\frac{5}{6})} &\equiv& 
m_1 + \frac{\lambda_1 \Phi_2 }{3 \sqrt{2}} - \frac{\lambda_1 \Phi_3}{6} 
- \frac{2 \lambda_7 A_1}{5} - \frac{1}{4} \sqrt{\frac{3}{5}} \lambda_{10} E.  
\eea
\begin{minipage}{16cm}
$\left[{\bf{(3,2}}, \frac{1}{6}) +c.c. \right]$\\[.2cm]
{\bf c:}
$\widehat{A}^{(3,2,\frac{1}{6})}_{(6,2,2)}$,
$\widehat{E}^{(3,2,\frac{1}{6})}_{(6,2,2)}$,
$\widehat{D}^{(3,2,\frac{1}{6})}_{(15,2,2)}$,
$\widehat{\Delta}^{(3,2,\frac{1}{6})}_{(15,2,2)}$,
$\widehat{\overline{\Delta}}^{(3,2,\frac{1}{6})}_{(15,2,2)}$,
$\widehat{\Phi}^{(3,2,\frac{1}{6})}_{(6,2,2)}$,
$\widehat{\Phi}^{(3,2,\frac{1}{6})}_{(10,2,2)}$\\[.15cm]
{\bf r:}
$\widehat{A}^{(\overline{3},2,-\frac{1}{6})}_{(6,2,2)}$,
$\widehat{E}^{(\overline{3},2,-\frac{1}{6})}_{(6,2,2)}$,
$\widehat{D}^{(\overline{3},2,-\frac{1}{6})}_{(15,2,2)}$,
$\widehat{\overline{\Delta}}^{(\overline{3},2,-\frac{1}{6})}_{(15,2,2)}$,
$\widehat{\Delta}^{(\overline{3},2,-\frac{1}{6})}_{(15,2,2)}$,
$\widehat{\Phi}^{(\overline{3},2,-\frac{1}{6})}_{(6,2,2)}$,
$\widehat{\Phi}^{(\overline{3},2,-\frac{1}{6})}_{(\overline{10},2,2)}$\\[.2cm]
\be
\left(
\begin{array}{ccccccc}
\begin{array}{c}
m_{11}^{(3,2,\frac{1}{6})} \\
\frac{\lambda_9 A_1}{2} - \frac{\lambda_9 A_2}{\sqrt{6}} \\
\frac{i \lambda_{18} v_R}{\sqrt{10}} \\
- \frac{\lambda_{6} v_R}{5} \\
0 \\
m_{61}^{(3,2,\frac{1}{6})} \\
m_{71}^{(3,2,\frac{1}{6})} \\
\end{array}
\begin{array}{c}
\frac{\lambda_9 A_1}{2} - \frac{\lambda_9 A_2}{\sqrt{6}}  \\
m_5 + \frac{1}{2}\sqrt{\frac{3}{5}} \lambda_8 E \\
0 \\
0 \\
-\frac{2 \lambda_{11} v_R}{5} \\
\frac{\lambda_{10} \Phi_3}{4 \sqrt{3}} - \frac{\lambda_{10} \Phi_1}{2 \sqrt{2}} \\
\frac{\lambda_{10} \Phi_2}{2 \sqrt{3}} - \frac{\lambda_{10} \Phi_3}{2 \sqrt{6}} 
\end{array}
\begin{array}{c}
-\frac{i \lambda_{19} \overline{v_R}}{\sqrt{10}} \\ 
0 \\
m_{33}^{(3,2,\frac{1}{6})} \\
m_{43}^{(3,2,\frac{1}{6})} \\
m_{53}^{(3,2,\frac{1}{6})} \\
- \frac{\lambda_{21} \overline{v_R}}{2 \sqrt{30}} \\
- \frac{\lambda_{21} \overline{v_R}}{2 \sqrt{15}} 
\end{array}
\begin{array}{c}
- \frac{\lambda_{6} \overline{v_R}}{5} \\
0 \\
m_{34}^{(3,2,\frac{1}{6})} \\
m_{44}^{(3,2,\frac{1}{6})} \\
\frac{\lambda_{11} E}{\sqrt{15}} \\
- \frac{\lambda_{2} \overline{v_R}}{10 \sqrt{3}} \\
- \frac{\lambda_{2} \overline{v_R}}{5 \sqrt{6}} \\
\end{array}
\begin{array}{c}
0 \\
-\frac{2 \lambda_{12} v_R}{5} \\
m_{35}^{(3,2,\frac{1}{6})} \\
\frac{\lambda_{12} E}{\sqrt{15}} \\
m_{55}^{(3,2,\frac{1}{6})} \\
0 \\
0
\end{array}
\begin{array}{c}
m_{61}^{(3,2,\frac{1}{6})} \\
\frac{\lambda_{10} \Phi_3}{4 \sqrt{3}} - \frac{\lambda_{10} \Phi_1}{2 \sqrt{2}}  \\
- \frac{\lambda_{20} v_R}{2 \sqrt{30}} \\
- \frac{\lambda_{2} v_R}{10 \sqrt{3}} \\
0 \\
m_{66}^{(3,2,\frac{1}{6})} \\
\frac{\lambda_1 \Phi_3}{3 \sqrt{2}} + \frac{4 \lambda_7 A_2}{5 \sqrt{3}}
\end{array}
\begin{array}{c}
m_{71}^{(3,2,\frac{1}{6})} \\
\frac{\lambda_{10} \Phi_2}{2 \sqrt{3}} - \frac{\lambda_{10} \Phi_3}{2 \sqrt{6}} \\
- \frac{\lambda_{20} v_R}{2 \sqrt{15}} \\
- \frac{\lambda_{2} v_R}{5 \sqrt{6}} \\
0 \\
\frac{\lambda_1 \Phi_3}{3 \sqrt{2}} + \frac{4 \lambda_7 A_2}{5 \sqrt{3}} \\
m_{77}^{(3,2,\frac{1}{6})} 

\end{array}
\end{array}
\right),
\ee
\end{minipage}\\[.2cm]
where 
\bea
m_{11}^{(3,2,\frac{1}{6})} &\equiv& 
m_4 + \frac{\lambda_5 \Phi_3}{3} + \frac{\lambda_9 E}{2 \sqrt{15}}, 
\nonumber\\
m_{33}^{(3,2,\frac{1}{6})} &\equiv& 
m_6 - \frac{\lambda_{14} E}{3 \sqrt{15}} 
+ \frac{\sqrt{2} \lambda_{15} \Phi_2}{9} 
+ \frac{2 \lambda_{15} \Phi_3}{9}, 
\nonumber\\
m_{34}^{(3,2,\frac{1}{6})} &\equiv& 
- \frac{i \lambda_{18} A_1}{2 \sqrt{10}} 
- \frac{i \lambda_{18} A_2}{2 \sqrt{15}} 
- \frac{\lambda_{20} \Phi_1}{4 \sqrt{15}} 
- \frac{\lambda_{20} \Phi_2}{6 \sqrt{5}} 
- \frac{\lambda_{20} \Phi_3}{4 \sqrt{10}}, 
\nonumber\\
m_{35}^{(3,2,\frac{1}{6})} &\equiv& 
- \frac{i \lambda_{19} A_1}{2 \sqrt{10}} 
+ \frac{i \lambda_{19} A_2}{2 \sqrt{15}} 
- \frac{\lambda_{21} \Phi_1}{4 \sqrt{15}} 
+ \frac{\lambda_{21} \Phi_2}{6 \sqrt{5}} 
- \frac{\lambda_{21} \Phi_3}{12 \sqrt{10}}, 
\nonumber\\
m_{43}^{(3,2,\frac{1}{6})} &\equiv& 
\frac{i \lambda_{19} A_1}{2 \sqrt{10}} 
+ \frac{i \lambda_{19} A_2}{2 \sqrt{15}} 
- \frac{\lambda_{21} \Phi_1}{4 \sqrt{15}} 
- \frac{\lambda_{21} \Phi_2}{6 \sqrt{5}} 
- \frac{\lambda_{21} \Phi_3}{4 \sqrt{10}}, 
\nonumber\\
m_{44}^{(3,2,\frac{1}{6})} &\equiv& 
m_2 
+ \frac{\lambda_2 \Phi_2}{30 \sqrt{2}}
+ \frac{\lambda_2 \Phi_3}{20}
- \frac{\lambda_6 A_1}{10} 
- \frac{1}{5} \sqrt{\frac{2}{3}} \lambda_6 A_2, 
\nonumber\\
m_{53}^{(3,2,\frac{1}{6})} &\equiv& 
\frac{i \lambda_{18} A_1}{2 \sqrt{10}} 
- \frac{i \lambda_{18} A_2}{2 \sqrt{15}} 
- \frac{\lambda_{20} \Phi_1}{4 \sqrt{15}} 
+ \frac{\lambda_{20} \Phi_2}{6 \sqrt{5}} 
- \frac{\lambda_{20} \Phi_3}{12 \sqrt{10}},
\nonumber\\ 
m_{55}^{(3,2,\frac{1}{6})} &\equiv& 
m_2 
+ \frac{\lambda_2 \Phi_2}{30 \sqrt{2}}
+ \frac{\lambda_2 \Phi_3}{60}
+ \frac{\lambda_6 A_1}{10} 
+ \frac{1}{5} \sqrt{\frac{2}{3}} \lambda_6 A_2, 
\nonumber\\
m_{61}^{(3,2,\frac{1}{6})} &\equiv& 
- \frac{\lambda_5 A_1}{\sqrt{3}} - \frac{2}{5} \sqrt{\frac{2}{3}} \lambda_7 \Phi_2,
\nonumber\\
m_{66}^{(3,2,\frac{1}{6})} &\equiv&
m_1 + \frac{\lambda_1 \Phi_3}{6} + \frac{7 \lambda_{10} E}{4 \sqrt{15}},
\nonumber\\
m_{71}^{(3,2,\frac{1}{6})} &\equiv&
- \frac{2 \lambda_5 A_2}{3} - \frac{2}{5} \sqrt{\frac{2}{3}} \lambda_7 \Phi_3,
\nonumber\\
m_{77}^{(3,2,\frac{1}{6})} &\equiv&
m_1 + \frac{\lambda_1 \Phi_2}{3 \sqrt{2}} + \frac{\lambda_1 \Phi_3}{6}
+ \frac{2 \lambda_7 A_1}{5} - \frac{1}{4} \sqrt{\frac{3}{5}} \lambda_{10} E
\eea
\begin{minipage}{16cm}
$\left[{\bf{(1,2}}, \frac{1}{2}) +c.c. \right]$\\[.2cm] 
{\bf c:}
$\widehat{H}^{(1,2,\frac{1}{2})}_{(1,2,2)}$,
$\widehat{D}^{(1,2,\frac{1}{2})}_{(1,2,2)}$,
$\widehat{D}^{(1,2,\frac{1}{2})}_{(15,2,2)}$,
$\widehat{\Delta}^{(1,2,\frac{1}{2})}_{(15,2,2)}$,
$\widehat{\overline{\Delta}}^{(1,2,\frac{1}{2})}_{(15,2,2)}$,
$\widehat{\Phi}^{(1,2,\frac{1}{2})}_{(6,2,2)}$\\[.15cm]
{\bf r:}
$\widehat{H}^{(1,2,-\frac{1}{2})}_{(1,2,2)}$,
$\widehat{D}^{(1,2,-\frac{1}{2})}_{(1,2,2)}$,
$\widehat{D}^{(1,2,-\frac{1}{2})}_{(15,2,2)}$,
$\widehat{\overline{\Delta}}^{(1,2,-\frac{1}{2})}_{(15,2,2)}$,
$\widehat{\Delta}^{(1,2,-\frac{1}{2})}_{(15,2,2)}$,
$\widehat{\Phi}^{(1,2,-\frac{1}{2})}_{(6,2,2)}$\\[.2cm]
\be
\left(
\begin{array}{cccccc}
\begin{array}{c}
m_3 + \sqrt{\frac{3}{5}} \lambda_{13} E \\
\frac{i \lambda_{16} A_1}{\sqrt{6}} - \frac{\lambda_{17} \Phi_1}{2} \\
\frac{i \lambda_{16} A_2}{\sqrt{3}} - \frac{\lambda_{17} \Phi_3}{2 \sqrt{2}} \\
\frac{\lambda_{4} \Phi_2}{\sqrt{10}} - \frac{\lambda_{4} \Phi_3}{2 \sqrt{5}} \\
- \frac{\lambda_{3} \Phi_2}{\sqrt{10}} - \frac{\lambda_{3} \Phi_3}{2 \sqrt{5}} \\
-\frac{\lambda_3 v_R}{\sqrt{5}}
\end{array}
\begin{array}{c}
-\frac{i \lambda_{16} A_1}{\sqrt{6}} - \frac{\lambda_{17} \Phi_1}{2} \\
m_6 + \sqrt{\frac{3}{5}} \lambda_{14} E \\
\frac{\lambda_{15} \Phi_3}{3 \sqrt{3}} \\
\frac{i \lambda_{19} A_2}{2 \sqrt{5}} + \frac{\lambda_{21} \Phi_3}{4 \sqrt{30}} \\
\frac{i \lambda_{18} A_2}{2 \sqrt{5}} + \frac{\lambda_{20} \Phi_3}{4 \sqrt{30}} \\
-\frac{\lambda_{20} v_R}{2\sqrt{30}}
\end{array}
\begin{array}{c}
-\frac{i \lambda_{16} A_2}{\sqrt{3}} - \frac{\lambda_{17} \Phi_3}{2 \sqrt{2}} \\
\frac{\lambda_{15} \Phi_3}{3 \sqrt{3}} \\
m_{33}^{(1,2,\frac{1}{2})} \\
m_{43}^{(1,2,\frac{1}{2})} \\
m_{53}^{(1,2,\frac{1}{2})} \\
-\frac{\lambda_{20} v_R}{2\sqrt{10}}
\end{array}
\begin{array}{c}
\frac{\lambda_{3} \Phi_2}{\sqrt{10}} - \frac{\lambda_{3} \Phi_3}{2 \sqrt{5}} \\
-\frac{i \lambda_{18} A_2}{2 \sqrt{5}} + \frac{\lambda_{20} \Phi_3}{4 \sqrt{30}} \\
m_{34}^{(1,2,\frac{1}{2})} \\
m_{44}^{(1,2,\frac{1}{2})} \\
\frac{\lambda_{11} E}{\sqrt{15}} \\
0
\end{array}
\begin{array}{c}
- \frac{\lambda_{4} \Phi_2}{\sqrt{10}} - \frac{\lambda_{4} \Phi_3}{2 \sqrt{5}} \\
-\frac{i \lambda_{19} A_2}{2 \sqrt{5}} + \frac{\lambda_{21} \Phi_3}{4 \sqrt{30}} \\
m_{35}^{(1,2,\frac{1}{2})} \\
\frac{\lambda_{12} E}{\sqrt{15}} \\
m_{55}^{(1,2,\frac{1}{2})} \\
\frac{\lambda_2 v_R}{10}
\end{array}
\begin{array}{c}
-\frac{\lambda_4 \overline{v_R}}{\sqrt{5}} \\
-\frac{\lambda_{21} \overline{v_R}}{2\sqrt{30}} \\
-\frac{\lambda_{21} \overline{v_R}}{2\sqrt{10}} \\
0 \\
\frac{\lambda_2 \overline{v_R}}{10} \\
m_{66}^{(1,2,\frac{1}{2})}
\end{array}
\end{array}
\right),  
\ee
\end{minipage}\\[.2cm]
where 
\bea
m_{33}^{(1,2,\frac{1}{2})} &\equiv& 
m_6 - \frac{\lambda_{14} E}{3\sqrt{15}} 
+ \frac{2 \sqrt{2} \lambda_{15} \Phi_2}{9}, 
\nonumber\\
m_{34}^{(1,2,\frac{1}{2})} &\equiv& 
- \frac{i \lambda_{18} A_1}{2 \sqrt{10}} 
+ \frac{i \lambda_{18} A_2}{\sqrt{15}}
+ \frac{\lambda_{20} \Phi_1}{4 \sqrt{15}} 
- \frac{\lambda_{20} \Phi_3}{6 \sqrt{10}}, 
\nonumber\\
m_{35}^{(1,2,\frac{1}{2})} &\equiv& 
- \frac{i \lambda_{19} A_1}{2 \sqrt{10}} 
- \frac{i \lambda_{19} A_2}{\sqrt{15}}
+ \frac{\lambda_{21} \Phi_1}{4 \sqrt{15}} 
+ \frac{\lambda_{21} \Phi_3}{6 \sqrt{10}}, 
\nonumber\\
m_{43}^{(1,2,\frac{1}{2})} &\equiv& 
\frac{i \lambda_{19} A_1}{2 \sqrt{10}} 
- \frac{i \lambda_{19} A_2}{\sqrt{15}}
+ \frac{\lambda_{21} \Phi_1}{4 \sqrt{15}} 
- \frac{\lambda_{21} \Phi_3}{6 \sqrt{10}}, 
\nonumber\\
m_{44}^{(1,2,\frac{1}{2})} &\equiv& 
m_2 + \frac{\lambda_2 \Phi_2}{15\sqrt{2}} 
- \frac{\lambda_2 \Phi_3}{30} +\frac{\lambda_6 A_1}{10}, 
\nonumber\\
m_{53}^{(1,2,\frac{1}{2})} &\equiv& 
\frac{i \lambda_{18} A_1}{2 \sqrt{10}} 
+ \frac{i \lambda_{18} A_2}{\sqrt{15}}
+ \frac{\lambda_{20} \Phi_1}{4 \sqrt{15}} 
+ \frac{\lambda_{20} \Phi_3}{6 \sqrt{10}}, 
\nonumber\\
m_{55}^{(1,2,\frac{1}{2})} &\equiv& 
m_2 + \frac{\lambda_2 \Phi_2}{15\sqrt{2}} 
+ \frac{\lambda_2 \Phi_3}{30} - \frac{\lambda_6 A_1}{10}, 
\nonumber\\
m_{66}^{(1,2,\frac{1}{2})} &\equiv& 
m_1 + \frac{\lambda_1 \Phi_2}{\sqrt{2}} 
+ \frac{\lambda_1 \Phi_3}{2} 
+ \frac{2 \lambda_7 A_1}{5} 
- \frac{1}{4} \sqrt{\frac{3}{5}} \lambda_{10} E.  
\eea

\begin{minipage}{16cm}
$\left[{\bf{(3,1}}, -\frac{1}{3}) +c.c. \right]$\\[.2cm] 
{\bf c:}
$\widehat{H}^{(3,1,-\frac{1}{3})}_{(6,1,1)}$,
$\widehat{D}^{(3,1,-\frac{1}{3})}_{(6,1,3)}$,
$\widehat{D}^{(3,1,-\frac{1}{3})}_{(10,1,1)}$,
$\widehat{\Delta}^{(3,1,-\frac{1}{3})}_{(6,1,1)}$,
$\widehat{\overline{\Delta}}^{(3,1,-\frac{1}{3})}_{(6,1,1)}$,
$\widehat{\overline{\Delta}}^{(3,1,-\frac{1}{3})}_{(10,1,3)}$,
$\widehat{\Phi}^{(3,1,-\frac{1}{3})}_{(15,1,3)}$\\[.15cm]
{\bf r:}
$\widehat{H}^{(\overline{3},1,\frac{1}{3})}_{(6,1,1)}$,
$\widehat{D}^{(\overline{3},1,\frac{1}{3})}_{(6,1,3)}$,
$\widehat{D}^{(\overline{3},1,\frac{1}{3})}_{(\overline{10},1,1)}$,
$\widehat{\overline{\Delta}}^{(\overline{3},1,\frac{1}{3})}_{(6,1,1)}$,
$\widehat{\Delta}^{(\overline{3},1,\frac{1}{3})}_{(6,1,1)}$,
$\widehat{\Delta}^{(\overline{3},1,\frac{1}{3})}_{(\overline{10},1,3)}$,
$\widehat{\Phi}^{(\overline{3},1,\frac{1}{3})}_{(15,1,3)}$\\[.2cm]
\be
\left(
\begin{array}{ccccccc}
\begin{array}{c}
m_3 - \frac{2 \lambda_{13} E}{\sqrt{15}} \\
m_{21}^{(3,1,-\frac{1}{3})} \\
m_{31}^{(3,1,-\frac{1}{3})} \\
m_{41}^{(3,1,-\frac{1}{3})} \\
m_{51}^{(3,1,-\frac{1}{3})} \\
-\sqrt{\frac{2}{15}} \lambda_3 \Phi_3 \\
\frac{\lambda_3 v_R}{\sqrt{5}}
\end{array}
\begin{array}{c}
m_{12}^{(3,1,-\frac{1}{3})} \\
m_{22}^{(3,1,-\frac{1}{3})} \\
\frac{2 \lambda_{15} \Phi_3}{9} \\
m_{42}^{(3,1,-\frac{1}{3})} \\
m_{52}^{(3,1,-\frac{1}{3})} \\
m_{62}^{(3,1,-\frac{1}{3})} \\
\frac{\lambda_{20} v_R}{2 \sqrt{15}} 
\end{array}
\begin{array}{c}
m_{13}^{(3,1,-\frac{1}{3})} \\
\frac{2 \lambda_{15} \Phi_3}{9} \\
m_{33}^{(3,1,-\frac{1}{3})} \\
m_{43}^{(3,1,-\frac{1}{3})} \\
m_{53}^{(3,1,-\frac{1}{3})} \\
m_{63}^{(3,1,-\frac{1}{3})} \\
\frac{\lambda_{20} v_R}{2 \sqrt{15}} 
\end{array}
\begin{array}{c}
m_{14}^{(3,1,-\frac{1}{3})} \\
m_{24}^{(3,1,-\frac{1}{3})} \\
m_{34}^{(3,1,-\frac{1}{3})} \\
m_2 + \frac{\lambda_6 A_2}{5 \sqrt{6}} \\
\frac{2 \lambda_{11} E}{\sqrt{15}} \\
0 \\
0
\end{array}
\begin{array}{c}
m_{15}^{(3,1,-\frac{1}{3})} \\
m_{25}^{(3,1,-\frac{1}{3})} \\
m_{35}^{(3,1,-\frac{1}{3})} \\
\frac{2 \lambda_{12} E}{\sqrt{15}} \\
m_2 - \frac{\lambda_6 A_2}{5 \sqrt{6}} \\
\frac{\lambda_2 \Phi_3}{15 \sqrt{2}} \\
-\frac{\lambda_{2} v_R}{10 \sqrt{3}} 
\end{array}
\begin{array}{c}
-\sqrt{\frac{2}{15}} \lambda_4 \Phi_3 \\
m_{26}^{(3,1,-\frac{1}{3})} \\
m_{36}^{(3,1,-\frac{1}{3})} \\
0 \\
\frac{\lambda_2 \Phi_3}{15 \sqrt{2}} \\
m_{66}^{(3,1,-\frac{1}{3})} \\
-\frac{\lambda_{2} v_R}{5 \sqrt{6}}
\end{array}
\begin{array}{c}
\frac{\lambda_4 \overline{v_R}}{\sqrt{5}} \\
\frac{\lambda_{21} \overline{v_R}}{2 \sqrt{15}} \\
\frac{\lambda_{21} \overline{v_R}}{2 \sqrt{15}} \\
0 \\
-\frac{\lambda_{2} \overline{v_R}}{10 \sqrt{3}} \\
-\frac{\lambda_{2} \overline{v_R}}{5 \sqrt{6}} \\
m_{77}^{(3,1,-\frac{1}{3})}
\end{array}
\end{array}
\right), 
\ee
\end{minipage}\\[.2cm]
where 
\bea
m_{12}^{(3,1,-\frac{1}{3})} &\equiv& 
- \frac{i \lambda_{16} A_1}{\sqrt{3}} - \frac{\lambda_{17} \Phi_3}{2 \sqrt{3}}, 
\nonumber\\
m_{13}^{(3,1,-\frac{1}{3})} &\equiv& 
-\frac{i \sqrt{2} \lambda_{16} A_2}{3} - \frac{\lambda_{17} \Phi_2}{\sqrt{6}}, 
\nonumber\\
m_{14}^{(3,1,-\frac{1}{3})} &\equiv& 
\frac{\lambda_{3} \Phi_2}{\sqrt{30}} - \frac{\lambda_{3} \Phi_1}{\sqrt{10}}, 
\nonumber\\
m_{15}^{(3,1,-\frac{1}{3})} &\equiv& 
-\frac{\lambda_{4} \Phi_1}{\sqrt{10}} - \frac{\lambda_{4} \Phi_2}{\sqrt{30}}, 
\nonumber\\
m_{21}^{(3,1,-\frac{1}{3})} &\equiv&
\frac{i \lambda_{16} A_1}{\sqrt{3}} - \frac{\lambda_{17} \Phi_3}{2 \sqrt{3}}, 
\nonumber\\
m_{22}^{(3,1,-\frac{1}{3})} &\equiv& 
m_6 + \frac{4 \lambda_{14} E}{3 \sqrt{15}} 
+ \frac{1}{3}\sqrt{\frac{2}{3}} \lambda_{15} \Phi_1, 
\nonumber\\
m_{24}^{(3,1,-\frac{1}{3})} &\equiv& 
-\frac{i \lambda_{18} A_1}{2 \sqrt{5}} + \frac{\lambda_{20} \Phi_3}{12 \sqrt{5}}, 
\nonumber\\
m_{25}^{(3,1,-\frac{1}{3})} &\equiv& 
-\frac{i \lambda_{19} A_1}{2 \sqrt{5}} + \frac{\lambda_{21} \Phi_3}{12 \sqrt{5}}, 
\nonumber\\
m_{26}^{(3,1,-\frac{1}{3})} &\equiv& 
-\frac{i \lambda_{19} A_2}{\sqrt{15}} + \frac{\lambda_{21} \Phi_2}{6 \sqrt{5}}, 
\nonumber\\
m_{31}^{(3,1,-\frac{1}{3})} &\equiv& 
\frac{i \sqrt{2} \lambda_{16} A_2}{3} - \frac{\lambda_{17} \Phi_2}{\sqrt{6}}, 
\nonumber\\
m_{33}^{(3,1,-\frac{1}{3})} &\equiv&
m_6 - \frac{2 \lambda_{14} E}{\sqrt{15}} + \frac{\sqrt{2} \lambda_{15} \Phi_2}{9}
\nonumber\\
m_{34}^{(3,1,-\frac{1}{3})} &\equiv& 
\frac{i \lambda_{18} A_2}{\sqrt{30}} - \frac{\lambda_{20} \Phi_2}{6 \sqrt{10}}, 
\nonumber\\
m_{35}^{(3,1,-\frac{1}{3})} &\equiv& 
-\frac{i \lambda_{19} A_2}{\sqrt{30}} + \frac{\lambda_{21} \Phi_2}{6 \sqrt{10}}, 
\nonumber\\
m_{36}^{(3,1,-\frac{1}{3})} &\equiv& 
-\frac{i \lambda_{19} A_1}{\sqrt{10}} + \frac{\lambda_{21} \Phi_3}{6 \sqrt{10}}, 
\nonumber\\
m_{41}^{(3,1,-\frac{1}{3})} &\equiv& 
\frac{\lambda_{4} \Phi_2}{\sqrt{30}} - \frac{\lambda_{4} \Phi_1}{\sqrt{10}}, 
\nonumber\\
m_{42}^{(3,1,-\frac{1}{3})} &\equiv& 
\frac{i \lambda_{19} A_1}{2 \sqrt{5}} + \frac{\lambda_{21} \Phi_3}{12 \sqrt{5}}, 
\nonumber\\
m_{43}^{(3,1,-\frac{1}{3})} &\equiv& 
-\frac{i \lambda_{19} A_2}{\sqrt{30}} - \frac{\lambda_{21} \Phi_2}{6 \sqrt{10}}, 
\nonumber\\
m_{51}^{(3,1,-\frac{1}{3})} &\equiv& 
-\frac{\lambda_{3} \Phi_1}{\sqrt{10}} - \frac{\lambda_{3} \Phi_2}{\sqrt{30}}, 
\nonumber\\
m_{52}^{(3,1,-\frac{1}{3})} &\equiv& 
\frac{i \lambda_{18} A_1}{2 \sqrt{5}} + \frac{\lambda_{20} \Phi_3}{12 \sqrt{5}}, 
\nonumber\\
m_{53}^{(3,1,-\frac{1}{3})} &\equiv& 
\frac{i \lambda_{18} A_2}{\sqrt{30}} + \frac{\lambda_{20} \Phi_2}{6 \sqrt{10}}, 
\nonumber\\
m_{62}^{(3,1,-\frac{1}{3})} &\equiv& 
\frac{i \lambda_{18} A_2}{\sqrt{15}} + \frac{\lambda_{20} \Phi_2}{6 \sqrt{5}}, 
\nonumber\\
m_{63}^{(3,1,-\frac{1}{3})} &\equiv& 
\frac{i \lambda_{18} A_1}{\sqrt{10}} + \frac{\lambda_{20} \Phi_3}{6 \sqrt{10}}, 
\nonumber\\
m_{66}^{(3,1,-\frac{1}{3})} &\equiv& 
m_2 
+ \frac{\lambda_2 \Phi_1}{10 \sqrt{6}} + \frac{\lambda_2 \Phi_2}{30 \sqrt{2}} 
- \frac{\lambda_6 A_2}{5 \sqrt{6}}, 
\nonumber\\
m_{77}^{(3,1,-\frac{1}{3})} &\equiv& 
m_1 
+ \frac{\lambda_1 \Phi_1}{\sqrt{6}}
+ \frac{\lambda_1 \Phi_2}{3 \sqrt{2}} 
+ \frac{2 \lambda_1 \Phi_3}{3} 
+ \frac{2}{5} \sqrt{\frac{2}{3}} \lambda_7 A_2
+ \frac{\lambda_{10} E}{2 \sqrt{15}}.
\eea  

\begin{minipage}{16cm}
$\left[{\bf{(1,1}}, 2) +c.c. \right]$\\[.2cm]
{\bf c:}
$\widehat{\Delta}^{(1,1,2)}_{(\overline{10},1,3)}$\\[.15cm]
{\bf r:}
$\widehat{\overline{\Delta}}^{(1,1,-2)}_{(10,1,3)}$\\[.2cm]
\be
m_2 
+ \frac{\lambda_2 \Phi_1}{10 \sqrt{6}} 
+ \frac{\lambda_2 \Phi_2}{10 \sqrt{2}}
- \frac{\lambda_2 \Phi_3}{10}
+ \frac{\lambda_6 A_1}{5} 
- \frac{1}{5} \sqrt{\frac{3}{2}} \lambda_6 A_2.  
\ee
\end{minipage}\\[.2cm]
\begin{minipage}{16cm}
$\left[{\bf{(1,2}}, \frac{3}{2}) +c.c. \right]$\\[.2cm]
{\bf c:}
$\widehat{\Phi}^{(1,2,\frac{3}{2})}_{(\overline{10},2,2)}$\\[.15cm]
{\bf r:}
$\widehat{\Phi}^{(1,2,-\frac{3}{2})}_{(10,2,2)}$\\[.2cm]
\be
m_1 
+ \frac{\lambda_1 \Phi_2}{\sqrt{2}}
- \frac{\lambda_1 \Phi_3}{2} 
- \frac{2}{5} \lambda_7 A_1
- \frac{1}{4} \sqrt{\frac{3}{5}} \lambda_{10} E.  
\ee
\end{minipage}\\[.2cm]
\begin{minipage}{16cm}
${\bf{(1,3}}, 0)$\\[.2cm]
{\bf c:}
$\widehat{A}^{(1,3,0)}_{(1,3,1)}$,
$\widehat{E}^{(1,3,0)}_{(1,3,3)}$,
$\widehat{\Phi}^{(1,3,0)}_{(15,3,1)}$\\[.15cm]
{\bf r:}
$\widehat{A}^{(1,3,0)}_{(1,3,1)}$,
$\widehat{E}^{(1,3,0)}_{(1,3,3)}$,
$\widehat{\Phi}^{(1,3,0)}_{(15,3,1)}$\\[.2cm]
\be
\left(
\begin{array}{ccc}
\begin{array}{c}
m_4 - \sqrt{\frac{2}{3}} \lambda_5 \Phi_1 + \sqrt{\frac{3}{5}} \lambda_9 E \\
\lambda_9 A_1 \\
- \sqrt{\frac{2}{3}} \lambda_5 A_2 + \frac{2}{5} \sqrt{2} \lambda_7 \Phi_2
\end{array}
\begin{array}{c}
\lambda_9 A_1 \\
m_5 + 3 \sqrt{\frac{3}{5}} \lambda_8 E \\
- \frac{\lambda_{10} \Phi_3}{2} 
\end{array}
\begin{array}{c}
- \sqrt{\frac{2}{3}} \lambda_5 A_2 + \frac{2}{5} \sqrt{2} \lambda_7 \Phi_2 \\
- \frac{\lambda_{10} \Phi_3}{2} \\ 
m_1 
- \frac{\lambda_1 \Phi_1}{\sqrt{6}}
+ \frac{\sqrt{2} \lambda_1 \Phi_2}{3}
- \frac{4}{5} \sqrt{\frac{2}{3}} \lambda_7 A_2
+ \frac{\lambda_{10} E}{2 \sqrt{15}} 
\end{array}
\end{array}
\right).  
\ee
\end{minipage}\\[.2cm]
\begin{minipage}{16cm}
$\left[{\bf{(1,3}}, 1) +c.c. \right]$\\[.2cm]
{\bf c:}
$\widehat{E}^{(1,3,1)}_{(1,3,3)}$,
$\widehat{\overline{\Delta}}^{(1,3,1)}_{(\overline{10},3,1)}$\\[.15cm]
{\bf r:}
$\widehat{E}^{(1,3,-1)}_{(1,3,3)}$,
$\widehat{\Delta}^{(1,3,-1)}_{(10,3,1)}$\\[.2cm]
\be
\label{M131}
\left(
\begin{array}{cc}
\begin{array}{c}
m_5 + 3 \sqrt{\frac{3}{5}} \lambda_{8} E \\
\frac{2 \lambda_{11} v_R}{5} 
\end{array}
\begin{array}{c}
\frac{2 \lambda_{12} \overline{v_R}}{5} \\
m_2 
- \frac{\lambda_2 \Phi_1}{10 \sqrt{6}} 
+ \frac{\lambda_2 \Phi_2}{10 \sqrt{2}}
+ \frac{1}{5} \sqrt{\frac{3}{2}} \lambda_6 A_2
\end{array}
\end{array}
\right).  
\ee
\end{minipage}\\[.2cm]
\begin{minipage}{16cm}
$\left[{\bf{(3,1}}, -\frac{4}{3}) +c.c. \right]$\\[.2cm]
{\bf c:}
$\widehat{D}^{(3,1,-\frac{4}{3})}_{(6,1,3)}$,
$\widehat{\overline{\Delta}}^{(3,1,-\frac{4}{3})}_{(10,1,3)}$\\[.15cm]
{\bf r:}
$\widehat{\overline{D}}^{(\overline{3},1,\frac{4}{3})}_{(6,1,3)}$,
$\widehat{\Delta}^{(\overline{3},1,\frac{4}{3})}_{(\overline{10},1,3)}$\\[.2cm]
\be
\left(
\begin{array}{cc}
\begin{array}{c}
m_6 + \frac{4 \lambda_{14} E}{3 \sqrt{15}} 
+ \frac{1}{3}\sqrt{\frac{2}{3}} \lambda_{15} \Phi_1 
- \frac{2 \lambda_{15} \Phi_3 }{9} \\
\frac{i \lambda_{18} A_2}{\sqrt{15}} 
+ \frac{\lambda_{20} \Phi_2}{6 \sqrt{5}} 
- \frac{\lambda_{20} \Phi_3}{3 \sqrt{10}} 
\end{array}
\begin{array}{c}
- \frac{i \lambda_{19} A_2}{\sqrt{15}} 
+ \frac{\lambda_{21} \Phi_2}{6 \sqrt{5}} 
- \frac{\lambda_{21} \Phi_3}{3 \sqrt{10}} \\
m_2 
+ \frac{\lambda_2 \Phi_1}{10 \sqrt{6}} 
+ \frac{\lambda_2 \Phi_2}{30 \sqrt{2}}
- \frac{\lambda_2 \Phi_3}{30} 
+ \frac{\lambda_6 A_1}{5} 
- \frac{\lambda_6 A_2}{5 \sqrt{6}} 
\end{array}
\end{array}
\right).  
\ee
\end{minipage}\\[.2cm]
\begin{minipage}{16cm}
$\left[{\bf{(3,1}}, \frac{5}{3}) +c.c. \right]$\\[.2cm]
{\bf c:}
$\widehat{\Phi}^{(3,1,\frac{5}{3})}_{(15,1,3)}$\\[.15cm]
{\bf r:}
$\widehat{\Phi}^{(\overline{3},1,-\frac{5}{3})}_{(15,1,3)}$\\[.2cm]
\be
m_1 
+ \frac{\lambda_1 \Phi_1}{\sqrt{6}}
+ \frac{\lambda_1 \Phi_2}{3 \sqrt{2}}
- \frac{2 \lambda_1 \Phi_3}{3}
+ \frac{2}{5} \sqrt{\frac{2}{3}} \lambda_7 A_2
+ \frac{\lambda_{10} E}{2 \sqrt{15}}.  
\ee
\end{minipage}\\[.2cm]
\begin{minipage}{16cm}
$\left[{\bf{(3,2}}, \frac{7}{6}) +c.c. \right]$\\[.2cm]
{\bf c:}
$\widehat{D}^{(3,2,\frac{7}{6})}_{(15,2,2)}$,
$\widehat{\Delta}^{(3,2,\frac{7}{6})}_{(15,2,2)}$,
$\widehat{\overline{\Delta}}^{(3,2,\frac{7}{6})}_{(15,2,2)}$\\[.15cm]
{\bf r:}
$\widehat{D}^{(\overline{3},2,-\frac{7}{6})}_{(15,2,2)}$,
$\widehat{\overline{\Delta}}^{(\overline{3},2,-\frac{7}{6})}_{(15,2,2)}$,
$\widehat{\Delta}^{(\overline{3},2,-\frac{7}{6})}_{(15,2,2)}$\\[.2cm]
\be
\left(
\begin{array}{ccc}
\begin{array}{c}
m_{11}^{(3,2,\frac{7}{6})} \\
m_{21}^{(3,2,\frac{7}{6})} \\
m_{31}^{(3,2,\frac{7}{6})}
\end{array}
\begin{array}{c}
m_{12}^{(3,2,\frac{7}{6})} \\
m_{22}^{(3,2,\frac{7}{6})} \\
\frac{\lambda_{11} E}{\sqrt{15}} 
\end{array}
\begin{array}{c}
m_{13}^{(3,2,\frac{7}{6})} \\
\frac{\lambda_{12} E}{\sqrt{15}} \\ 
m_{33}^{(3,2,\frac{7}{6})} 
\end{array}
\end{array}
\right),  
\ee
\end{minipage}\\[.2cm]
where 
\bea
m_{11}^{(3,2,\frac{7}{6})} &\equiv& 
m_6 - \frac{\lambda_{14} E}{3 \sqrt{15}} 
+ \frac{\sqrt{2} \lambda_{15} \Phi_2}{9} 
- \frac{2 \lambda_{15} \Phi_3}{9}, 
\nonumber\\
m_{12}^{(3,2,\frac{7}{6})} &\equiv& 
- \frac{i \lambda_{18} A_1}{2 \sqrt{10}} 
+ \frac{i \lambda_{18} A_2}{2 \sqrt{15}} 
+ \frac{\lambda_{20} \Phi_1}{4 \sqrt{15}} 
+ \frac{\lambda_{20} \Phi_2}{6 \sqrt{5}} 
- \frac{\lambda_{20} \Phi_3}{4 \sqrt{10}}, 
\nonumber\\
m_{13}^{(3,2,\frac{7}{6})} &\equiv& 
- \frac{i \lambda_{19} A_1}{2 \sqrt{10}} 
- \frac{i \lambda_{19} A_2}{2 \sqrt{15}} 
+ \frac{\lambda_{21} \Phi_1}{4 \sqrt{15}} 
- \frac{\lambda_{21} \Phi_2}{6 \sqrt{5}} 
- \frac{\lambda_{21} \Phi_3}{12 \sqrt{10}}, 
\nonumber\\
m_{21}^{(3,2,\frac{7}{6})} &\equiv& 
\frac{i \lambda_{19} A_1}{2 \sqrt{10}} 
- \frac{i \lambda_{19} A_2}{2 \sqrt{15}} 
+ \frac{\lambda_{21} \Phi_1}{4 \sqrt{15}} 
+ \frac{\lambda_{21} \Phi_2}{6 \sqrt{5}} 
- \frac{\lambda_{21} \Phi_3}{4 \sqrt{10}}, 
\nonumber\\
m_{22}^{(3,2,\frac{7}{6})} &\equiv& 
m_2 + \frac{\lambda_2 \Phi_2}{30 \sqrt{2}}
- \frac{\lambda_2 \Phi_3}{20}
+ \frac{\lambda_6 A_1}{10} 
- \frac{1}{5} \sqrt{\frac{2}{3}} \lambda_6 A_2, 
\nonumber\\
m_{31}^{(3,2,\frac{7}{6})} &\equiv& 
\frac{i \lambda_{18} A_1}{2 \sqrt{10}} 
+ \frac{i \lambda_{18} A_2}{2 \sqrt{15}} 
+ \frac{\lambda_{20} \Phi_1}{4 \sqrt{15}} 
- \frac{\lambda_{20} \Phi_2}{6 \sqrt{5}} 
- \frac{\lambda_{20} \Phi_3}{12 \sqrt{10}}, 
\nonumber\\
m_{33}^{(3,2,\frac{7}{6})} &\equiv& 
m_2 
+ \frac{\lambda_2 \Phi_2}{30 \sqrt{2}}
- \frac{\lambda_2 \Phi_3}{60}
- \frac{\lambda_6 A_1}{10} 
+ \frac{1}{5} \sqrt{\frac{2}{3}} \lambda_6 A_2.  
\eea
\begin{minipage}{16cm}
$\left[{\bf{(3,3}}, -\frac{1}{3}) +c.c. \right]$\\[.2cm]
{\bf c:}
$\widehat{D}^{(3,3,-\frac{1}{3})}_{(6,3,1)}$,
$\widehat{\Delta}^{(3,3,-\frac{1}{3})}_{(10,3,1)}$\\[.15cm]
{\bf r:}
$\widehat{D}^{(\overline{3},3,\frac{1}{3})}_{(6,3,1)}$
$\widehat{\overline{\Delta}}^{(\overline{3},3,\frac{1}{3})}_
{(\overline{10},3,1)}$\\[.2cm]
\be
\left(
\begin{array}{cc}
\begin{array}{c}
m_6 + \frac{4 \lambda_{14} E}{3 \sqrt{15}} 
- \frac{1}{3} \sqrt{\frac{2}{3}} \lambda_{15} \Phi_1 \\
\frac{i \lambda_{19} A_2}{\sqrt{15}} 
+ \frac{\lambda_{21} \Phi_2}{6 \sqrt{5}} 
\end{array}
\begin{array}{c}
- \frac{i \lambda_{18} A_2}{\sqrt{15}} 
+ \frac{\lambda_{20} \Phi_2}{6 \sqrt{5}} \\
m_2 
- \frac{\lambda_2 \Phi_1}{10 \sqrt{6}} 
+ \frac{\lambda_2 \Phi_2}{30 \sqrt{2}}
+ \frac{\lambda_6 A_2}{5 \sqrt{6}} 
\end{array}
\end{array}
\right).  
\ee
\end{minipage}\\[.2cm]
\begin{minipage}{16cm}
$\left[{\bf{(3,3}}, \frac{2}{3}) +c.c. \right]$\\[.2cm]
{\bf c:}
$\widehat{\Phi}^{(3,3,\frac{2}{3})}_{(15,3,1)}$\\[.15cm]
{\bf r:}
$\widehat{\Phi}^{(\overline{3},3,-\frac{2}{3})}_{(15,3,1)}$\\[.2cm]
\be
m_1 
- \frac{\lambda_1 \Phi_1}{\sqrt{6}}
+ \frac{\lambda_1 \Phi_2}{3 \sqrt{2}} 
- \frac{2}{5} \sqrt{\frac{2}{3}} \lambda_7 A_2
+ \frac{\lambda_{10} E}{2 \sqrt{15}}.  
\ee
\end{minipage}\\[.2cm]
\begin{minipage}{16cm}
$\left[{\bf{(6,1}}, -\frac{2}{3}) +c.c. \right]$\\[.2cm]
{\bf c:}
$\widehat{E}^{(6,1,-\frac{2}{3})}_{(20',1,1)}$,
$\widehat{\overline{\Delta}}^{(6,1,-\frac{2}{3})}_{(10,1,3)}$\\[.15cm]
{\bf r:}
$\widehat{E}^{(6,1,\frac{2}{3})}_{(20',1,1)}$,
$\widehat{\Delta}^{(6,1,\frac{2}{3})}_{(\overline{10},1,3)}$\\[.2cm]
\be
\left(
\begin{array}{cc}
\begin{array}{c}
m_5 - 2 \sqrt{\frac{3}{5}} \lambda_{8} E \\
\frac{2 \lambda_{11} v_R}{5} 
\end{array}
\begin{array}{c}
\frac{2 \lambda_{12} \overline{v_R}}{5} \\
m_2 
+ \frac{\lambda_2 \Phi_1}{10 \sqrt{6}} 
- \frac{\lambda_2 \Phi_2}{30 \sqrt{2}}
+ \frac{\lambda_2 \Phi_3}{30}
+ \frac{\lambda_6 A_1}{5} 
+ \frac{\lambda_6 A_2}{5 \sqrt{6}} 
\end{array}
\end{array}
\right).  
\ee
\end{minipage}\\[.2cm]
\begin{minipage}{16cm}
$\left[{\bf{(6,1}}, \frac{1}{3}) +c.c. \right]$\\[.2cm]
{\bf c:}
$\widehat{D}^{(6,1,\frac{1}{3})}_{(10,1,1)}$,
$\widehat{\overline{\Delta}}^{(6,1,\frac{1}{3})}_{(10,1,3)}$\\[.15cm]
{\bf r:}
$\widehat{D}^{(6,1,-\frac{1}{3})}_{(\overline{10},1,1)}$,
$\widehat{\Delta}^{(6,1,-\frac{1}{3})}_{(\overline{10},1,3)}$\\[.2cm]
\be
\left(
\begin{array}{cc}
\begin{array}{c}
m_6 - \frac{2 \lambda_{14} E}{\sqrt{15}} 
- \frac{\sqrt{2} \lambda_{15} \Phi_2}{9} \\
\frac{i \lambda_{18} A_1}{\sqrt{10}} 
- \frac{\lambda_{20} \Phi_3}{6 \sqrt{10}} 
\end{array}
\begin{array}{c}
- \frac{i \lambda_{19} A_1}{\sqrt{10}} 
- \frac{\lambda_{21} \Phi_3}{6 \sqrt{10}} \\
m_2 
+ \frac{\lambda_2 \Phi_1}{10 \sqrt{6}} 
- \frac{\lambda_2 \Phi_2}{30 \sqrt{2}}
+ \frac{\lambda_6 A_2}{5 \sqrt{6}} 
\end{array}
\end{array}
\right).  
\ee
\end{minipage}\\[.2cm]
\begin{minipage}{16cm}
$\left[{\bf{(6,1}}, \frac{4}{3}) +c.c. \right]$\\[.2cm]
{\bf c:}
$\widehat{\overline{\Delta}}^{(6,1,\frac{4}{3})}_{(10,1,3)}$\\[.15cm]
{\bf r:}
$\widehat{\Delta}^{(6,1,-\frac{4}{3})}_{(\overline{10},1,3)}$\\[.2cm]
\be
m_2 
+ \frac{\lambda_2 \Phi_1}{10 \sqrt{6}} 
- \frac{\lambda_2 \Phi_2}{30 \sqrt{2}}
- \frac{\lambda_2 \Phi_3}{30}
- \frac{\lambda_6 A_1}{5} 
+ \frac{\lambda_6 A_2}{5 \sqrt{6}}.  
\ee
\end{minipage}\\[.2cm]
\begin{minipage}{16cm}
$\left[{\bf{(6,2}}, -\frac{1}{6}) +c.c. \right]$\\[.2cm]
{\bf c:}
$\widehat{\Phi}^{(6,2,-\frac{1}{6})}_{(10,2,2)}$\\[.15cm]
{\bf r:}
$\widehat{\Phi}^{(6,2,\frac{1}{6})}_{(\overline{10},2,2)}$\\[.2cm]
\be
m_1 
- \frac{\lambda_1 \Phi_2}{3 \sqrt{2}}
+ \frac{\lambda_1 \Phi_3}{6} 
- \frac{2}{5} \lambda_7 A_1
- \frac{1}{4} \sqrt{\frac{3}{5}} \lambda_{10} E.  
\ee
\end{minipage}\\[.2cm]
\begin{minipage}{16cm}
$\left[{\bf{(6,2}}, \frac{5}{6}) +c.c. \right]$\\[.2cm]
{\bf c:}
$\widehat{\Phi}^{(6,2,\frac{5}{6})}_{(10,2,2)}$\\[.15cm]
{\bf r:}
$\widehat{\Phi}^{(6,2,-\frac{5}{6})}_{(\overline{10},2,2)}$\\[.2cm]
\be
m_1 
- \frac{\lambda_1 \Phi_2}{3 \sqrt{2}}
- \frac{\lambda_1 \Phi_3}{6} 
+ \frac{2}{5} \lambda_7 A_1
- \frac{1}{4} \sqrt{\frac{3}{5}} \lambda_{10} E.  
\ee
\end{minipage}\\[.2cm]
\begin{minipage}{16cm}
$\left[{\bf{(6,3}}, \frac{1}{3}) +c.c. \right]$\\[.2cm]
{\bf c:}
$\widehat{\Delta}^{(6,3,\frac{1}{3})}_{(10,3,1)}$\\[.15cm]
{\bf r:}
$\widehat{\overline{\Delta}}^{(6,3,-\frac{1}{3})}_{(\overline{10},3,1)}$\\[.2cm]
\be
m_2 
- \frac{\lambda_2 \Phi_1}{10 \sqrt{6}}
- \frac{\lambda_2 \Phi_2}{30 \sqrt{2}}
- \frac{\lambda_6 A_2}{5 \sqrt{6}}.  
\ee
\end{minipage}\\[.2cm]
\begin{minipage}{16cm}
${\bf{(8,1}}, 0)$\\[.2cm]
{\bf c:}
$\widehat{A}^{(8,1,0)}_{(15,1,1)}$,
$\widehat{E}^{(8,1,0)}_{(20',1,1)}$,
$\widehat{\Phi}^{(8,1,0)}_{(15,1,1)}$,
$\widehat{\Phi}^{(8,1,0)}_{(15,1,3)}$\\[.15cm]
{\bf r:}
$\widehat{A}^{(8,1,0)}_{(15,1,1)}$,
$\widehat{E}^{(8,1,0)}_{(20',1,1)}$,
$\widehat{\Phi}^{(8,1,0)}_{(15,1,1)}$,
$\widehat{\Phi}^{(8,1,0)}_{(15,1,3)}$\\[.2cm]
\be
\left(
\begin{array}{cccc}
\begin{array}{c}
m_4 - \frac{\sqrt{2} \lambda_5 \Phi_2}{3} - \frac{2 \lambda_9 E}{\sqrt{15}} \\
\sqrt{\frac{2}{3}} \lambda_9 A_2 \\
- \frac{\sqrt{2} \lambda_5 A_2}{3} + \frac{2}{5} \sqrt{2} \lambda_7 \Phi_1 \\
- \sqrt{\frac{2}{3}} \lambda_5 A_1 + \frac{2}{5} \sqrt{\frac{2}{3}} \lambda_7 \Phi_3 \\
\end{array}
\begin{array}{c}
\sqrt{\frac{2}{3}} \lambda_9 A_2 \\
m_5 - 2 \sqrt{\frac{3}{5}} \lambda_8 E \\
- \frac{\lambda_{10} \Phi_2}{\sqrt{6}} \\
- \frac{\lambda_{10} \Phi_3}{\sqrt{6}} 
\end{array}
\begin{array}{c}
- \frac{\sqrt{2} \lambda_5 A_2}{3} + \frac{2 \sqrt{2} \lambda_7 \Phi_1}{5} \\
- \frac{\lambda_{10} \Phi_2}{\sqrt{6}} \\
m_1 - \frac{\lambda_1 \Phi_2}{3 \sqrt{2}} - \frac{2 \lambda_{10} E}{\sqrt{15}} \\
\frac{\lambda_1 \Phi_3}{3 \sqrt{2}} - \frac{2 \sqrt{2} \lambda_7 A_1}{5} 
\end{array}
\begin{array}{c}
- \sqrt{\frac{2}{3}} \lambda_5 A_1 + \frac{2}{5} \sqrt{\frac{2}{3}} \lambda_7 \Phi_3 \\
- \frac{\lambda_{10} \Phi_3}{\sqrt{6}} \\ 
\frac{\lambda_1 \Phi_3}{3 \sqrt{2}} - \frac{2 \sqrt{2} \lambda_7 A_1}{5} \\
m_{44}^{(8,1,0)}
\end{array}
\end{array}
\right),  
\ee
\end{minipage}\\[.2cm]
where 
\be
m_{44}^{(8,1,0)} \equiv
m_1 + \frac{\lambda_1 \Phi_1}{\sqrt{6}}
- \frac{\lambda_1 \Phi_2}{3 \sqrt{2}}
- \frac{2}{5} \sqrt{\frac{2}{3}} \lambda_7 A_2
+ \frac{\lambda_{10} E}{2 \sqrt{15}}.  
\ee

\begin{minipage}{16cm}
$\left[{\bf{(8,1}}, 1) +c.c. \right]$\\[.2cm]
{\bf c:}
$\widehat{\Phi}^{(8,1,1)}_{(15,1,3)}$\\[.15cm]
{\bf r:}
$\widehat{\Phi}^{(8,1,-1)}_{(15,1,3)}$\\[.2cm]
\be
m_1 
+ \frac{\lambda_1 \Phi_1}{\sqrt{6}}
- \frac{\lambda_1 \Phi_2}{3 \sqrt{2}}
- \frac{2}{5} \sqrt{\frac{2}{3}} \lambda_7 A_2
+ \frac{\lambda_{10} E}{2 \sqrt{15}}.  
\ee
\end{minipage}\\[.2cm]
\begin{minipage}{16cm}
$\left[{\bf{(8,2}}, -\frac{1}{2}) +c.c. \right]$\\[.2cm]
{\bf c:}
$\widehat{D}^{(8,2,\frac{1}{2})}_{(15,2,2)}$,
$\widehat{\Delta}^{(8,2,\frac{1}{2})}_{(15,2,2)}$,
$\widehat{\overline{\Delta}}^{(8,2,\frac{1}{2})}_{(15,2,2)}$\\[.15cm]
{\bf r:}
$\widehat{D}^{(8,2,-\frac{1}{2})}_{(15,2,2)}$,
$\widehat{\overline{\Delta}}^{(8,2,-\frac{1}{2})}_{(15,2,2)}$,
$\widehat{\Delta}^{(8,2,-\frac{1}{2})}_{(15,2,2)}$\\[.2cm]
\\[.2cm]
\be
\left(
\begin{array}{ccc}
\begin{array}{c}
m_{11}^{(8,2,-\frac{1}{2})} \\
m_{21}^{(8,2,-\frac{1}{2})}  \\
m_{31}^{(8,2,-\frac{1}{2})}
\end{array}
\begin{array}{c}
m_{12}^{(8,2,-\frac{1}{2})} \\
m_{22}^{(8,2,-\frac{1}{2})} \\
\frac{\lambda_{11} E}{\sqrt{15}} 
\end{array}
\begin{array}{c}
m_{13}^{(8,2,-\frac{1}{2})} \\
\frac{\lambda_{12} E}{\sqrt{15}} \\ 
m_{33}^{(8,2,-\frac{1}{2})} 
\end{array}
\end{array}
\right),  
\ee
\end{minipage}\\[.2cm]
where 
\bea
m_{11}^{(8,2,-\frac{1}{2})} &\equiv& 
m_6 - \frac{\lambda_{14} E}{3 \sqrt{15}} 
- \frac{\sqrt{2} \lambda_{15} \Phi_2}{9}, 
\nonumber\\
m_{12}^{(8,2,-\frac{1}{2})} &\equiv& 
-\frac{i \lambda_{18} A_1}{2 \sqrt{10}} 
- \frac{i \lambda_{18} A_2}{2 \sqrt{15}} 
+ \frac{\lambda_{20} \Phi_1}{4 \sqrt{15}} 
+ \frac{\lambda_{20} \Phi_3}{12 \sqrt{10}}, 
\nonumber\\
m_{13}^{(8,2,-\frac{1}{2})} &\equiv& 
- \frac{i \lambda_{19} A_1}{2 \sqrt{10}} 
+ \frac{i \lambda_{19} A_2}{2 \sqrt{15}} 
+ \frac{\lambda_{21} \Phi_1}{4 \sqrt{15}} 
- \frac{\lambda_{21} \Phi_3}{12 \sqrt{10}}, 
\nonumber\\
m_{21}^{(8,2,-\frac{1}{2})} &\equiv& 
\frac{i \lambda_{19} A_1}{2 \sqrt{10}} 
+ \frac{i \lambda_{19} A_2}{2 \sqrt{15}} 
+ \frac{\lambda_{21} \Phi_1}{4 \sqrt{15}} 
+ \frac{\lambda_{21} \Phi_3}{12 \sqrt{10}}, 
\nonumber\\
m_{22}^{(8,2,-\frac{1}{2})} &\equiv& 
m_2 - \frac{\lambda_2 \Phi_2}{30 \sqrt{2}}
+ \frac{\lambda_2 \Phi_3}{60}
+ \frac{\lambda_6 A_1}{10}, 
\nonumber\\
m_{31}^{(8,2,-\frac{1}{2})} &\equiv& 
\frac{i \lambda_{18} A_1}{2 \sqrt{10}} 
- \frac{i \lambda_{18} A_2}{2 \sqrt{15}} 
+ \frac{\lambda_{20} \Phi_1}{4 \sqrt{15}} 
- \frac{\lambda_{20} \Phi_3}{12 \sqrt{10}}, 
\nonumber\\
m_{33}^{(8,2,-\frac{1}{2})} &\equiv& 
m_2 
- \frac{\lambda_2 \Phi_2}{30 \sqrt{2}}
- \frac{\lambda_2 \Phi_3}{60}
- \frac{\lambda_6 A_1}{10}.  
\eea\\[.2cm]
\begin{minipage}{16cm}
${\bf{(8,3}}, 0)$\\[.2cm]
{\bf c:}
$\widehat{\Phi}^{(8,3,0)}_{(15,3,1)}$\\[.15cm]
{\bf r:}
$\widehat{\Phi}^{(8,3,0)}_{(15,3,1)}$\\[.2cm]
\be
m_1 
- \frac{\lambda_1 \Phi_1}{\sqrt{6}}
- \frac{\lambda_1 \Phi_2}{3 \sqrt{2}} 
+ \frac{2}{5} \sqrt{\frac{2}{3}} \lambda_7 A_2
+ \frac{\lambda_{10} E}{2 \sqrt{15}}.  
\ee
\end{minipage}
\subsection{Tests and consistency checks of the total mass matrix}

The following consistency checks of the total mass matrix have been performed
(see also Ref. (\cite{fuku})):
\begin{enumerate}
\item
The CG coefficients appearing in the total mass matrix have been found to satisfy the
hermiticity relation (\ref{herm}).
\item 
The trace of the total mass matrix has been evaluated and it has been found that it satisfies 
Eq. (\ref{TrMtot}).
\item 
For the $G_{321}$ symmetric 
vacuum, the number of the would-be NG modes have been found to be equal to the 
number of the broken generators i.e. 
massive gauge bosons, $45-12\ =\ 33$. Here, $45$ represents the number 
of gauge bosons in the adjoint irrep of the $SO(10)$ group, and $12$ represents
the number of the gauge bosons in the standard model ($G_{321}$ group). The check
has been performed numerically for several tens of randomly chosen sets of
the parameters of the superpotential $\lambda_1,\cdots,\lambda_{21}$ and 
$m_1,\cdots,m_6$, constrained by VEV Eqs. (\ref{VEVeqs}).
\item 
For the $SU(5)$ symmetric vacuum, it has been found that the number of the different
mass eigenvalues is $21$. The number of independent $SU(5)$ irreps contained 
in the total representation of the model $R$ is $22$
(see Appendix \ref{decomp}), and the would-be NG bosons are in two multiplets
$(({\bf 1})+({\bf 10+\overline{10}}))$. Therefore, there are in total $21$ mass eigenvalues: 
$20$ different from zero and one equal to zero.
All other states have to be accommodated in the $SU(5)$ irreps.
Further, for the $SU(5)$ symmetric vacuum, 
all eigenvalues of the doublet Higgs matrix $({\bf 1,2,}\frac{1}{2})$ 
are contained in the spectrum 
of the triplet Higgs matrix $({\bf 3,1,}-\frac{1}{3})$. The only mass eigenvalue of the
triplet Higgs matrix spectrum not contained in the doublet mass spectrum 
belongs to the $SU(5)$ multiplet ${\bf 50}$, leading to the
following relation between determinants of these two matrices,
\be
\label{checkDT}
\det (M^{(3,1,-\frac{1}{3})} - \lambda\times\mathbf{1})
\ =\
(m^{(3,1,-\frac{1}{3})}_{50}-\lambda)\det (M^{(1,2,\frac{1}{2})} - \lambda\times\mathbf{1}).
\ee
The above relation gives a very strong test for these two matrices. 
All above checks have been performed numerically, in the same way as explained in the previous item.
Our mass matrices passed all checks. 
For the $SU(5)$ symmetric vacuum, all VEVs, except $E$, are different
from zero, and they are non trivially correlated through the VEV Eqs. 
(\ref{VEVeqs}) and $SU(5)$ symmetry conditions (\ref{G5vac}). 
Therefore, these tests serve as a strong check of the total mass matrix up to
terms which depend on $E$.
\item
For the $G_{422}$ symmetric vacuum, it has been  found that the number of different mass eigenvalues
is $27$, what is just the number of the independent $G_{422}$ irreps 
(see Appendix \ref{decomp}). The 
number of NG states, contained in $G_{422}$ irrep $({\bf 6,2,2})$ is $45-21\ =\ 24$. 
These checks have been performed numerically as explained above. 
All other states have been found to be accommodated in $G_{422}$ irreps.
For the $G_{422}$ symmetric vacuum,
VEVs different from zero are $\Phi_1$ and $E$, and therefore 
this serves as a check of the $E$-dependent
parts of the total mass matrix.
\item
Similar tests are satisfied for other higher symmetries 
$G_{51}$,
$G_{421}$, $G_{3221}$ and $G_{3211}$.
\end{enumerate}

We stress that all mass matrices in Ref. \cite{fuku}, derived for minimal 
$SO(10)$ model ($R={\bf 10}+{\bf 126}+{\bf \overline{126}}+{\bf 210}$),
satisfy all above consistency checks,
and are just a special case of the mass matrices in this paper. Further, all results 
including CG coefficients and mass matrices 
were obtained analytically including checks.
 
In the recent calculations in Refs. \cite{senj,aul}, that appeared after 
Ref. \cite{fuku}, the necessary 
condition (\ref{checkDT}) between doublets and triplets is not satisfied. 
The hermiticity condition (\ref{herm}) and the total trace relation (\ref{TrMtot}) are
also not satisfied in these references. Further, none of the higher symmetry tests 
is satisfied. That is a consequence of different phase conventions in Ref. \cite{fuku} 
and phase conventions in Refs. \cite{senj} and \cite{aul}.

In this paper, all results for the CG coefficients and mass matrices have been also obtained 
analytically. Checks have been performed numerically and it has been found that 
the mass matrices satisfy ALL 
consistency checks. 
\section{Mass Matrices of Quarks and Leptons} \label{fermion}
After the symmetry breaking down to the $G_{321}$ subgroup, the electroweak 
symmetry breaking $SU(2)_L \times U(1)_Y \to U(1)_{\mathrm{em}}$ can be achieved 
by the VEVs of doublets included in the fields 
$H$, $D$, $\Delta$, $\overline{\Delta}$ and $\Phi$.  
%
%
These fields are given by (see Table \ref{tabd})
$\widetilde{H}_{u/d}
= H_{(1,2,2)}^{({1,2}, \pm \frac{1}{2})}$,
$D_{u/d}
= D_{(1,2,2)}^{({1,2}, \pm \frac{1}{2})}$,
$\widetilde{D}_{u/d}
= D_{(15,2,2)}^{({1,2}, \pm \frac{1}{2})}$,
$\overline{\Delta}_{u/d}
= \overline{\Delta}_{(15,2,2)}^{({1,2}, \pm \frac{1}{2})}$,
$\Delta_{u/d}
= \Delta_{(15,2,2)}^{({1,2}, \pm \frac{1}{2})}$,
$\Phi_{u/d}
= \Phi_{(\overline{10},2,2)}^{({1,2}, \pm \frac{1}{2})}$.
The Yukawa couplings of Eq. (\ref{Y}) including these doublets 
can be written as follows: 
\bea
W_{Y} &=& 
U_i^{c} \left( 
Y_{10}^{ij} \, \widetilde{H}_{u} + Y_{120}^{ij} \,D_{u} 
+ Y_{120}^{ij} \, \widetilde{D}_{u} 
+ Y_{126}^{ij} \,\overline{\Delta}_{u} \right) Q_j 
\nonumber\\
&+& 
D_i^{c} \left( 
Y_{10}^{ij} \, \widetilde{H}_{d} + Y_{120}^{ij} \, D_{d} 
+ Y_{120}^{ij} \, \widetilde{D}_{d} 
+ Y_{126}^{ij} \,\overline{\Delta}_{d} \right) Q_j 
\nonumber\\
&+& 
N_i^{c} \left( 
Y_{10}^{ij} \, \widetilde{H}_{u} + Y_{120}^{ij} \, D_{u} 
-3 \,Y_{120}^{ij} \, \widetilde{D}_{u} 
-3 \,Y_{126}^{ij} \,\overline{\Delta}_{u} \right) L_j 
\nonumber\\
&+& 
E_i^{c} \left( 
Y_{10}^{ij} \, \widetilde{H}_{d} + Y_{120}^{ij} \, D_{d} 
-3 \,Y_{120}^{ij} \, \widetilde{D}_{d} 
-3 \,Y_{126}^{ij} \,\overline{\Delta}_{d} \right) L_j.  
\eea
where $U^c$, $D^c$, $N^c$ and $E^c$ are the right-handed $SU(2)_L$ 
singlet quark and lepton superfields, $Q$ and $L$ are the left-handed $SU(2)_L$ 
doublet quark and lepton superfields, respectively. This is a generalization 
of the renormalizable minimal $SO(10)$ model \cite{f-o,goh-mohapatra-ng},
including ${\bf 120}$.  
Note that the successful gauge-couplings unification is realized 
with only the MSSM particle contents.  
This means that only one pair of Higgs doublets remains light 
and others should be heavy ($\geq M_G$).  
Here we accept the simple picture that the low-energy superpotential 
is described by only one pair of light Higgs doublets ($H_u$ and $H_d$) 
in the MSSM.  
But, in general, these Higgs fields are admixtures of all Higgs doublets 
having the same quantum numbers in the original model such as:  
\begin{eqnarray} 
H_u &=& 
\widetilde{\alpha}_u^{1} \, \widetilde{H}_u 
+ \widetilde{\alpha}_u^{2} \, D_u 
+ \widetilde{\alpha}_u^{3} \, \widetilde{D}_u 
+ \widetilde{\alpha}_u^{4} \, \overline{\Delta}_u 
+ \widetilde{\alpha}_u^{5} \, \Delta_u 
+ \widetilde{\alpha}_u^{6} \, \Phi_u, 
\nonumber \\
H_d &=& 
\widetilde{\alpha}_d^{1} \, \widetilde{H}_d 
+ \widetilde{\alpha}_d^{2} \, D_d 
+ \widetilde{\alpha}_d^{3} \, \widetilde{D}_d 
+ \widetilde{\alpha}_d^{4} \, \overline{\Delta}_d 
+ \widetilde{\alpha}_d^{5} \, \Delta_d 
+ \widetilde{\alpha}_d^{6} \, \Phi_d, 
\label{mix}
\end{eqnarray} 
where $\widetilde{\alpha}_{u,d}^{i}$ $(i = 1,2,\cdots,5,6)$ denote elements 
of the unitary matrix which rotate the flavor basis in the original model 
into the (SUSY) mass eigenstates.  As mentioned above, the low-energy superpotential 
is described only by the light Higgs doublets $H_u$ and $H_d$ 
\begin{eqnarray}
W_Y &=& 
U_i^c \left( 
\alpha_u^{1} Y_{10}^{ij} + \alpha_u^{2} Y_{120}^{ij} 
+ \alpha_u^{3} Y_{120}^{ij} + \alpha_u^{4} Y_{126}^{ij} 
\right) H_u \, Q_j 
\nonumber\\
&+&
D_i^c \left( 
\alpha_d^{1} Y_{10}^{ij} + \alpha_d^{2} Y_{120}^{ij} 
+ \alpha_d^{3} Y_{120}^{ij} + \alpha_d^{4} Y_{126}^{ij} 
\right) H_d \, Q_j 
\nonumber\\
&+&
N_i^c \left( 
\alpha_u^{1} Y_{10}^{ij} + \alpha_u^{2} Y_{120}^{ij} 
-3 \, \alpha_u^{3} Y_{120}^{ij} -3 \, \alpha_u^{4} Y_{126}^{ij} 
\right) H_u \, L_j 
\nonumber\\
&+&
E_i^c \left( 
\alpha_d^{1} Y_{10}^{ij} + \alpha_d^{2} Y_{120}^{ij} 
-3 \, \alpha_d^{3} Y_{120}^{ij} -3 \, \alpha_d^{4} Y_{126}^{ij} 
\right) H_d \, L_j, 
\label{Yukawa3}
\end{eqnarray} 
where the formulae of the inverse unitary transformation 
of Eq. (\ref{mix}), 
\bea
\widetilde{H}_u &=& \alpha_u^{1} \, H_u, \quad \quad
D_u \ =\ \alpha_u^{2} \,H_u, 
\nonumber\\
\widetilde{D}_u  &=& \alpha_u^{3} \, H_u, \quad \quad 
\overline{\Delta}_u \ =\ \alpha_u^{4} \, H_u, 
\nonumber \\
\widetilde{H}_d &=& \alpha_d^{1} \, H_d, \quad \quad 
D_d \ =\ \alpha_d^{2} \,H_d, 
\nonumber\\
\widetilde{D}_d  &=& \alpha_d^{3} \, H_d, \quad \quad
\overline{\Delta}_d \ =\ \alpha_d^{4} \, H_d, 
\eea
have been used.  

Providing the Higgs VEVs, 
$H_u = v \sin \beta$ and $H_d = v \cos \beta$ 
with $v \simeq 174.1 \,{\mathrm{[GeV]}}$, 
the quark and lepton mass matrices can be read off as 
\bea
\label{MudDeR}
M_u &=& c_{10}\, M_{10} + c_{120} \, M_{120} 
+ \widetilde{c}_{120} \,\widetilde{M}_{120} 
+ c_{126} \,M_{126}, 
\nonumber\\
M_d &=& M_{10} + M_{120} + \widetilde{M}_{120} 
+ M_{126}, 
\nonumber\\
M_D &=& c_{10} \,M_{10} + c_{120} \,M_{120} 
- 3 \,\widetilde{c}_{120} \,\widetilde{M}_{120} 
- 3 \,c_{126} \,M_{126}, 
\nonumber\\
M_e &=& M_{10} + M_{120} - 3 \,\widetilde{M}_{120} 
- 3 \,M_{126}, 
\nonumber\\
M_R &=& c_R \,M_{126},
\eea
where $M_u$, $M_d$, $M_D$, $M_e$, and $M_R$ denote the up-type quark,
down-type quark, neutrino Dirac, charged-lepton, and right-handed neutrino
Majorana mass matrices, respectively.
Note that the mass matrices at the right-hand side of Eq. (\ref{MudDeR}) are defined as
$M_{10}= Y_{10}\, \alpha_d^1 v \cos\beta$,
$M_{120}= Y_{120}\, \alpha_d^2 v \cos\beta$,
$\widetilde{M}_{120}= Y_{120}\, \alpha_d^3 v \cos\beta$,
and $M_{126} = Y_{126}\, \alpha_d^4 v \cos\beta$,
respectively, and the coefficients are defined as
$c_{10}= (\alpha_u^1/\alpha_d^1) \tan \beta$,
$c_{120}= (\alpha_u^2/\alpha_d^2) \tan \beta$,
$\widetilde{c}_{120} = (\alpha_u^3/\alpha_d^3) \tan \beta$,
$c_{126}= (\alpha_u^4/\alpha_d^4) \tan \beta $,
and $c_R = v_R/(\alpha_d^4 v \cos \beta)$.
These mass matrices are directly connected with low-energy
observations and are crucial to model judgement.
\section{Conclusion} \label{conclusion}
We have presented a simple method for the calculation of CG coefficients. 
We have constructed all states for all antisymmetric 
and symmetric $SO(10)$ tensor irreps.
We list 
all tables for the CG coefficients for the SO(10) irreps ${\bf 10}$, ${\bf 45}$,
${\bf 54}$, ${\bf 120}$, ${\bf 126}$, ${\bf \overline{126}}$ and ${\bf 210}$,
for all possible cubic invariants. 
We have constructed all mass matrices for the corresponding
Higgs-Higgsino sector is SUSY GUT $SO(10)$ models. 
We have found a set of consistency checks for the CG coefficients and 
mass matrices which proved the correctness of all our results. 
The results obtained here are useful for a wide class of GUT models
based on the $SO(10)$ group.
\section*{Acknowledgements}
The work of A.I. and S.M. is supported by the Ministry of the Science 
and Technology of the Republic of Croatia under contracts No. 00119261 and No. 0098003, 
respectively. The work of T.F. , T.K. and N.O. is supported 
by the Grant in Aid for Scientific Research from the Ministry of Education, 
Science and Culture. The work of T.K. is also supported by the Research Fellowship 
of the Japan Society for the Promotion of Science for Young Scientists.  
T.F would like to thank G. Senjanovi\'c and A.Y. Smirnov for their hospitality 
at ICTP.
\appendix
\section*{Appendix}
\section{Decomposition of Representations under $G_{321}$}\label{decomp}
Here we list the decompositions of ${\bf 10}$, ${\bf 16}$, ${\bf 45}$, ${\bf 54}$, 
${\bf 120}$, ${\bf \overline{126}}$, and ${\bf 210}$ representations 
under the chain of subgroups 
$G_{422} \supset G_{3122} \supset G_{3121} \supset G_{321}$ 
in Tables \ref{10}, \ref{16}, \ref{45}, \ref{54}, \ref{120}, \ref{126b} and \ref{210}, 
where $U(1)$ groups are related to $SU(4) \to SU(3)_{C} \times U(1)_{B-L}$, 
$SU(2)_{L} \times SU(2)_{R} \to SU(2)_{L} \times U(1)_{R}$.  
\footnote{We use the same notation as Slansky \cite{slansky} but with the proper 
$U(1)$ normalizations.}
Note also that we may consider another chain of subgroups 
$SO(10) \to SU(5) \times U(1)_{X}$ and $SU(5) \to G_{321}$.  
The relations between the generators of $U(1)_{X,Y}$ and $U(1)_{B-L, R}$ are: 
\bea
\frac{1}{10} \left(-X + 4 \, Y \right) &=& B-L , 
\nonumber\\
Y &=& B-L + T^{3}_{R}, 
\eea
where $T^{3}_{R}$ denotes the $U(1)_{R}$ generator.  
\begin{table}[p]
\caption{Decomposition of the representation ${\bf 10}$}
\label{10}
\begin{center}
\begin{tabular}{|c|c|c|c||c|}
\hline \hline
$\left(4,2_L,2_R \right)$ & $\left(3_C,2_L,2_R,1_{B-L} \right)$ 
& $\left(3_C,2_L,1_R,1_{B-L} \right)$ & $\left(3_C,2_L,1_Y \right)$ 
& $\left(5,1_{X} \right)$ \\
\hline
$\left({\bf6,1,1} \right)$ 
& $\left({\bf 3,1,1};-\frac{1}{3} \right)$ 
& $\left({\bf 3,1};0,-\frac{1}{3} \right)$ 
& $\left({\bf 3,1};-\frac{1}{3} \right)$ 
& $\left({\bf5},2 \right)$ 
\\
& $\left({\bf \overline{3},1,1};\frac{1}{3} \right)$ 
& $\left({\bf \overline{3},1};0,\frac{1}{3} \right)$ 
& $\left({\bf \overline{3},1};\frac{1}{3} \right)$ 
& $\left({\bf \overline{5}},-2 \right)$ 
\\
\hline
$\left({\bf 1,2,2} \right)$ 
& $\left({\bf 1,2,2};0 \right)$ 
& $\left({\bf 1,2};\frac{1}{2},0 \right)$ 
& $\left({\bf 1,2};\frac{1}{2} \right)$ 
& $\left({\bf 5},2 \right)$ 
\\
& & $\left({\bf 1,2};-\frac{1}{2},0 \right)$ 
& $\left({\bf 1,2};-\frac{1}{2} \right)$ 
& $\left({\bf \overline{5}},-2 \right)$ 
\\
\hline \hline
\end{tabular}
\end{center}
\end{table}
\begin{table}[p]
\caption{Decomposition of the representation ${\bf 16}$}
\label{16}
\begin{center}
\begin{tabular}{|c|c|c|c||c|}
\hline \hline
$\left(4,2_L,2_R \right)$ & $\left(3_C,2_L,2_R,1_{B-L} \right)$ 
& $\left(3_C,2_L,1_R,1_{B-L} \right)$ & $\left(3_C,2_L,1_Y \right)$ 
& $\left(5,1_{X} \right)$ \\
\hline
$\left({\bf4,2,1} \right)$ 
& $\left({\bf 3,2,1};\frac{1}{6} \right)$ 
& $\left({\bf 3,2};0,\frac{1}{6} \right)$ 
& $\left({\bf 3,2};\frac{1}{6} \right)$ 
& $\left({\bf 10},-1 \right)$ 
\\
& $\left({\bf 1,2,1};-\frac{1}{2} \right)$ 
& $\left({\bf 1,2};0,-\frac{1}{2} \right)$ 
& $\left({\bf 1,2};-\frac{1}{2} \right)$ 
& $\left({\bf \overline{5}},3 \right)$ 
\\
\hline
$\left({\bf \overline{4},1,2} \right)$ 
& $\left({\bf \overline{3},1,2};-\frac{1}{6} \right)$ 
& $\left({\bf \overline{3},1};\frac{1}{2},-\frac{1}{6} \right)$ 
& $\left({\bf \overline{3},1};\frac{1}{3} \right)$ 
& $\left({\bf \overline{5}},3 \right)$ 
\\
& & $\left({\bf \overline{3},1};-\frac{1}{2},-\frac{1}{6} \right)$ 
& $\left({\bf \overline{3},1};-\frac{2}{3} \right)$ 
& $\left({\bf 10},-1 \right)$ 
\\
& $\left({\bf 1,1,2};\frac{1}{2} \right)$ 
& $\left({\bf 1,1};\frac{1}{2},\frac{1}{2} \right)$ 
& $\left({\bf 1,1};1 \right)$ 
& $\left({\bf 10},-1 \right)$ 
\\
& & $\left({\bf 1,1};-\frac{1}{2},\frac{1}{2} \right)$ 
& $\left({\bf 1,1};0 \right)$ 
& $\left({\bf 1},-5 \right)$ 
\\
\hline \hline
\end{tabular}
\end{center}
\end{table}
\begin{table}[p]
\caption{Decomposition of the representation ${\bf 45}$}
\label{45}
\begin{center}
\begin{tabular}{|c|c|c|c||c|}
\hline \hline
$\left(4,2_L,2_R \right)$ & $\left(3_C,2_L,2_R,1_{B-L} \right)$ 
& $\left(3_C,2_L,1_R,1_{B-L} \right)$ & $\left(3_C,2_L,1_Y \right)$ 
& $\left(5,1_{X} \right)$ \\
\hline
$\left({\bf 1,1,3} \right)$ 
& $\left({\bf 1,1,3};0 \right)$ 
& $\left({\bf 1,1};1,0 \right)$ 
& $\left({\bf 1,1};1 \right)$ 
& $\left({\bf10},4 \right)$ 
\\
& & $\left({\bf 1,1};0,0 \right)$ 
& $\left({\bf 1,1};0 \right)$ 
& $\left({\bf1},0 \right)$ 
\\
& & $\left({\bf 1,1};-1,0 \right)$ 
& $\left({\bf 1,1};-1 \right)$ 
& $\left({\bf \overline{10}},-4 \right)$
\\
\hline
$\left({\bf 1,3,1} \right)$ 
& $\left({\bf 1,3,1};0 \right)$ 
& $\left({\bf 1,3};0,0 \right)$ 
& $\left({\bf 1,3};0 \right)$ 
& $\left({\bf 24},0 \right)$
\\
\hline
$\left({\bf 6,2,2} \right)$ 
& $\left({\bf 3,2,2};-\frac{1}{3} \right)$ 
& $\left({\bf 3,2};\frac{1}{2},-\frac{1}{3} \right)$ 
& $\left({\bf 3,2};\frac{1}{6} \right)$ 
& $\left({\bf10},4 \right)$ 
\\
& & $\left({\bf 3,2};-\frac{1}{2},-\frac{1}{3} \right)$ 
& $\left({\bf 3,2};-\frac{5}{6} \right)$ 
& $\left({\bf 24},0 \right)$
\\
& $\left({\bf \overline{3},2,2};\frac{1}{3} \right)$ 
& $\left({\bf \overline{3},2};\frac{1}{2},\frac{1}{3} \right)$ 
& $\left({\bf \overline{3},2};\frac{5}{6} \right)$ 
& $\left({\bf 24},0 \right)$
\\
\hline \hline
\end{tabular}
\end{center}
\end{table}
\begin{table}[p]
\caption{Decomposition of the representation ${\bf 54}$}
\label{54}
\begin{center}
\begin{tabular}{|c|c|c|c||c|}
\hline \hline
$\left(4,2_L,2_R \right)$ & $\left(3_C,2_L,2_R,1_{B-L} \right)$ 
& $\left(3_C,2_L,1_R,1_{B-L} \right)$ & $\left(3_C,2_L,1_Y \right)$ 
& $\left(5, 1_{X} \right)$ \\
\hline
$\left({\bf1,1,1} \right)$ 
& $\left({\bf 1,1,1};0 \right)$ 
& $\left({\bf 1,1};0,0 \right)$ 
& $\left({\bf 1,1};0 \right)$ 
& $\left({\bf 24},0 \right)$
\\ 
\hline
$\left({\bf 1,3,3} \right)$ 
& $\left({\bf 1,3,3};0 \right)$ 
& $\left({\bf 1,3};1,0 \right)$ 
& $\left({\bf 1,3};1 \right)$ 
& $\left({\bf15},4 \right)$ 
\\
& & $\left({\bf 1,3};0,0 \right)$ 
& $\left({\bf 1,3};0 \right)$ 
& $\left({\bf 24},0 \right)$ 
\\
& & $\left({\bf 1,3};-1,0 \right)$ 
& $\left({\bf 1,3};-1 \right)$ 
& $\left({\bf \overline{15}},-4 \right)$ 
\\
\hline
$\left({\bf 20^{\prime},1,1} \right)$ 
& $\left({\bf \overline{6},1,1};\frac{2}{3} \right)$ 
& $\left({\bf \overline{6},1};0,\frac{2}{3} \right)$ 
& $\left({\bf \overline{6},1};\frac{2}{3} \right)$ 
& $\left({\bf \overline{15}},-4 \right)$
\\
& $\left({\bf 6,1,1};-\frac{2}{3} \right)$ 
& $\left({\bf 6,1};0,-\frac{2}{3} \right)$ 
& $\left({\bf 6,1};-\frac{2}{3} \right)$ 
& $\left({\bf15},4 \right)$
\\
& $\left({\bf 8,1,1};0 \right)$ 
& $\left({\bf 8,1};0,0 \right)$ 
& $\left({\bf 8,1};0 \right)$ 
& $\left({\bf 24},0 \right)$ 
\\
\hline \hline
\end{tabular}
\end{center}
\end{table}
\begin{table}[p]
\caption{Decomposition of the representation ${\bf 120}$}
\label{120}
\begin{center}
\begin{tabular}{|c|c|c|c||c|}
\hline \hline
$\left(4,2_L,2_R \right)$ & $\left(3_C,2_L,2_R,1_{B-L} \right)$ 
& $\left(3_C,2_L,1_R,1_{B-L} \right)$ & $\left(3_C,2_L,1_Y \right)$ 
& $\left(5,1_{X} \right)$
\\
\hline
$\left({\bf 1,2,2} \right)$ 
& $\left({\bf 1,2,2};0 \right)$ 
& $\left({\bf 1,2};\frac{1}{2},0 \right)$ 
& $\left({\bf 1,2};\frac{1}{2} \right)$ 
& $\left({\bf5},2 \right)$ 
\\
& & $\left({\bf 1,2};-\frac{1}{2},0 \right)$ 
& $\left({\bf 1,2};-\frac{1}{2} \right)$ 
& $\left({\bf \overline{5}},-2 \right)$ 
\\
\hline
$\left({\bf 10,1,1} \right)$ 
& $\left({\bf 1,1,1};-1 \right)$ 
& $\left({\bf 1,1};0,-1 \right)$ 
& $\left({\bf 1,1};-1 \right)$ 
& $\left({\bf \overline{10}},6 \right)$ 
\\
& $\left({\bf 3,1,1}; -\frac{1}{3} \right)$ 
& $\left({\bf 3,1};0,-\frac{1}{3}  \right)$ 
& $\left({\bf 3,1};-\frac{1}{3}  \right)$ 
& $\left({\bf5},2 \right)$ 
\\
& $\left({\bf 6,1,1};\frac{1}{3} \right)$ 
& $\left({\bf 6,1};0,\frac{1}{3}  \right)$ 
& $\left({\bf 6,1};\frac{1}{3}  \right)$ 
& $\left({\bf \overline{45}},-2 \right)$ 
\\
\hline
$\left({\bf \overline{10},1,1} \right)$ 
& $\left({\bf 1,1,1};1 \right)$ 
& $\left({\bf 1,1};0,1 \right)$ 
& $\left({\bf 1,1};1 \right)$ 
& $\left({\bf10},-6 \right)$ 
\\
& $\left({\bf \overline{3},1,1}; \frac{1}{3} \right)$ 
& $\left({\bf \overline{3},1};0,\frac{1}{3}  \right)$ 
& $\left({\bf \overline{3},1};\frac{1}{3}  \right)$ 
& $\left({\bf \overline{5}},-2 \right)$ 
\\
\hline \hline
\end{tabular}
\end{center}
\end{table}
\begin{table}[p]
\caption{Decomposition of the representation ${\bf \overline{126}}$}
\label{126b}
\begin{center}
\begin{tabular}{|c|c|c|c||c|}
\hline \hline
$\left(4,2_L,2_R \right)$ & $\left(3_C,2_L,2_R,1_{B-L} \right)$ 
& $\left(3_C,2_L,1_R,1_{B-L} \right)$ & $\left(3_C,2_L,1_Y \right)$ 
& $\left(5,1_{X} \right)$ 
\\
\hline
$\left({\bf6,1,1} \right)$ 
& $\left({\bf 3,1,1};-\frac{1}{3} \right)$ 
& $\left({\bf 3,1};0,-\frac{1}{3} \right)$ 
& $\left({\bf 3,1};-\frac{1}{3} \right)$ 
& $\left({\bf 5},2 \right)$ 
\\
& $\left({\bf \overline{3},1,1};\frac{1}{3} \right)$ 
& $\left({\bf \overline{3},1};0,\frac{1}{3} \right)$ 
& $\left({\bf \overline{3},1};\frac{1}{3} \right)$ 
& $\left({\bf \overline{45}},-2 \right)$ 
\\
\hline
$\left({\bf \overline{10},3,1} \right)$ 
& $\left({\bf 1,3,1};1 \right)$ 
& $\left({\bf 1,3};0,1 \right)$ 
& $\left({\bf 1,3};1 \right)$ 
& $\left({\bf 15},-6 \right)$
\\
& $\left({\bf \overline{3},1,3};\frac{1}{3} \right)$ 
& $\left({\bf \overline{3},3};0,\frac{1}{3} \right)$ 
& $\left({\bf \overline{3},3};\frac{1}{3} \right)$ 
& $\left({\bf \overline{45}},-2 \right)$ 
\\
& $\left({\bf \overline{6},1,3};-\frac{1}{3} \right)$ 
& $\left({\bf \overline{6},3};0,-\frac{1}{3} \right)$ 
& $\left({\bf \overline{6},3};-\frac{1}{3} \right)$ 
& $\left({\bf 50},2 \right)$ 
\\
\hline
$\left({\bf 10,1,3} \right)$ 
& $\left({\bf 1,1,3};-1 \right)$ 
& $\left({\bf 1,1};1,-1 \right)$ 
& $\left({\bf 1,1};0 \right)$ 
& $\left({\bf1},10 \right)$ 
\\
& & $\left({\bf 1,1};0,-1 \right)$ 
& $\left({\bf 1,1};-1 \right)$ 
& $\left({\bf \overline{10}},6 \right)$ 
\\
\hline \hline
\end{tabular}
\end{center}
\end{table}
\begin{table}[p]
\begin{center}
\caption{Decomposition of the representation ${\bf 210}$}
\label{210}
\begin{tabular}{|c|c|c|c||c|}
\hline \hline
$\left(4,2_L,2_R \right)$ 
& $\left(3_C,2_L,2_R,1_{B-L} \right)$ 
& $\left(3_C,2_L,1_R,1_{B-L} \right)$ 
& $\left(3_C,2_L,1_Y \right)$ 
& $\left( 5,1_{X} \right)$
\\
\hline
$\left({\bf1,1,1} \right)$ 
& $\left({\bf 1,1,1};0 \right)$ 
& $\left({\bf 1,1};0,0 \right)$ 
& $\left({\bf 1,1};0 \right)$ 
& $\left({\bf1},0 \right)$ 
\\ \hline
$\left({\bf 15,1,1} \right)$ 
& $\left({\bf 1,1,1};0 \right)$ 
& $\left({\bf 1,1};0,0 \right)$ 
& $\left({\bf 1,1};0 \right)$ 
& $\left({\bf 24},0 \right)$ 
\\
& $\left({\bf 3,1,1};\frac{2}{3} \right)$ 
& $\left({\bf 3,1};0,\frac{2}{3} \right)$ 
& $\left({\bf 3,1};\frac{2}{3} \right)$ 
& $\left({\bf \overline{10}},-4 \right)$ 
\\
& $\left({\bf \overline{3},1,1};-\frac{2}{3} \right)$ 
& $\left({\bf \overline{3},1};0,-\frac{2}{3} \right)$ 
& $\left({\bf \overline{3},1};-\frac{2}{3} \right)$ 
& $\left({\bf 10},4 \right)$ 
\\
& $\left({\bf 8,1,1};0 \right)$ 
& $\left({\bf 8,1};0,0 \right)$ 
& $\left({\bf 8,1};0 \right)$ 
& $\left({\bf 24},0 \right)$ 
\\
\hline
$\left({\bf 6,2,2} \right)$ 
& $\left({\bf 3,2,2};-\frac{1}{3} \right)$ 
& $\left({\bf 3,2};\frac{1}{2},-\frac{1}{3} \right)$ 
& $\left({\bf 3,2};\frac{1}{6} \right)$ 
& $\left({\bf10},4 \right)$
\\
& & $\left({\bf 3,2};-\frac{1}{2},-\frac{1}{3} \right)$ 
& $\left({\bf 3,2};-\frac{5}{6} \right)$ 
& $\left({\bf 24},0 \right)$
\\
\hline
\end{tabular}
\end{center}
\end{table}
\begin{table}[h]
\begin{center}
\begin{tabular}{|c|c|c|c||c|}
\hline
$\left({\bf 10,2,2} \right)$ 
& $\left({\bf 1,2,2};-1 \right)$ 
& $\left({\bf 1,2};\frac{1}{2},-1 \right)$ 
& $\left({\bf 1,2};-\frac{1}{2} \right)$ 
& $\left({\bf \overline{5}},8 \right)$ 
\\
& & $\left({\bf 1,2};-\frac{1}{2},-1 \right)$ 
& $\left({\bf 1,2};-\frac{3}{2} \right)$ 
& $\left({\bf \overline{40}},4 \right)$ 
\\
& $\left({\bf 3,2,2};-\frac{1}{3} \right)$ 
& $\left({\bf 3,2};\frac{1}{2},\frac{1}{3} \right)$ 
& $\left({\bf 3,2};\frac{1}{6} \right)$ 
& $\left({\bf \overline{40}},4 \right)$ 
\\
& & $\left({\bf 3,2};-\frac{1}{2},-\frac{1}{3} \right)$ 
& $\left({\bf 3,2};-\frac{5}{6} \right)$ 
& $\left({\bf 75},0 \right)$ 
\\
& $\left({\bf 6,2,2};\frac{1}{3} \right)$ 
& $\left({\bf 6,2};\frac{1}{2},\frac{1}{3} \right)$ 
& $\left({\bf 6,2};\frac{5}{6} \right)$ 
& $\left({\bf 75},0 \right)$ 
\\
& & $\left({\bf 6,2};-\frac{1}{2},\frac{1}{3} \right)$ 
& $\left({\bf 6,2};-\frac{1}{6} \right)$ 
& $\left({\bf 40},-4 \right)$ 
\\
\hline
$\left({\bf \overline{10},2,2} \right)$ 
& $\left({\bf 1,2,2};1 \right)$ 
& $\left({\bf 1,2};\frac{1}{2},1 \right)$ 
& $\left({\bf 1,2};\frac{3}{2} \right)$ 
& $\left({\bf 40},-4 \right)$ 
\\
& & $\left({\bf 1,2};-\frac{1}{2},1 \right)$ 
& $\left({\bf 1,2};\frac{1}{2} \right)$ 
& $\left({\bf5},-8 \right)$
\\
\hline \hline
\end{tabular}
\end{center}
\end{table}
\newpage
\section{CG coefficients}\label{CGC}
The CG coefficients for $HH$, $AA$, $EE$, $DD$, $\Delta \Delta$, 
$\overline{\Delta} \overline{\Delta}$, $\overline{\Delta} \Delta$, 
$\Phi \Phi$, $E A$, $A \Phi$, $E \Delta$, $E \overline{\Delta}$, 
$E \Phi$, $\Delta H$, $\overline{\Delta} H$, $H \Phi$, $\Phi \Delta$, 
$\Phi \overline{\Delta}$, $A \Delta$, $A \overline{\Delta}$, $D H$, 
$D A$, $D \Delta$, $D \overline{\Delta}$ and $D \Phi$ 
combinations are listed in Tables 
\ref{CGC1}, \ref{CGC2}, \ref{CGC3}, \ref{CGC4}, \ref{CGC5}, \ref{CGC6}, 
\ref{CGC7}, \ref{CGC8a}, \ref{CGC9}, \ref{CGC10}, \ref{CGC11}, 
\ref{CGC12}, \ref{CGC13}, \ref{CGC14}, \ref{CGC15}, \ref{CGC16}, \ref{CGC17}, 
\ref{CGC18}, \ref{CGC19}, \ref{CGC21}, \ref{CGC22}, \ref{CGC23}, \ref{CGC24}, 
and \ref{CGC25}, respectively.  
%
%
\begin{table}[h]
\caption{CG coefficients for the $HH$ combination of fields}
\label{CGC1}
\begin{center}
\begin{tabular}{|cc|c|}
\hline \hline
$H$ & $H$ & $\widehat{E}HH$
\\
\hline \hline
$\widehat{H}^{(1,2,\frac{1}{2})}_{(1,2,2)}$ 
& $\widehat{H}^{(1,2,-\frac{1}{2})}_{(1,2,2)}$
& $\frac{\sqrt{3}}{2\sqrt{5}}$
\\
$\widehat{H}^{(3,1,-\frac{1}{3})}_{(6,1,1)}$ 
& $\widehat{H}^{(\overline{3},1,\frac{1}{3})}_{(6,1,1)}$
& $-\frac{1}{\sqrt{15}}$
\\
\hline \hline
\end{tabular}
\end{center}
\end{table}
%
\begin{table}[p]
\caption{CG coefficients for the $AA$ combination of fields}
\label{CGC2}
\begin{center}
\begin{tabular}{|cc|c|ccc|}
\hline \hline
$A$ & $A$ & $\widehat{E}AA$ & $A A \widehat{\Phi}_1$ 
& $A A \widehat{\Phi}_2$& $A A \widehat{\Phi}_3$
\\
\hline \hline
$\widehat{A}^{(1,1,0)}_{(1,1,3)}$ & $\widehat{A}^{(1,1,0)}_{(1,1,3)}$
& $\frac{\sqrt{3}}{2\sqrt{5}}$
& $\frac{1}{\sqrt{6}}$ & $0$ & $0$
\\
$\widehat{A}^{(1,1,0)}_{(15,1,1)}$ & $\widehat{A}^{(1,1,0)}_{(15,1,1)}$
& $-\frac{1}{\sqrt{15}}$
& $0$ & $\frac{\sqrt{2}}{3}$ & $0$
\\
$\widehat{A}^{(1,1,0)}_{(1,1,3)}$ & $\widehat{A}^{(1,1,0)}_{(15,1,1)}$
& $0$
& $0$ & $0$ & $\frac{1}{\sqrt{6}}$
\\
$\widehat{A}^{(1,1,1)}_{(1,1,3)}$ & $\widehat{A}^{(1,1,-1)}_{(1,1,3)}$
& $\frac{\sqrt{3}}{2\sqrt{5}}$
& $\frac{1}{\sqrt{6}}$ & $0$ & $0$
\\
$\widehat{A}^{(1,3,0)}_{(1,3,1)}$ & $\widehat{A}^{(1,3,0)}_{(1,3,1)}$
& $\frac{\sqrt{3}}{2\sqrt{5}}$
& $-\frac{1}{\sqrt{6}}$ & $0$ & $0$
\\
$\widehat{A}^{(3,1,\frac{2}{3})}_{(15,1,1)}$ & $\widehat{A}^{(\overline{3},1,-\frac{2}{3})}_{(15,1,1)}$
& $-\frac{1}{\sqrt{15}}$
& $0$ & $\frac{1}{3\sqrt{2}}$ & $0$
\\
$\widehat{A}^{(3,2,-\frac{5}{6})}_{(6,2,2)}$ & $\widehat{A}^{(\overline{3},2,\frac{5}{6})}_{(6,2,2)}$
& $\frac{1}{4\sqrt{15}}$
& $0$ & $0$ & $-\frac{1}{6}$
\\
$\widehat{A}^{(3,2,\frac{1}{6})}_{(6,2,2)}$ & $\widehat{A}^{(\overline{3},2,-\frac{1}{6})}_{(6,2,2)}$
& $\frac{1}{4\sqrt{15}}$
& $0$ & $0$ & $\frac{1}{6}$
\\
$\widehat{A}^{(8,1,0)}_{(15,1,1)}$ & $\widehat{A}^{(8,1,0)}_{(15,1,1)}$
& $-\frac{1}{\sqrt{15}}$
& $0$ & $-\frac{1}{3\sqrt{2}}$ & $0$
\\
\hline \hline
\end{tabular}
\end{center}
\end{table}
%
%
\begin{table}[p]
\caption{CG coefficients for the $EE$ combination of fields}
\label{CGC3}
\begin{center}
\begin{tabular}{|cc|c|}
\hline\hline
$E$ & $E$ & $\widehat{E}EE$
\\
\hline \hline
$\widehat{E}^{(1,1,0)}_{(1,1,1)}$
& $\widehat{E}^{(1,1,0)}_{(1,1,1)}$
& $\frac{1}{2\sqrt{15}}$
\\
$\widehat{E}^{(1,3,0)}_{(1,3,3)}$
& $\widehat{E}^{(1,3,0)}_{(1,3,3)}$
& $\frac{\sqrt{3}}{2\sqrt{5}}$
\\
$\widehat{E}^{(1,3,1)}_{(1,3,3)}$
& $\widehat{E}^{(1,3,-1)}_{(1,3,3)}$
& $\frac{\sqrt{3}}{2\sqrt{5}}$
\\
$\widehat{E}^{(3,2,-\frac{5}{6})}_{(6,2,2)}$
& $\widehat{E}^{(\overline{3},2,\frac{5}{6})}_{(6,2,2)}$
& $\frac{1}{4\sqrt{15}}$
\\
$\widehat{E}^{(3,2,\frac{1}{6})}_{(6,2,2)}$
& $\widehat{E}^{(\overline{3},2,-\frac{1}{6})}_{(6,2,2)}$
& $\frac{1}{4\sqrt{15}}$
\\
$\widehat{E}^{(6,1,-\frac{2}{3})}_{(20',1,1)}$
& $\widehat{E}^{(\overline{6},1,\frac{2}{3})}_{(20',1,1)}$
& $-\frac{1}{\sqrt{15}}$
\\
$\widehat{E}^{(8,1,0)}_{(20',1,1)}$
& $\widehat{E}^{(8,1,0)}_{(20',1,1)}$
& $-\frac{1}{\sqrt{15}}$
\\
\hline \hline
\end{tabular}
\end{center}
\end{table}
%
%
\begin{table}[p]
\caption{CG coefficients for the $DD$ combination of fields}
\label{CGC4}
\begin{center}
\begin{tabular}{|cc|c|ccc|}
\hline \hline
$D$ & $D$ & $\widehat{E}DD$ & $DD\widehat{\Phi}_1$ & $DD\widehat{\Phi}_2$& $DD\widehat{\Phi}_3$
\\
\hline \hline
$\widehat{D}^{(1,1,1)}_{(\overline{10},1,1)}$ 
& $\widehat{D}^{(1,1,-1)}_{(10,1,1)}$ 

& $-\frac{1}{\sqrt{15}}$ 
& 0 & $\frac{1}{3\sqrt{2}}$ & $0$ 
\\
$\widehat{D}^{(1,2,\frac{1}{2})}_{(1,2,2)}$ 
& $\widehat{D}^{(1,2,-\frac{1}{2})}_{(1,2,2)}$ 
& $\frac{\sqrt{3}}{2\sqrt{5}}$ & 
$0$ & $0$ & $0$
\\
$\widehat{D}^{(1,2,\frac{1}{2})}_{(15,2,2)}$ 
& $\widehat{D}^{(1,2,-\frac{1}{2})}_{(15,2,2)}$ 
& $-\frac{1}{6\sqrt{15}}$ & 
$0$ & $\frac{\sqrt{2}}{9}$ & $0$
\\
$\widehat{D}^{(1,2,\frac{1}{2})}_{(1,2,2)}$ 
& $\widehat{D}^{(1,2,-\frac{1}{2})}_{(15,2,2)}$ 
& $0$  
& $0$ & $0$ & $\frac{1}{6\sqrt{3}}$
\\
$\widehat{D}^{(1,2,\frac{1}{2})}_{(15,2,2)}$ 
& $\widehat{D}^{(1,2,-\frac{1}{2})}_{(1,2,2)}$ 
& $0$ 
& $0$ & $0$ & $\frac{1}{6\sqrt{3}}$
\\
$\widehat{D}^{(3,1,-\frac{4}{3})}_{(6,1,3)}$ 
& $\widehat{D}^{(\overline{3},1,\frac{4}{3})}_{(6,1,3)}$ 
& $\frac{2}{3\sqrt{15}}$ 
& $\frac{1}{3\sqrt{6}}$ & $0$ & $-\frac{1}{9}$
\\
$\widehat{D}^{(3,1,-\frac{1}{3})}_{(6,1,3)}$ 
& $\widehat{D}^{(\overline{3},1,\frac{1}{3})}_{(6,1,3)}$ 
& $\frac{2}{3\sqrt{15}}$ 
& $\frac{1}{3\sqrt{6}}$ & $0$ & $0$
\\
\hline \hline
\end{tabular}
\end{center}
\end{table}
%
%
\begin{table}[p]
\caption{CG coefficients for the $\Delta \Delta$ combination of fields}
\label{CGC5}
\begin{center}
\begin{tabular}{|cc|c|}
\hline \hline
$\Delta$ & $\Delta$ & $\widehat{E}\Delta\Delta$
\\
\hline \hline
$\widehat{\Delta}^{(1,1,0)}_{(\overline{10},1,3)}$ 
& $\widehat{\Delta}^{(1,1,0)}_{(\overline{10},1,3))}$
& $0$
\\
$\widehat{\Delta}^{(1,2,\frac{1}{2})}_{(15,2,2)}$ 
& $\widehat{\Delta}^{(1,2,-\frac{1}{2})}_{(15,2,2)}$
& $\frac{1}{2\sqrt{15}}$
\\
$\widehat{\Delta}^{(3,1,-\frac{1}{3})}_{(6,1,1)}$
& $\widehat{\Delta}^{(\overline{3},1,\frac{1}{3})}_{(6,1,1)}$
& $\frac{1}{\sqrt{15}}$
\\
$\widehat{\Delta}^{(3,1,-\frac{1}{3})}_{(6,1,1)}$
& $\widehat{\Delta}^{(\overline{3},1,\frac{1}{3})}_{(\overline{10},1,3)}$
& $0$
\\
$\widehat{\Delta}^{(3,2,\frac{1}{6})}_{(15,2,2)}$ 
& $\widehat{\Delta}^{(\overline{3},2,-\frac{1}{6})}_{(15,2,2)}$
& $\frac{1}{2\sqrt{15}}$
\\
$\widehat{\Delta}^{(3,2,\frac{7}{6})}_{(15,2,2)}$ 
& $\widehat{\Delta}^{(\overline{3},2,-\frac{7}{6})}_{(15,2,2)}$
& $\frac{1}{2\sqrt{15}}$
\\
$\widehat{\Delta}^{(8,2,-\frac{1}{2})}_{(15,2,2)}$ 
& $\widehat{\Delta}^{(8,2,\frac{1}{2})}_{(15,2,2)}$
& $\frac{1}{2\sqrt{15}}$
\\
\hline \hline
\end{tabular}
\end{center}
\end{table}
%
%
\begin{table}[p]
\caption{CG coefficients for the 
$\overline{\Delta} \overline{\Delta}$ combination of fields}
\label{CGC6}
\begin{center}
\begin{tabular}{|cc|c|}
\hline \hline
$\overline{\Delta}$ & $\overline{\Delta}$ 
& $\widehat{E}\overline{\Delta}\overline{\Delta}$
\\
\hline \hline
$\widehat{\overline{\Delta}}^{(1,1,0)}_{(10,1,3)}$
& $\widehat{\overline{\Delta}}^{(1,1,0)}_{(10,1,3))}$
& $0$
\\
$\widehat{\overline{\Delta}}^{(1,2,\frac{1}{2})}_{(15,2,2)}$
& $\widehat{\overline{\Delta}}^{(1,2,-\frac{1}{2})}_{(15,2,2)}$
& $\frac{1}{2\sqrt{15}}$
\\
$\widehat{\overline{\Delta}}^{(3,1,-\frac{1}{3})}_{(6,1,1)}$
& $\widehat{\overline{\Delta}}^{(\overline{3},1,\frac{1}{3})}_{(6,1,1)}$
& $\frac{1}{\sqrt{15}}$
\\
$\widehat{\overline{\Delta}}^{(3,1,-\frac{1}{3})}_{(10,1,3)}$
& $\widehat{\overline{\Delta}}^{(\overline{3},1,\frac{1}{3})}_{(6,1,1)}$
& $0$
\\
$\widehat{\overline{\Delta}}^{(3,2,\frac{1}{6})}_{(15,2,2)}$
& $\widehat{\overline{\Delta}}^{(\overline{3},2,-\frac{1}{6})}_{(15,2,2)}$
& $\frac{1}{2\sqrt{15}}$
\\
$\widehat{\overline{\Delta}}^{(3,2,\frac{7}{6})}_{(15,2,2)}$
& $\widehat{\overline{\Delta}}^{(\overline{3},2,-\frac{7}{6})}_{(15,2,2)}$
& $\frac{1}{2\sqrt{15}}$
\\
$\widehat{\overline{\Delta}}^{(8,2,-\frac{1}{2})}_{(15,2,2)}$
& $\widehat{\overline{\Delta}}^{(8,2,\frac{1}{2})}_{(15,2,2)}$
& $\frac{1}{2\sqrt{15}}$
\\
\hline \hline
\end{tabular}
\end{center}
\end{table}
%
%
\begin{table}[p]
\caption{CG coefficients for the
$\overline{\Delta} \Delta$ combination of fields}
\label{CGC7}
\begin{center}
\begin{tabular}{|cc|cc|ccc|}
\hline \hline
$\overline{\Delta}$ & $\Delta$ &
$-i\widehat{A}_1 \overline{\Delta} \Delta$ & 
$-i\widehat{A}_2 \overline{\Delta} \Delta$ &
$\widehat{\Phi}_1 \overline{\Delta}\Delta$ &
$\widehat{\Phi}_2 \overline{\Delta}\Delta$ &
$\widehat{\Phi}_3 \overline{\Delta}\Delta$ 
\\
\hline \hline
$\widehat{\overline{\Delta}}^{(1,1,0)}_{(10,1,3)}$
& $\widehat{\Delta}^{(1,1,0)}_{(\overline{10},1,3)}$
& $-\frac{1}{5}$ & $-\frac{\sqrt{3}}{5\sqrt{2}}$ 
& $\frac{1}{10\sqrt{6}}$ & $\frac{1}{10\sqrt{2}}$ & $\frac{1}{10}$
\\
$\widehat{\overline{\Delta}}^{(1,1,-1)}_{(10,1,3)}$
& $\widehat{\Delta}^{(1,1,1)}_{(\overline{10},1,3)}$
& $0$ & $-\frac{\sqrt{3}}{5\sqrt{2}}$ 
& $\frac{1}{10\sqrt{6}}$ & $\frac{1}{10\sqrt{2}}$ & $0$
\\
$\widehat{\overline{\Delta}}^{(1,1,-2)}_{(10,1,3)}$
& $\widehat{\Delta}^{(1,1,2)}_{(\overline{10},1,3)}$
& $\frac{1}{5}$ & $-\frac{\sqrt{3}}{5\sqrt{2}}$
& $\frac{1}{10\sqrt{6}}$ & $\frac{1}{10\sqrt{2}}$ & $-\frac{1}{10}$
\\
$\widehat{\overline{\Delta}}^{(1,2,\frac{1}{2})}_{(15,2,2)}$
& $\widehat{\Delta}^{(1,2,-\frac{1}{2})}_{(15,2,2)}$
& $-\frac{1}{10}$ & $0$
& $0$ & $\frac{1}{15\sqrt{2}}$ & $\frac{1}{30}$
\\
$\widehat{\overline{\Delta}}^{(1,2,-\frac{1}{2})}_{(15,2,2)}$
& $\widehat{\Delta}^{(1,2,\frac{1}{2})}_{(15,2,2)}$
& $\frac{1}{10}$ & $0$ 
& $0$ & $\frac{1}{15\sqrt{2}}$ & $-\frac{1}{30}$
\\
$\widehat{\overline{\Delta}}^{(1,3,1)}_{(\overline{10},3,1)}$
& $\widehat{\Delta}^{(1,3,-1)}_{(10,3,1)}$
& $0$ & $\frac{\sqrt{3}}{5\sqrt{2}}$
& $-\frac{1}{10\sqrt{6}}$ & $\frac{1}{10\sqrt{2}}$ & $0$
\\
$\widehat{\overline{\Delta}}^{(3,1,-\frac{4}{3})}_{(10,1,3)}$
& $\widehat{\Delta}^{(\overline{3},1,\frac{4}{3})}_{(\overline{10},1,3)}$
& $\frac{1}{5}$ & $-\frac{1}{5\sqrt{6}}$
& $\frac{1}{10\sqrt{6}}$ & $\frac{1}{30\sqrt{2}}$ & $-\frac{1}{30}$
\\
\hline \hline
\end{tabular}
\end{center}
\end{table}
%
%
\begin{table}[p]
\caption{CG coefficients for the
$\Phi \Phi$ combination of fields}
\label{CGC8a}
\begin{center}
\begin{tabular}{|cc|cc|c|ccc|}
\hline\hline
$\Phi$ & $\Phi$ &
$\frac{1}{120}\varepsilon \widehat{A}_1 \Phi \Phi$ &
$\frac{1}{120}\varepsilon \widehat{A}_2 \Phi \Phi$ &
$\widehat{E}\Phi\Phi$ &
$\widehat{\Phi}_1 \Phi\Phi$ &
$\widehat{\Phi}_2 \Phi\Phi$ &
$\widehat{\Phi}_3 \Phi\Phi$ 
\\
\hline \hline
$\widehat{\Phi}^{(1,1,0)}_{(1,1,1)}$
& $\widehat{\Phi}^{(1,1,0)}_{(1,1,1)}$
& $0$ & $0$
& $\frac{\sqrt{3}}{2\sqrt{5}}$
& $0$ & $0$ & $0$
\\
$\widehat{\Phi}^{(1,1,0)}_{(1,1,1)}$
& $\widehat{\Phi}^{(1,1,0)}_{(15,1,1)}$
& $0$ & $\frac{\sqrt{2}}{5}$
& $0$
& $0$ & $0$ & $0$
\\
$\widehat{\Phi}^{(1,1,0)}_{(1,1,1)}$
& $\widehat{\Phi}^{(1,1,0)}_{(15,1,3)}$
& $0$ & $0$
& $0$
& $0$ & $0$ & $\frac{1}{6\sqrt{6}}$
\\
$\widehat{\Phi}^{(1,1,0)}_{(15,1,1)}$
& $\widehat{\Phi}^{(1,1,0)}_{(15,1,1)}$
& $0$ & $0$
& $-\frac{1}{\sqrt{15}}$
& $0$ & $\frac{1}{9\sqrt{2}}$ & $0$
\\
$\widehat{\Phi}^{(1,1,0)}_{(15,1,1)}$
& $\widehat{\Phi}^{(1,1,0)}_{(15,1,3)}$
& $\frac{\sqrt{2}}{5}$ & $0$
& $0$
& $0$ & $0$ & $\frac{1}{9\sqrt{2}}$
\\
$\widehat{\Phi}^{(1,1,0)}_{(15,1,3)}$
& $\widehat{\Phi}^{(1,1,0)}_{(15,1,3)}$
& $0$ & $\frac{2\sqrt{2}}{5\sqrt{3}}$
& $\frac{1}{4\sqrt{15}}$
& $\frac{1}{6\sqrt{6}}$ & $\frac{1}{9\sqrt{2}}$ & $0$
\\
$\widehat{\Phi}^{(1,1,1)}_{(15,1,3)}$
& $\widehat{\Phi}^{(1,1,-1)}_{(15,1,3)}$
& $0$ & $\frac{2\sqrt{2}}{5\sqrt{3}}$
& $\frac{1}{4\sqrt{15}}$
& $\frac{1}{6\sqrt{6}}$ & $\frac{1}{9\sqrt{2}}$ & $0$
\\
\hline \hline
\end{tabular}
\end{center}
\end{table}
%
%
\begin{table}[p]
\label{CGC8b}
\begin{center}
\begin{tabular}{|cc|cc|c|ccc|}
\hline\hline
$\Phi$ & $\Phi$ &
$\frac{1}{120} \varepsilon\widehat{A}_1 \Phi \Phi$ &
$\frac{1}{120} \varepsilon\widehat{A}_1 \Phi \Phi$ &
$\widehat{E} \Phi \Phi$ &
$\widehat{\Phi}_1 \Phi \Phi$ &
$\widehat{\Phi}_2 \Phi \Phi$ &
$\widehat{\Phi}_3 \Phi \Phi$
\\
\hline \hline
$\widehat{\Phi}^{(6,2,-\frac{1}{6})}_{(10,2,2)}$
& $\widehat{\Phi}^{(\overline{6},2,\frac{1}{6})}_{(\overline{10},2,2)}$
& $-\frac{1}{5}$ & $0$
& $-\frac{\sqrt{3}}{8\sqrt{5}}$
& $0$ & $-\frac{1}{18\sqrt{2}}$ & $\frac{1}{36}$
\\
$\widehat{\Phi}^{(6,2,\frac{5}{6})}_{(10,2,2)}$
& $\widehat{\Phi}^{(\overline{6},2,-\frac{5}{6})}_{(\overline{10},2,2)}$
& $\frac{1}{5}$ & $0$     
& $-\frac{\sqrt{3}}{8\sqrt{5}}$
& $0$ & $-\frac{1}{18\sqrt{2}}$ & $-\frac{1}{36}$
\\
$\widehat{\Phi}^{(8,1,0)}_{(15,1,1)}$
& $\widehat{\Phi}^{(8,1,0)}_{(15,1,1)}$
& $0$ & $0$
& $-\frac{1}{\sqrt{15}}$
& $0$ & $-\frac{1}{18\sqrt{2}}$ & $0$
\\
$\widehat{\Phi}^{(8,1,0)}_{(15,1,1)}$
& $\widehat{\Phi}^{(8,1,0)}_{(15,1,3)}$
& $-\frac{\sqrt{2}}{5}$ & $0$
& $0$
& $0$ & $0$ & $\frac{1}{18\sqrt{2}}$
\\
$\widehat{\Phi}^{(8,1,0)}_{(15,1,3)}$
& $\widehat{\Phi}^{(8,1,0)}_{(15,1,3)}$
& $0$ & $-\frac{\sqrt{2}}{5\sqrt{3}}$
& $\frac{1}{4\sqrt{15}}$
& $\frac{1}{6\sqrt{6}}$ & $-\frac{1}{18\sqrt{2}}$ & $0$
\\
$\widehat{\Phi}^{(8,1,1)}_{(15,1,3)}$
& $\widehat{\Phi}^{(8,1,-1)}_{(15,1,3)}$
& $0$ & $-\frac{\sqrt{2}}{5\sqrt{3}}$
& $\frac{1}{4\sqrt{15}}$
& $\frac{1}{6\sqrt{6}}$ & $-\frac{1}{18\sqrt{2}}$ & $0$
\\
$\widehat{\Phi}^{(8,3,0)}_{(15,3,1)}$
& $\widehat{\Phi}^{(8,3,0)}_{(15,3,1)}$
& $0$ & $\frac{\sqrt{2}}{5\sqrt{3}}$
& $\frac{1}{4\sqrt{15}}$
& $-\frac{1}{6\sqrt{6}}$ & $-\frac{1}{18\sqrt{2}}$ & $0$
\\
\hline \hline
\end{tabular}
\end{center}
\end{table}
%
%
\begin{table}[p]
\caption{CG coefficients for the
$E A$ combination of fields}
\label{CGC9}
\begin{center}
\begin{tabular}{|cc|cc|}
\hline\hline
$E$ & $A$ & $EA\widehat{A}_1$ & $EA\widehat{A}_2$
\\
\hline \hline
$\widehat{E}^{(1,1,0)}_{(1,1,1)}$
& $\widehat{A}^{(1,1,0)}_{(1,1,3)}$
& $\frac{\sqrt{3}}{2\sqrt{5}}$ & $0$
\\
$\widehat{E}^{(1,1,0)}_{(1,1,1)}$
& $\widehat{A}^{(1,1,0)}_{(15,1,1)}$
& $0$ & $-\frac{1}{\sqrt{15}}$
\\
$\widehat{E}^{(1,3,0)}_{(1,3,3)}$
& $\widehat{A}^{(1,3,0)}_{(1,3,1)}$
& $\frac{1}{2}$ & $0$
\\
$\widehat{E}^{(3,2,-\frac{5}{6})}_{(6,2,2)}$
& $\widehat{A}^{(\overline{3},2,\frac{5}{6})}_{(6,2,2)}$
& $-\frac{1}{4}$ & $-\frac{1}{2\sqrt{6}}$
\\
$\widehat{E}^{(\overline{3},2,\frac{5}{6})}_{(6,2,2)}$
& $\widehat{A}^{(3,2,-\frac{5}{6})}_{(6,2,2)}$
& $-\frac{1}{4}$ & $-\frac{1}{2\sqrt{6}}$
\\
$\widehat{E}^{(3,2,\frac{1}{6})}_{(6,2,2)}$
& $\widehat{A}^{(\overline{3},2,-\frac{1}{6})}_{(6,2,2)}$
& $\frac{1}{4}$ & $-\frac{1}{2\sqrt{6}}$
\\
$\widehat{E}^{(\overline{3},2,-\frac{1}{6})}_{(6,2,2)}$
& $\widehat{A}^{(3,2,\frac{1}{6})}_{(6,2,2)}$
& $\frac{1}{4}$ & $-\frac{1}{2\sqrt{6}}$
\\
$\widehat{E}^{(8,1,0)}_{(20',1,1)}$
& $\widehat{A}^{(8,1,0)}_{(15,1,1)}$
& $0$ & $\frac{1}{\sqrt{6}}$
\\
\hline \hline
\end{tabular}
\end{center}
\end{table}
%
%
\begin{table}[p]
\caption{CG coefficients for the
$A \Phi$ combination of fields}
\label{CGC10}
\begin{center}
\begin{tabular}{|cc|cc|ccc|}
\hline \hline
$A$ & $\Phi$ & 
$\widehat{A}_1A\Phi$ & $\widehat{A}_2A\Phi$ & 
$\frac{1}{120}\varepsilon A\Phi\widehat{\Phi}_1$ & 
$\frac{1}{120}\varepsilon A\Phi\widehat{\Phi}_2$ & 
$\frac{1}{120}\varepsilon A\Phi\widehat{\Phi}_3$
\\
\hline \hline
$\widehat{A}^{(1,1,0)}_{(1,1,3)}$
& $\widehat{\Phi}^{(1,1,0)}_{(1,1,1)}$
& $\frac{1}{\sqrt{6}}$ & $0$
& $0$ & $0$ & $0$
\\
$\widehat{A}^{(1,1,0)}_{(1,1,3)}$
& $\widehat{\Phi}^{(1,1,0)}_{(15,1,1)}$
& $0$ & $0$
& $0$ & $0$ & $\frac{\sqrt{2}}{5}$
\\
$\widehat{A}^{(1,1,0)}_{(1,1,3)}$
& $\widehat{\Phi}^{(1,1,0)}_{(15,1,3)}$
& $0$ & $\frac{1}{\sqrt{6}}$
& $0$ & $\frac{\sqrt{2}}{5}$ & $0$
\\
$\widehat{A}^{(1,1,0)}_{(15,1,1)}$
& $\widehat{\Phi}^{(1,1,0)}_{(1,1,1)}$
& $0$ & $0$
& $0$ & $\frac{\sqrt{2}}{5}$ & $0$
\\
$\widehat{A}^{(1,1,0)}_{(15,1,1)}$
& $\widehat{\Phi}^{(1,1,0)}_{(15,1,1)}$
& $0$ & $\frac{\sqrt{2}}{3}$
& $\frac{\sqrt{2}}{5}$ & $0$ & $0$
\\
$\widehat{A}^{(1,1,0)}_{(15,1,1)}$
& $\widehat{\Phi}^{(1,1,0)}_{(15,1,3)}$
& $\frac{1}{\sqrt{6}}$ & $0$
& $0$ & $0$ & $\frac{2\sqrt{2}}{5\sqrt{3}}$
\\
$\widehat{A}^{(1,1,1)}_{(1,1,3)}$
& $\widehat{\Phi}^{(1,1,-1)}_{(15,1,3)}$
& $0$ & $-\frac{1}{\sqrt{6}}$
& $0$ & $-\frac{\sqrt{2}}{5}$ & $0$
\\
\hline \hline
\end{tabular}
\end{center}
\end{table}
%
%
\begin{table}[p]
\caption{CG coefficients for the
$E \Delta$ combination of fields}
\label{CGC11}
\begin{center}
\begin{tabular}{|cc|c|}
\hline \hline
$E$ & $\Delta$ & $E\Delta\widehat{\Delta}$
\\
\hline \hline
$\widehat{E}^{(1,1,0)}_{(1,1,1)}$
& $\widehat{\Delta}^{(1,1,0)}_{(\overline{10},1,3)}$
& $0$ 
\\
$\widehat{E}^{(1,3,1)}_{(1,3,3)}$
& $\widehat{\Delta}^{(1,3,-1)}_{(10,3,1)}$
& $\frac{1}{5}$
\\
$\widehat{E}^{(3,2,\frac{1}{6})}_{(6,2,2)}$
& $\widehat{\Delta}^{(\overline{3},2,-\frac{1}{6})}_{(15,2,2)}$
& $-\frac{1}{5}$
\\
$\widehat{E}^{(\overline{3},2,-\frac{1}{6})}_{(6,2,2)}$
& $\widehat{\Delta}^{(3,2,\frac{1}{6})}_{(15,2,2)}$
& $0$
\\
$\widehat{E}^{(6,1,-\frac{2}{3})}_{(20',1,1)}$
& $\widehat{\Delta}^{(\overline{6},1,\frac{2}{3})}_{(\overline{10},1,3)}$
& $\frac{1}{5}$
\\
\hline \hline
\end{tabular}
\end{center}
\end{table}
%
%
\begin{table}[p]
\caption{CG coefficients for the
$E \overline{\Delta}$ combination of fields}
\label{CGC12}
\begin{center}
\begin{tabular}{|cc|c|}
\hline \hline
$E$ & $\overline{\Delta}$ & $E\overline{\Delta}\widehat{\overline{\Delta}}$
\\
\hline \hline
$\widehat{E}^{(1,1,0)}_{(1,1,1)}$
& $\widehat{\overline{\Delta}}^{(1,1,0)}_{(10,1,3)}$
& $0$
\\
$\widehat{E}^{(1,3,-1)}_{(1,3,3)}$
& $\widehat{\overline{\Delta}}^{(1,3,1)}_{(\overline{10},3,1)}$
& $\frac{1}{5}$
\\
$\widehat{E}^{(\overline{3},2,-\frac{1}{6})}_{(6,2,2)}$
& $\widehat{\overline{\Delta}}^{(3,2,\frac{1}{6})}_{(15,2,2)}$
& $-\frac{1}{5}$
\\
$\widehat{E}^{(3,2,\frac{1}{6})}_{(6,2,2)}$
& $\widehat{\overline{\Delta}}^{(\overline{3},2,-\frac{1}{6})}_{(15,2,2)}$
& $0$
\\
$\widehat{E}^{(\overline{6},1,\frac{2}{3})}_{(20',1,1)}$
& $\widehat{\overline{\Delta}}^{(6,1,-\frac{2}{3})}_{(10,1,3)}$
& $\frac{1}{5}$
\\
\hline \hline
\end{tabular}
\end{center}
\end{table}
%
%
\begin{table}[p]
\caption{CG coefficients for the
$E \Phi$ combination of fields}
\label{CGC13}
\begin{center}
\begin{tabular}{|cc|ccc|}
\hline \hline
$E$ & $\Phi$ & 
$E\Phi\widehat{\Phi}_1$ & 
$E\Phi\widehat{\Phi}_2$ & 
$E\Phi\widehat{\Phi}_3$
\\
\hline \hline
$\widehat{E}^{(1,1,0)}_{(1,1,1)}$
& $\widehat{\Phi}^{(1,1,0)}_{(1,1,1)}$
& $\frac{\sqrt{3}}{2\sqrt{5}}$ & $0$ & $0$
\\
$\widehat{E}^{(1,1,0)}_{(1,1,1)}$
& $\widehat{\Phi}^{(1,1,0)}_{(15,1,1)}$
& $0$ & $-\frac{1}{\sqrt{15}}$ & $0$
\\
$\widehat{E}^{(1,1,0)}_{(1,1,1)}$
& $\widehat{\Phi}^{(1,1,0)}_{(15,1,3)}$
& $0$ & $0$ & $\frac{1}{4\sqrt{15}}$
\\
$\widehat{E}^{(1,3,0)}_{(1,3,3)}$
& $\widehat{\Phi}^{(1,3,0)}_{(15,3,1)}$
& $0$ & $0$ & $-\frac{1}{4}$ 
\\
$\widehat{E}^{(3,2,-\frac{5}{6})}_{(6,2,2)}$
& $\widehat{\Phi}^{(\overline{3},2,\frac{5}{6})}_{(\overline{10},2,2)}$
& $0$ & $\frac{1}{4\sqrt{3}}$ & $\frac{1}{4\sqrt{6}}$
\\
$\widehat{E}^{(\overline{3},2,\frac{5}{6})}_{(6,2,2)}$
& $\widehat{\Phi}^{(3,2,-\frac{5}{6})}_{(10,2,2)}$
& $0$ & $\frac{1}{4\sqrt{3}}$ & $\frac{1}{4\sqrt{6}}$
\\
$\widehat{E}^{(3,2,-\frac{5}{6})}_{(6,2,2)}$
& $\widehat{\Phi}^{(\overline{3},2,\frac{5}{6})}_{(6,2,2)}$
& $\frac{1}{4\sqrt{2}}$ & $0$ & $\frac{1}{8\sqrt{3}}$
\\
\hline \hline
\end{tabular}
\end{center}
\end{table}
%
%
\begin{table}[p]
\caption{CG coefficients for the
$\Delta H$ combination of fields}
\label{CGC14}
\begin{center}
\begin{tabular}{|cc|ccc|}
\hline \hline
$\Delta$ & $H$ & 
$\Delta H\widehat{\Phi}_1$ & 
$\Delta H\widehat{\Phi}_2$ & 
$\Delta H\widehat{\Phi}_3$
\\
\hline \hline
$\widehat{\Delta}^{(1,2,\frac{1}{2})}_{(15,2,2)}$
& $\widehat{H}^{(1,2,-\frac{1}{2})}_{(1,2,2)}$
& $0$ & $\frac{1}{\sqrt{10}}$ & $-\frac{1}{2\sqrt{5}}$
\\
$\widehat{\Delta}^{(1,2,-\frac{1}{2})}_{(15,2,2)}$
& $\widehat{H}^{(1,2,\frac{1}{2})}_{(1,2,2)}$
& $0$ & $-\frac{1}{\sqrt{10}}$ & $-\frac{1}{2\sqrt{5}}$
\\
$\widehat{\Delta}^{(3,1,-\frac{1}{3})}_{(6,1,1)}$
& $\widehat{H}^{(\overline{3},1,\frac{1}{3})}_{(6,1,1)}$
& $-\frac{1}{\sqrt{10}}$ & $\frac{1}{\sqrt{30}}$ & $0$
\\
$\widehat{\Delta}^{(\overline{3},1,\frac{1}{3})}_{(6,1,1)}$
& $\widehat{H}^{(3,1,-\frac{1}{3})}_{(6,1,1)}$
& $-\frac{1}{\sqrt{10}}$ & $-\frac{1}{\sqrt{30}}$ & $0$
\\
$\widehat{\Delta}^{(\overline{3},1,\frac{1}{3})}_{(\overline{10},1,3)}$
& $\widehat{H}^{(3,1,-\frac{1}{3})}_{(6,1,1)}$
& $0$ & $0$ & $-\frac{\sqrt{2}}{\sqrt{15}}$
\\
\hline \hline
\end{tabular}
\end{center}
\end{table}
%
%
\begin{table}[p]
\caption{CG coefficients for the
$\overline{\Delta} H$ combination of fields}
\label{CGC15}
\begin{center}
\begin{tabular}{|cc|ccc|}
\hline \hline
$\overline{\Delta}$ & $H$ & 
$\overline{\Delta}H\widehat{\Phi}_1$ & 
$\overline{\Delta}H\widehat{\Phi}_2$ & 
$\overline{\Delta}H\widehat{\Phi}_3$
\\
\hline \hline
$\widehat{\overline{\Delta}}^{(1,2,-\frac{1}{2})}_{(15,2,2)}$
& $\widehat{H}^{(1,2,\frac{1}{2})}_{(1,2,2)}$
& $0$ & $\frac{1}{\sqrt{10}}$ & $-\frac{1}{2\sqrt{5}}$
\\
$\widehat{\overline{\Delta}}^{(1,2,\frac{1}{2})}_{(15,2,2)}$
& $\widehat{H}^{(1,2,-\frac{1}{2})}_{(1,2,2)}$
& $0$ & $-\frac{1}{\sqrt{10}}$ & $-\frac{1}{2\sqrt{5}}$
\\
$\widehat{\overline{\Delta}}^{(\overline{3},1,\frac{1}{3})}_{(6,1,1)}$
& $\widehat{H}^{(3,1,-\frac{1}{3})}_{(6,1,1)}$
& $-\frac{1}{\sqrt{10}}$ & $\frac{1}{\sqrt{30}}$ & $0$
\\
$\widehat{\overline{\Delta}}^{(3,1,-\frac{1}{3})}_{(6,1,1)}$
& $\widehat{H}^{(\overline{3},1,\frac{1}{3})}_{(6,1,1)}$
& $-\frac{1}{\sqrt{10}}$ & $-\frac{1}{\sqrt{30}}$ & $0$
\\
$\widehat{\overline{\Delta}}^{(3,1,-\frac{1}{3})}_{(10,1,3)}$
& $\widehat{H}^{(\overline{3},1,\frac{1}{3})}_{(6,1,1)}$
& $0$ & $0$ & $-\frac{\sqrt{2}}{\sqrt{15}}$
\\
\hline \hline
\end{tabular}
\end{center}
\end{table}
%
%
\begin{table}[p]
\caption{CG coefficients for the $H \Phi$ combination of fields}
\label{CGC16}
\begin{center}
\begin{tabular}{|cc|c|c|}
\hline \hline
$H$ & $\Phi$ & 
$\widehat{\Delta} H\Phi$ & $\widehat{\overline{\Delta}} H\Phi$ 
\\
\hline \hline
$\widehat{H}^{(1,2,\frac{1}{2})}_{(1,2,2)}$
& $\widehat{\Phi}^{(1,2,-\frac{1}{2})}_{(10,2,2)}$
& $-\frac{1}{\sqrt{5}}$
& $0$
\\
$\widehat{H}^{(1,2,-\frac{1}{2})}_{(1,2,2)}$
& $\widehat{\Phi}^{(1,2,\frac{1}{2})}_{(\overline{10},2,2)}$
& $0$
& $-\frac{1}{\sqrt{5}}$
\\
$\widehat{H}^{(3,1,-\frac{1}{3})}_{(6,1,1)}$
& $\widehat{\Phi}^{(\overline{3},1,\frac{1}{3})}_{(15,1,3)}$
& $\frac{1}{\sqrt{5}}$
& $0$
\\
$\widehat{H}^{(\overline{3},1,\frac{1}{3})}_{(6,1,1)}$
& $\widehat{\Phi}^{(3,1,-\frac{1}{3})}_{(15,1,3)}$
& $0$
& $\frac{1}{\sqrt{5}}$
\\
\hline \hline
\end{tabular}
\end{center}
\end{table}
\newpage
%
\begin{table}[h]
\caption{CG coefficients for the $\Phi \Delta$ combination of fields}
\label{CGC17}
\begin{center}
\begin{tabular}{|cc|c|}
\hline \hline
$\Phi$ & $\Delta$ & $\Phi \widehat{\overline{\Delta}} \Delta$
\\
\hline \hline
$\widehat{\Phi}^{(1,1,0)}_{(1,1,1)}$
& $\widehat{\Delta}^{(1,1,0)}_{(\overline{10},1,3)}$
& $\frac{1}{10\sqrt{6}}$ 
\\
$\widehat{\Phi}^{(1,1,0)}_{(15,1,1)}$
& $\widehat{\Delta}^{(1,1,0)}_{(\overline{10},1,3)}$
& $\frac{1}{10\sqrt{2}}$
\\
$\widehat{\Phi}^{(1,1,0)}_{(15,1,3)}$
& $\widehat{\Delta}^{(1,1,0)}_{(\overline{10},1,3)}$
& $\frac{1}{10}$
\\
$\widehat{\Phi}^{(1,1,-1)}_{(15,1,3)}$
& $\widehat{\Delta}^{(1,1,1)}_{(\overline{10},1,3)}$
& $-\frac{1}{10}$
\\
$\widehat{\Phi}^{(1,2,-\frac{1}{2})}_{(10,2,2)}$
& $\widehat{\Delta}^{(1,2,\frac{1}{2})}_{(15,2,2)}$
& $0$
\\
$\widehat{\Phi}^{(1,2,\frac{1}{2})}_{(\overline{10},2,2)}$
& $\widehat{\Delta}^{(1,2,-\frac{1}{2})}_{(15,2,2)}$
& $\frac{1}{10}$
\\
$\widehat{\Phi}^{(3,1,-\frac{1}{3})}_{(15,1,3)}$
& $\widehat{\Delta}^{(\overline{3},1,\frac{1}{3})}_{(6,1,1)}$
& $-\frac{1}{10\sqrt{3}}$
\\
\hline \hline
\end{tabular}
\end{center}
\end{table}
%
%
\begin{table}[p]
\caption{CG coefficients for the
$\Phi \overline{\Delta}$ combination of fields}
\label{CGC18}
\begin{center}
\begin{tabular}{|cc|c|}
\hline \hline
$\Phi$ & $\overline{\Delta}$ & $\Phi\widehat{\Delta}\overline{\Delta}$
\\
\hline \hline
$\widehat{\Phi}^{(1,1,0)}_{(1,1,1)}$
& $\widehat{\overline{\Delta}}^{(1,1,0)}_{(10,1,3)}$
& $\frac{1}{10\sqrt{6}}$
\\
$\widehat{\Phi}^{(1,1,0)}_{(15,1,1)}$
& $\widehat{\overline{\Delta}}^{(1,1,0)}_{(10,1,3)}$
& $\frac{1}{10\sqrt{2}}$
\\
$\widehat{\Phi}^{(1,1,0)}_{(15,1,3)}$
& $\widehat{\overline{\Delta}}^{(1,1,0)}_{(10,1,3)}$
& $\frac{1}{10}$
\\
$\widehat{\Phi}^{(1,1,1)}_{(15,1,3)}$
& $\widehat{\overline{\Delta}}^{(1,1,-1)}_{(10,1,3)}$
& $-\frac{1}{10}$
\\
$\widehat{\Phi}^{(1,2,\frac{1}{2})}_{(\overline{10},2,2)}$
& $\widehat{\overline{\Delta}}^{(1,2,-\frac{1}{2})}_{(15,2,2)}$
& $0$
\\
$\widehat{\Phi}^{(1,2,-\frac{1}{2})}_{(10,2,2)}$
& $\widehat{\overline{\Delta}}^{(1,2,\frac{1}{2})}_{(15,2,2)}$
& $\frac{1}{10}$
\\
$\widehat{\Phi}^{(\overline{3},1,\frac{1}{3})}_{(15,1,3)}$
& $\widehat{\overline{\Delta}}^{(3,1,-\frac{1}{3})}_{(6,1,1)}$
& $-\frac{1}{10\sqrt{3}}$
\\
\hline \hline
\end{tabular}
\end{center}
\end{table}
%
%
\begin{table}[p]
\caption{CG coefficients for the
$A \Delta$ combination of fields}
\label{CGC19}
\begin{center}
\begin{tabular}{|cc|c|}
\hline \hline
$A$ & $\Delta$ & $-iA\widehat{\overline{\Delta}}\Delta$
\\
\hline \hline
$\widehat{A}^{(1,1,0)}_{(1,1,3)}$
& $\widehat{\Delta}^{(1,1,0)}_{(\overline{10},1,3)}$
& $-\frac{1}{5}$
\\
$\widehat{A}^{(1,1,0)}_{(15,1,1)}$
& $\widehat{\Delta}^{(1,1,0)}_{(\overline{10},1,3)}$
& $-\frac{\sqrt{3}}{5\sqrt{2}}$
\\
$\widehat{A}^{(1,1,-1)}_{(1,1,3)}$
& $\widehat{\Delta}^{(1,1,1)}_{(\overline{10},1,3)}$
& $-\frac{1}{5}$
\\
$\widehat{A}^{(3,1,\frac{2}{3})}_{(15,1,1)}$
& $\widehat{\Delta}^{(\overline{3},1,-\frac{2}{3})}_{(\overline{10},1,3)}$
& $-\frac{1}{5}$
\\
$\widehat{A}^{(3,2,\frac{1}{6})}_{(6,2,2)}$
& $\widehat{\Delta}^{(\overline{3},2,-\frac{1}{6})}_{(15,2,2)}$
& $0$
\\
$\widehat{A}^{(\overline{3},2,-\frac{1}{6})}_{(6,2,2)}$
& $\widehat{\Delta}^{(3,2,\frac{1}{6})}_{(15,2,2)}$
& $-\frac{1}{5}$
\\
\hline \hline
\end{tabular}
\end{center}
\end{table}
%
%
\begin{table}[p]
\caption{CG coefficients for the
$A \overline{\Delta}$ combination of fields}
\label{CGC20}
\begin{center}
\begin{tabular}{|cc|c|}
\hline \hline
$A$ & $\overline{\Delta}$ & $-iA\overline{\Delta}\widehat{\Delta}$
\\
\hline \hline
$\widehat{A}^{(1,1,0)}_{(1,1,3)}$
& $\widehat{\overline{\Delta}}^{(1,1,0)}_{(10,1,3)}$
& $-\frac{1}{5}$
\\
$\widehat{A}^{(1,1,0)}_{(15,1,1)}$
& $\widehat{\overline{\Delta}}^{(1,1,0)}_{(10,1,3)}$
& $-\frac{\sqrt{3}}{5\sqrt{2}}$
\\
$\widehat{A}^{(1,1,1)}_{(1,1,3)}$
& $\widehat{\overline{\Delta}}^{(1,1,-1)}_{(10,1,3)}$
& $-\frac{1}{5}$
\\
$\widehat{A}^{(\overline{3},1,-\frac{2}{3})}_{(15,1,1)}$
& $\widehat{\overline{\Delta}}^{(3,1,\frac{2}{3})}_{(10,1,3)}$
& $-\frac{1}{5}$
\\
$\widehat{A}^{(\overline{3},2,-\frac{1}{6})}_{(6,2,2)}$
& $\widehat{\overline{\Delta}}^{(3,2,\frac{1}{6})}_{(15,2,2)}$
& $0$
\\
$\widehat{A}^{(3,2,\frac{1}{6})}_{(6,2,2)}$
& $\widehat{\overline{\Delta}}^{(\overline{3},2,-\frac{1}{6})}_{(15,2,2)}$
& $-\frac{1}{5}$
\\
\hline \hline
\end{tabular}
\end{center}
\end{table}
%
%
\begin{table}[p]
\caption{CG coefficients for the 
$D H$ combination of fields}
\label{CGC21}
\begin{center}
\begin{tabular}{|cc|cc|ccc|}
\hline \hline
$D$ & $H$ & 
$DH\widehat{A}_1$ & $DH\widehat{A}_2$ & 
$DH\widehat{\Phi}_1$ & $DH\widehat{\Phi}_2$ & $DH\widehat{\Phi}_3$
\\
\hline \hline
$\widehat{D}^{(1,2,\frac{1}{2})}_{(1,2,2)}$
& $\widehat{H}^{(1,2,-\frac{1}{2})}_{(1,2,2)}$
& $-\frac{i}{\sqrt{6}}$ & $0$
& $-\frac{1}{2}$ & $0$ & $0$
\\
$\widehat{D}^{(1,2,-\frac{1}{2})}_{(1,2,2)}$
& $\widehat{H}^{(1,2,\frac{1}{2})}_{(1,2,2)}$
& $\frac{i}{\sqrt{6}}$ & $0$
& $-\frac{1}{2}$ & $0$ & $0$
\\
$\widehat{D}^{(1,2,\frac{1}{2})}_{(15,2,2)}$
& $\widehat{H}^{(1,2,-\frac{1}{2})}_{(1,2,2)}$
& $0$ & $-\frac{i}{\sqrt{3}}$
& $0$ & $0$ & $-\frac{1}{2\sqrt{2}}$
\\
$\widehat{D}^{(1,2,-\frac{1}{2})}_{(15,2,2)}$
& $\widehat{H}^{(1,2,\frac{1}{2}))}_{(1,2,2)}$
& $0$ & $\frac{i}{\sqrt{3}}$
& $0$ & $0$ & $-\frac{1}{2\sqrt{2}}$
\\
$\widehat{D}^{(3,1,-\frac{1}{3})}_{(6,1,3)}$
& $\widehat{H}^{(\overline{3},1,\frac{1}{3})}_{(6,1,1)}$
& $-\frac{i}{\sqrt{3}}$ & $0$
& $0$ & $0$ & $-\frac{1}{2\sqrt{3}}$
\\
$\widehat{D}^{(\overline{3},1,\frac{1}{3})}_{(6,1,3)}$
& $\widehat{H}^{(3,1,-\frac{1}{3})}_{(6,1,1)}$
& $\frac{i}{\sqrt{3}}$ & $0$
& $0$ & $0$ & $-\frac{1}{2\sqrt{3}}$ 
\\
$\widehat{D}^{(3,1,-\frac{1}{3})}_{(10,1,1)}$
& $\widehat{H}^{(\overline{3},1,\frac{1}{3})}_{(6,1,1)}$
& $0$ & $-\frac{i\sqrt{2}}{3}$
& $0$ & $-\frac{1}{\sqrt{6}}$ & $0$
\\
$\widehat{D}^{(\overline{3},1,\frac{1}{3})}_{(\overline{10},1,1)}$
& $\widehat{H}^{(3,1,-\frac{1}{3})}_{(6,1,1)}$
& $0$ & $\frac{i\sqrt{2}}{3}$
& $0$ & $-\frac{1}{\sqrt{6}}$ & $0$
\\
\hline \hline
\end{tabular}
\end{center}
\end{table}
%
%
\begin{table}[p]
\caption{CG coefficients for the
$D A$ combination of fields}
\label{CGC22}
\begin{center}
\begin{tabular}{|cc|c|c|}
\hline \hline
$D$ & $A$ & $D\widehat{\Delta}A$ & $D\widehat{\overline{\Delta}}A$ 
\\
\hline \hline
$\widehat{D}^{(1,1,1)}_{(\overline{10},1,1)}$
& $\widehat{A}^{(1,1,-1)}_{(1,1,3)}$
& $0$ 
& $-\frac{i}{10}$ 
\\
$\widehat{D}^{(1,1,-1)}_{(10,1,1)}$
& $\widehat{A}^{(1,1,1)}_{(1,1,3)}$
& $\frac{i}{10}$
& $0$
\\
$\widehat{D}^{(3,1,\frac{2}{3})}_{(6,1,3)}$
& $\widehat{A}^{(\overline{3},1,-\frac{2}{3})}_{(15,1,1)}$
& $-\frac{i}{10}$
& $0$
\\
$\widehat{D}^{(\overline{3},1,-\frac{2}{3})}_{(6,1,3)}$
& $\widehat{A}^{(3,1,\frac{2}{3})}_{(15,1,1)}$
& $0$
& $\frac{i}{10}$
\\
$\widehat{D}^{(3,2,\frac{1}{6})}_{(15,2,2)}$
& $\widehat{A}^{(\overline{3},2,-\frac{1}{6})}_{(6,2,2)}$
& $0$
& $-\frac{i}{10}$
\\
$\widehat{D}^{(\overline{3},2,-\frac{1}{6})}_{(15,2,2)}$
& $\widehat{A}^{(3,2,\frac{1}{6})}_{(6,2,2)}$
& $\frac{i}{10}$
& $0$
\\
\hline \hline
\end{tabular}
\end{center}
\end{table}
%
%
\begin{table}[p]
\caption{CG coefficients for the
$D \Delta$ combination of fields}
\label{CGC23}
\begin{center}
\begin{tabular}{|cc|cc|ccc|}
\hline \hline
$D$ & $\Delta$ & 
$D\Delta\widehat{A}_1$ & $D\Delta\widehat{A}_2$ & 
$D\Delta\widehat{\Phi}_1$ & 
$D\Delta\widehat{\Phi}_2$ & 
$D\Delta\widehat{\Phi}_3$
\\
\hline \hline
$\widehat{D}^{(1,1,-1)}_{(10,1,1)}$
& $\widehat{\Delta}^{(1,1,1)}_{(\overline{10},1,3)}$
& $-\frac{i}{\sqrt{10}}$ & $0$
& $0$ & $0$ & $-\frac{1}{2\sqrt{10}}$
\\
$\widehat{D}^{(1,2,\frac{1}{2})}_{(1,2,2)}$
& $\widehat{\Delta}^{(1,2,-\frac{1}{2})}_{(15,2,2)}$
& $0$ & $\frac{i}{2\sqrt{5}}$
& $0$ & $0$ & $\frac{1}{4\sqrt{30}}$
\\
$\widehat{D}^{(1,2,-\frac{1}{2})}_{(1,2,2)}$
& $\widehat{\Delta}^{(1,2,\frac{1}{2})}_{(15,2,2)}$
& $0$ & $-\frac{i}{2\sqrt{5}}$
& $0$ & $0$ & $\frac{1}{4\sqrt{30}}$
\\
$\widehat{D}^{(1,2,\frac{1}{2})}_{(15,2,2)}$
& $\widehat{\Delta}^{(1,2,-\frac{1}{2})}_{(15,2,2)}$
& $\frac{i}{2\sqrt{10}}$ & $\frac{i}{\sqrt{15}}$
& $\frac{1}{4\sqrt{15}}$ & $0$ & $\frac{1}{6\sqrt{10}}$
\\
$\widehat{D}^{(1,2,-\frac{1}{2})}_{(15,2,2)}$
& $\widehat{\Delta}^{(1,2,\frac{1}{2})}_{(15,2,2)}$
& $-\frac{i}{2\sqrt{10}}$ & $\frac{i}{\sqrt{15}}$
& $\frac{1}{4\sqrt{15}}$ & $0$ & $-\frac{1}{6\sqrt{10}}$
\\
$\widehat{D}^{(3,1,-\frac{4}{3})}_{(6,1,3)}$
& $\widehat{\Delta}^{(\overline{3},1,\frac{4}{3})}_{(\overline{10},1,3)}$
& $0$ & $\frac{i}{\sqrt{15}}$
& $0$ & $\frac{1}{6\sqrt{5}}$ & $-\frac{1}{3\sqrt{10}}$
\\
$\widehat{D}^{(\overline{3},1,\frac{1}{3})}_{(6,1,3)}$
& $\widehat{\Delta}^{(3,1,-\frac{1}{3})}_{(6,1,1)}$
& $-\frac{i}{2\sqrt{5}}$ & $0$
& $0$ & $0$ & $\frac{1}{12\sqrt{5}}$
\\
\hline \hline
\end{tabular}
\end{center}
\end{table}
%
%
\begin{table}[p]
\caption{CG coefficients for the
$D \overline{\Delta}$ combination of fields}
\label{CGC24}
\begin{center}
\begin{tabular}{|cc|cc|ccc|}
\hline \hline
$D$ & $\Delta$ &
$D\overline{\Delta}\widehat{A}_1$ & $D\overline{\Delta}\widehat{A}_2$ &
$D\overline{\Delta}\widehat{\Phi}_1$ &
$D\overline{\Delta}\widehat{\Phi}_2$ &
$D\overline{\Delta}\widehat{\Phi}_3$
\\
\hline \hline
$\widehat{D}^{(1,1,1)}_{(\overline{10},1,1)}$
& $\widehat{\overline{\Delta}}^{(1,1,-1)}_{(10,1,3)}$
& $\frac{i}{\sqrt{10}}$ & $0$
& $0$ & $0$ & $-\frac{1}{2\sqrt{10}}$
\\
$\widehat{D}^{(1,2,-\frac{1}{2})}_{(1,2,2)}$
& $\widehat{\overline{\Delta}}^{(1,2,\frac{1}{2})}_{(15,2,2)}$
& $0$ & $-\frac{i}{2\sqrt{5}}$
& $0$ & $0$ & $\frac{1}{4\sqrt{30}}$
\\
$\widehat{D}^{(1,2,\frac{1}{2})}_{(1,2,2)}$
& $\widehat{\overline{\Delta}}^{(1,2,-\frac{1}{2})}_{(15,2,2)}$
& $0$ & $\frac{i}{2\sqrt{5}}$
& $0$ & $0$ & $\frac{1}{4\sqrt{30}}$
\\
$\widehat{D}^{(1,2,-\frac{1}{2})}_{(15,2,2)}$
& $\widehat{\overline{\Delta}}^{(1,2,\frac{1}{2})}_{(15,2,2)}$
& $-\frac{i}{2\sqrt{10}}$ & $-\frac{i}{\sqrt{15}}$
& $\frac{1}{4\sqrt{15}}$ & $0$ & $\frac{1}{6\sqrt{10}}$
\\
$\widehat{D}^{(1,2,\frac{1}{2})}_{(15,2,2)}$
& $\widehat{\overline{\Delta}}^{(1,2,-\frac{1}{2})}_{(15,2,2)}$
& $\frac{i}{2\sqrt{10}}$ & $-\frac{i}{\sqrt{15}}$
& $\frac{1}{4\sqrt{15}}$ & $0$ & $-\frac{1}{6\sqrt{10}}$
\\
$\widehat{D}^{(\overline{3},1,\frac{4}{3})}_{(6,1,3)}$
& $\widehat{\overline{\Delta}}^{(3,1,-\frac{4}{3})}_{(10,1,3)}$
& $0$ & $-\frac{i}{\sqrt{15}}$
& $0$ & $\frac{1}{6\sqrt{5}}$ & $-\frac{1}{3\sqrt{10}}$
\\
$\widehat{D}^{(3,1,-\frac{1}{3})}_{(6,1,3)}$
& $\widehat{\overline{\Delta}}^{(\overline{3},1,\frac{1}{3})}_{(6,1,1)}$
& $\frac{i}{2\sqrt{5}}$ & $0$
& $0$ & $0$ & $\frac{1}{12\sqrt{5}}$
\\
\hline \hline
\end{tabular}
\end{center}
\end{table}
%
%
\begin{table}[p]
\caption{CG coefficients for the
$D \Phi$ combination of fields}
\label{CGC25}
\begin{center}
\begin{tabular}{|cc|c|c|}
\hline \hline
$D$ & $\Phi$ & $D\widehat{\Delta}\Phi$ & 
$D\widehat{\overline{\Delta}}\Phi$
\\
\hline \hline
$\widehat{D}^{(1,1,1)}_{(\overline{10},1,1)}$
& $\widehat{\Phi}^{(1,1,-1)}_{(15,1,3)}$
& $0$ 
& $-\frac{1}{2\sqrt{10}}$
\\
$\widehat{D}^{(1,1,-1)}_{(10,1,1)}$
& $\widehat{\Phi}^{(1,1,1)}_{(15,1,3)}$
& $-\frac{1}{2\sqrt{10}}$
& $0$
\\
$\widehat{D}^{(1,2,\frac{1}{2})}_{(1,2,2)}$
& $\widehat{\Phi}^{(1,2,-\frac{1}{2})}_{(10,2,2)}$
& $-\frac{1}{2\sqrt{30}}$
& $0$
\\
$\widehat{D}^{(1,2,-\frac{1}{2})}_{(1,2,2)}$
& $\widehat{\Phi}^{(1,2,\frac{1}{2})}_{(\overline{10},2,2)}$
& $0$ 
& $-\frac{1}{2\sqrt{30}}$
\\
$\widehat{D}^{(1,2,\frac{1}{2})}_{(15,2,2)}$
& $\widehat{\Phi}^{(1,2,-\frac{1}{2})}_{(10,2,2)}$
& $-\frac{1}{2\sqrt{10}}$
& $0$
\\
$\widehat{D}^{(1,2,-\frac{1}{2})}_{(15,2,2)}$
& $\widehat{\Phi}^{(1,2,\frac{1}{2})}_{(\overline{10},2,2)}$
& $0$ 
& $-\frac{1}{2\sqrt{10}}$
\\
$\widehat{D}^{(3,1,-\frac{1}{3})}_{(6,1,3)}$
& $\widehat{\Phi}^{(\overline{3},1,\frac{1}{3})}_{(15,1,3)}$
& $-\frac{1}{2\sqrt{15}}$
& $0$
\\
\hline \hline
\end{tabular}
\end{center}
\end{table}
\newpage

\end{document}